\documentclass[aps,pra,10pt,superscriptaddress,nofootinbib,longbibliography]{revtex4-1}

\usepackage[english]{babel}
\usepackage[T1]{fontenc} 
 \usepackage{times}

\usepackage{amsthm} 
\usepackage{amssymb} 
\usepackage{amsmath} 
\usepackage[normalem]{ulem}
\usepackage[usenames,dvipsnames]{xcolor}
\usepackage{graphicx}
\usepackage{overpic}
\usepackage{xifthen}
\usepackage{verbatim}
\usepackage{booktabs}
\usepackage{float}

\usepackage{paralist} 

\newtheorem{theorem}{Theorem}
\newtheorem{lemma}[theorem]{Lemma}

\usepackage{tikz}

\usepackage{mathtools} 

\usepackage{hyperref}
\hypersetup{
  colorlinks,
  citecolor=Blue,
  linkcolor=Red,
  urlcolor=Blue}

\usepackage{listings}
\newcommand{\pyline}[1]%
{
\definecolor{number}{gray}{0.6}
\definecolor{keywords}{rgb}{1.0,0.3,0.3}
\definecolor{comments}{rgb}{0.1,0.65,0.1}
\definecolor{strings}{rgb}{0.3,0.0,1.0}
\lstset{language=python,
        morekeywords={switch,case},
        sensitive=true,
        showspaces=false, 
        basicstyle=\ttfamily\small\mdseries,
      	keywordstyle=\bfseries\color{keywords},
 	      commentstyle=\color{comments},
 	      stringstyle=\color{strings},
 	      numbers=left,
 	      numberstyle=\scriptsize\color{number},
 	      stepnumber=1,
 	      breaklines=true,
 	      frame=none,
        showstringspaces=false,
 	      tabsize=1,
 	      xleftmargin=0pt,
 	      xrightmargin=0pt,
 	      aboveskip=\bigskipamount,
      	belowskip=0pt
}
\lstinline{#1}
}

\newcommand{\pycode}[3]%
{
\definecolor{number}{gray}{0.6}
\definecolor{keywords}{rgb}{1.0,0.3,0.3}
\definecolor{comments}{rgb}{0.1,0.65,0.1}
\definecolor{strings}{rgb}{0.3,0.0,1.0}
\lstset{language=python,
        morekeywords={switch,case},
        sensitive=true,
        showspaces=false, 
        basicstyle=\ttfamily\small\mdseries,
      	keywordstyle=\bfseries\color{keywords},
 	      commentstyle=\color{comments},
 	      stringstyle=\color{strings},
 	      numbers=left,
 	      numberstyle=\scriptsize\color{number},
 	      stepnumber=1,
 	      breaklines=true,
 	      frame=none,
        showstringspaces=false,
 	      tabsize=4,
 	      xleftmargin=\bigskipamount,
 	      xrightmargin=\bigskipamount,
 	      aboveskip=\bigskipamount,
      	belowskip=0pt
}
\lstinputlisting[caption=#2, label=#3, frame=tb]{#1}
}
\newcommand{\pyfrag}[5]%
{
\definecolor{number}{gray}{0.6}
\definecolor{keywords}{rgb}{1.0,0.3,0.3}
\definecolor{comments}{rgb}{0.1,0.65,0.1}
\definecolor{strings}{rgb}{0.3,0.0,1.0}
\lstset{language=python,
        morekeywords={switch,case},
        sensitive=true,
        showspaces=false, 
        basicstyle=\ttfamily\small\mdseries,
      	keywordstyle=\bfseries\color{keywords},
 	      commentstyle=\color{comments},
 	      stringstyle=\color{strings},
 	      numbers=left,
 	      numberstyle=\scriptsize\color{number},
 	      stepnumber=1,
 	      breaklines=true,
 	      frame=none,
        showstringspaces=false,
 	      tabsize=4,
 	      xleftmargin=\bigskipamount,
 	      xrightmargin=\bigskipamount,
 	      aboveskip=\bigskipamount,
      	belowskip=\bigskipamount
}
\lstinputlisting[caption=#2, label=#3, frame=tb, firstnumber=#4, firstline=#4,
  lastline=#5]{#1}
}

\usepackage[normalem]{ulem} 
\usepackage{cancel}

\newcommand{\py}{Python}

\newcommand{\cc}{\mathcal{F}}

\definecolor{c0}{HTML}{000000} 
\definecolor{c1}{HTML}{9EC05B} 
\definecolor{c2}{HTML}{3288BD} 
\definecolor{c3}{HTML}{D53E3F} 
\definecolor{c4}{HTML}{E8B42B} 
\definecolor{c5}{HTML}{B82EE6} 
\definecolor{c9}{HTML}{FFFFFF} 

\newcommand{\HLRQI}{H_{\mathrm{LRQI}}}
\newcommand{\OSMPS}{OSMPS}

\newcommand{\mioseventeenx}{\emph{2x(Intel Xeon E5-2680 Dodeca-core) 24 Cores 2.50GHz}} 
\newcommand{\hppaviliondv}{\emph{Intel(R) Core(TM) i7-3610QM CPU @ 2.30GHz, 8 GB RAM}}
\newcommand{\mionineteenx}{\emph{2x(Intel e5-2680 V4) 28 Cores 2.40 GHz}} 

\newcommand{\bpm}{\begin{pmatrix}}
\newcommand{\epm}{\end{pmatrix}}
\newcommand{\ZV}{\overrightarrow{0}}
\newcommand{\ZVT}{\overrightarrow{0}^T}
\newcommand{\ZM}{\overleftrightarrow{0}}

\newcommand{\Tcpu}{T_{\mathrm{CPU}}}

\newcommand{\erhoi}{\epsilon_{\mathrm{local}}}
\newcommand{\erhoij}{\epsilon_{\mathrm{corr}}}
\newcommand{\eener}{\epsilon_{\mathrm{E}}}
\newcommand{\eentr}{\epsilon_{\mathrm{S}}}
\newcommand{\entim}{\epsilon_{N(t)}}

\newcommand{\Tens}[4][]{\ifthenelse{\equal{#1}{}}
   {  
      #2_{#3}^{#4}
   }
   {  
      #2_{#3}^{\lbrack #1 \rbrack \, #4}
   }
}

\makeatletter
\newcommand{\pushright}[1]{\ifmeasuring@#1\else\omit\hfill$\displaystyle#1$\fi\ignorespaces}
\makeatother

\renewcommand{\i}{\mathrm{i}}
\newcommand{\1}{\mathbb{I}}

\newcommand{\ket}[1]{\left|{#1}\right\rangle}
\newcommand{\Lket}[1]{\left|{#1}\right\rangle\!\rangle}

\newcommand{\bra}[1]{\left\langle{#1}\right|}

\newcommand{\braket}[2]{\left\langle #1 \middle| #2 \right\rangle}

\definecolor{daniel}{rgb}{0,.1,1}

\definecolor{kenji}{RGB}{150,00,00} 

\definecolor{lincoln}{RGB}{40,180,40}

\definecolor{ulmcolor}{rgb}{.4,0,.6}

\definecolor{ulmmethod}{rgb}{0,.3,.6}

\definecolor{dangercolor}{rgb}{0.8,0.,0.}

\newcommand{\csm}{Department of Physics, Colorado School of Mines, Golden,
  Colorado 80401, USA}

\begin{document}

\title{Open source Matrix Product States: Exact diagonalization and other
  entanglement-accurate methods revisited in quantum systems}

\author{Daniel Jaschke}
\affiliation{\csm}
\author{Lincoln D.\ Carr}
\affiliation{\csm}


\begin{abstract}

%

  Tensor network methods as presented in our open source Matrix Product States
  software have opened up the possibility to study many-body quantum physics in
  one and quasi-one-dimensional systems in an easily accessible package similar
  to density functional theory codes but for strongly correlated dynamics. Here, we address methods
  which allow one to capture the full entanglement
  without truncation of the Hilbert space. Such methods are suitable for validation of
  and comparisons to tensor network algorithms, but especially useful in the case of new
  kinds of quantum states with high entanglement violating the truncation in
  tensor networks. Quantum cellular automata are one example for such a system,
  characterized by tunable complexity, entanglement, and a large spread
  over the Hilbert space. Beyond the
  evolution of pure states as a closed system, we adapt the techniques for open
  quantum systems simulated via the Lindblad master equation.
  We present three algorithms for solving closed-system many-body time evolution
  without truncation of the Hilbert space. Exact diagonalization methods have
  the advantage that they not only keep the full entanglement but also require
  no approximations to the propagator. Seeking the limits of a maximal
  number of qubits on a single core, we use Trotter decompositions or Krylov
  approximation to the
  exponential of the Hamiltonian. All three methods are also implemented for
  open systems represented via the Lindblad master equation built from local
  channels. We show their convergence parameters and focus on efficient
  schemes for their implementations including Abelian
  symmetries, e.g., $\mathcal{U}(1)$ symmetry used for number conservation in the
  Bose-Hubbard model or discrete $\mathbb{Z}_2$ symmetries in the quantum Ising
  model.
  We present the thermalization timescale in the long-range quantum Ising
  model as a key example of how exact diagonalization contributes to novel
  physics.

\end{abstract}

\maketitle

\tableofcontents

\section{Introduction                                                          \label{ed:sec:intro}}
Recently, exact diagonalization (ED) methods have seemed less important due to
many available simulation methods for many-body quantum systems including
the Density Matrix Renormalization Group (DMRG) \cite{White1992}, Quantum
Monte Carlo methods \cite{Sandvik1991,Prokofev1998}, dynamical mean field
theory \cite{Georges1996}, and the truncated Wigner approach
\cite{Polkovnikov2010,Schachenmayer2015}. The tensor network methods
themselves contain a variety of tailored approaches such as Matrix Product
States (MPS) \cite{Vidal2003,Schollwoeck2011}, Multi-scale Entanglement
Renormalization Ansatz (MERA) \cite{Vidal2008}, Tree Tensor Networks (TTN)
\cite{Shi2006}, and Projected Entangled Pair States (PEPS)
\cite{Verstraete2008}. But with
growing computing resources today, quantum states of larger and larger
systems can be simulated without entanglement truncation. The first limitation
of the simulation is the calculation of the propagator in the full Hilbert
space. This propagator can be approximated. The second limitation is the initial
state, which cannot be chosen as the ground state when systems are too large for
exact diagonalization methods. Quantum computing and quantum information theory
provide us with well-characterized states beyond the class of product or Fock
states. With these techniques, system sizes can be pushed to 27 qubits with
easily available computational resources, i.e., a single core on a computer.
The boundaries of such simulations have recently been pushed to $45$ qubits in
the context of quantum supremacy \cite{Haner2017}. Simulations,
using also quantum circuits, have achieved simulations for $49$ and $56$
qubits \cite{Pednault2017} shortly after, then $64$ qubits \cite{Chen2018}.
On the other hand, such methods become a valuable comparison for
benchmarking tensor network methods when exploring highly entangled
states as recently studied with Quantum Elementary Cellular Automata (QECA)
\cite{Hillberry2016} based on the original proposition in \cite{Watrous1995}, the Quantum Game
of Life \cite{Bleh2012,Vargas2016}, and for open quantum systems \cite{BreuerPetruccione}
among other physically important contexts. We foresee fruitful applications
to the developing field of quantum simulators. These systems exist on a variety
of platforms and form one significant part in the development of quantum technologies
as proposed in the quantum manifesto in Europe \cite{Gibney2016,QManifesto}. With
exact diagonalization methods, quantum simulators with large entanglement have
the possibility at hand to simulate systems up to a modest many-body system equivalent to 27 qubits. Furthermore,
the area law for entanglement \cite{Eisert2010} can be violated for
long-range interactions \cite{Vodola2015}, and tensor network methods are
likely to lose accuracy when the area law does not bound entanglement. Other
possible applications are the emerging field of synthetic quantum matter.

The inclusion of open system methods allows
us to approach thermalization of few-body quantum systems from different aspects,
e.g., exploring whether or not a subsystem follows a thermal state after taking the partial
trace of a pure state. A recent study considers the
case of six $^{87}$Rb atoms \cite{Kaufman2016} and is well within the
scope of exact diagonalization. Research on quantum computing
and quantum information can profit from such simulations provided through our
library, too. For example, one can study mutual information matrices as done
in a similar study with the MPS library \cite{Valdez2017}.
Moreover, exact diagonalization methods are
more accessible to new researchers in the field as they require less knowledge
of the methods and how to tune them with the set of convergence parameters,
e.g., associated with our MPS methods \cite{Wall2012,JaschkeMPS}.
Returning to exact diagonalization methods with the background of the
many-body simulations in tensor networks advances them with regards to various
aspects. For example, the usage of predefined rules to represent the Hamiltonian
not only simplifies the problem but allows one to multiply the Hamiltonian
with a state vector when the matrix of the Hamiltonian itself cannot be represented
due to memory limitations. The Krylov method employs precisely this
feature and overcomes limitations which are present in the Trotter
decomposition with the primary purpose of evolving models with nearest-neighbor
interactions. Starting with a certain number of predefined
rules for the Hamiltonian, sophisticated models can be represented
as well as with hand-built ED codes, but at a much lower cost, avoiding
development overhead. A preliminary version of this library was used for the
momentum-space study of the Bose-Hubbard model in \cite{Valdez2018}. We envision
that open systems will profit from this approach enormously in the future to
provide a variety of different Lindblad channels.

With these motivations in mind, we present three kinds of algorithms retaining
the complete Hilbert space and their efficient implementations for closed
systems. The first
approach calculates the exponential for the propagator exactly. Second, we
choose the Trotter decomposition for nearest-neighbor Hamiltonians allowing
for larger systems due to the local propagators. In the third algorithm, we
approximate the propagated state in the Krylov subspace allowing for the inclusion
of long-range terms in the Hamiltonian. The interfaces work
seamlessly with our package Open Source Matrix Product States (\OSMPS{})
\cite{JaschkeMPS,OSMPSPackage} and
therefore support both $\mathcal{U}(1)$ and $\mathbb{Z}_2$ symmetries in the
system. The measurements are adapted from tensor network methods and
have a many-body focus: For example, we include the bond entropy, a value typically
calculated for tensor networks obtained from the
Schmidt decomposition for two bipartitions of the system. Local observables and
two-site correlators are measured via the reduced density matrices, which
are a more efficient approach in comparison to representing the
observables on the complete Hilbert space.
Beyond closed systems, we include methods for the Lindblad master equation
\cite{Lindblad1976,BreuerPetruccione}, which describes
the dynamics of an open quantum system weakly coupled to its environment.
Methods such as the matrix exponential, Trotter decomposition, or Krylov
approximation can be adapted to density matrices. Quantum trajectories (QT)
\cite{Dalibard1992,Dum1992} are another approach to simulate the Lindblad
master equation. Quantum trajectories sample over a variety of pure state
systems and therefore all of our three time evolution methods can be used
for this approach.

In this Article, we discuss our versatile implementation of methods without entanglement
truncation and show their convergence. The scaling of computational resources
is compared for all three time evolution methods, where each method has its
advantage: matrix exponentials avoid errors in the propagator; Trotter
decompositions are ideal for nearest-neighbor interactions; and Krylov overcomes
the limitation of the matrix exponential on the complete space without being
limited to nearest-neighbor Hamiltonians. These approaches are applicable to
various systems. Exact diagonalization can simulate a significant fraction
of the number of ions of recent cold ions quantum simulator experiments,
i.e., around one-tenth of \cite{Bohnet2016}. This fraction is much
closer than, e.g., in systems
with ultracold atoms in optical lattices, where thousands to
millions of atoms \cite{Tang2015} are used in quantum simulators; the gap
between ED methods and the experimental number of particles is then much
larger. Looking at individual control of trapped ions, the
state-of-the-art experiments are simulating $53$ ions \cite{Zhang2017} and
ED methods are even more relevant. Moreover, the
development of new platforms for a universal quantum computer is likely to be
tested on small scales first, and the corresponding numerical methods can
contribute to a better understanding of those platforms.
Thus, we envision multiple applications of our exact diagonalization codes
beyond pure quantum computation and quantum information research. For
example, exact diagonalization can simulate systems according
to the full-spectrum Lindblad equation, where we consider a tensor network
implementation of the same technique as prohibitive. We show that the
thermalization timescale can both increase or decrease with the strength
of the long-range interactions in the long-range quantum Ising model.

We introduce the construction of the system
Hamiltonian in Sec.~\ref{ed:sec:statics}. In
Sec.~\ref{ed:sec:dynamics}, we discuss the three different time evolution schemes
before turning to the implementation of measurements in Sec.~\ref{ed:sec:meas}.
We give an overview of the convergence of these methods in Sec.~\ref{ed:sec:conv}
and discuss the necessary modification with regards to open quantum system
according to the Lindblad master equation in Sec.~\ref{ed:sec:open}. We
present the thermalization timescale with the full spectrum Lindblad master
equation for the long-range quantum Ising model as key application in
Sec.~\ref{ed:app:fullspec}. After a benchmarking study in
Sec.~\ref{ed:sec:benchmarking} of our package versus another standard
package, i.e., QuTip, we conclude in Sec.~\ref{ed:sec:conclusion}.

\section{Construction of Hamiltonians and Statics                              \label{ed:sec:statics}}

We first introduce the definition of systems in \OSMPS{}
and explain the exact diagonalization statics. The focus in this section
is on closed quantum systems governed by the Schr\"odinger equation
\begin{eqnarray}                                                                \label{ed:eq:schroedinger}
  \frac{\partial}{\partial t} \ket{\psi(t)}
  &=& - \frac{\i}{\hbar} H \ket{\psi(t)} \, ,
\end{eqnarray}
where $\ket{\psi}$ is the wave function depending on the time $t$. $H$
represents the Hamiltonian of the system. We consider in the following a
generic Hamiltonian with different terms defined as
\begin{eqnarray}                                                                \label{ed:eq:genericH}
  H &=& \underbrace{\sum_{\zeta=1}^{n_{S}} \sum_{k=1}^{L} c_k O_{k}^{[\zeta]}}_{\mathrm{Local~term}}
        + \underbrace{\sum_{\zeta=1}^{n_{B}} \sum_{k=1}^{L-1} c'_k O_{k}^{[B,\zeta]} O_{k+1}^{'[B,\zeta]}}_{\mathrm{Nearest-neighbor~term}}
        + \underbrace{\sum_{\zeta=1}^{n_{i}} \sum_{k=1}^{L-1} \sum_{k'=k+1}^{L} c_{k',k} O_{k}^{[I,\zeta]} O_{k'}^{'[I,\zeta]}}_{\mathrm{Long-range~term}} \, ,
\end{eqnarray}
where $L$ is the size of the one-dimensional system. We take a look at
the different terms one-by-one. The first term represents
a local Hamiltonian term acting on a single particle or site in a system. The first
sum over $\zeta$ allows us to have $n_{S}$ different local terms. The second sum runs
over the different sites in the system, and each local term might have a spatially dependent
coupling with the constants $c_k$. $O_{k}^{[\zeta]}$ is the corresponding
operator acting on site $k$; actually, identities are padded on the sites
$k' \neq k$ via the outer product, or Kronecker product, to represent the
operator acting on the full Hilbert space. This term is necessary for many different
systems. It represents the interaction with an external field in the quantum
Ising model, and encodes the on-site interaction and chemical potential in the
Bose-Hubbard model, just to name two examples. In detail, we take a look at the
Bose-Hubbard model with the Hamiltonian
\begin{eqnarray}                                                                \label{ed:eq:HBH}
  H_{\mathrm{BH}} &=&
  - J \sum_{k=1}^{L-1} \left(b_{k} b_{k+1}^{\dagger} + h.c. \right)
  + \frac{U}{2} \sum_{k=1}^{L} n_{k} (n_{k} - \1)
  - \mu \sum_{k=1}^{L} n_{k} \, ,
\end{eqnarray}
where $b$ and $b^{\dagger}$ are the bosonic annihilation and creation
operators and the number operator $n$ is defined as $n = b^{\dagger} b$. $J$
is the tunneling energy, and $U > 0$ is the repulsive interaction energy on
each site. Thus, the sum over the local terms runs over $\zeta \in \{1, 2\}$.
The first operators $O_{k}^{[\zeta = 1]} = n_{k} (n_{k} - \1)$ is a compound
operator combining the number operators and the identity, all acting on site
$k$. The chemical potential is the second local term, i.e.,
$O_{k}^{[\zeta = 2]} = n_{k}$. The next term is nearest-neighbor
interactions, where the operator $O_{k}^{[B,\zeta]}$ acting on site $k$ and
$O_{k+1}^{'[B,\zeta]}$ on site $k + 1$ can be different operators; hermiticity
has to be fulfilled for all terms together. Looking at the example of the
Bose-Hubbard in Eq.~\eqref{ed:eq:HBH}, the tunneling term is a nearest-neighbor
term. Specifically, we have $\zeta \in \{1, 2\}$ for the bond term with
$O_{k}^{[B,\zeta=1]} = b_{k}$, $O_{k+1}^{[B,\zeta=1]} = b_{k+1}^{\dagger}$,
$O_{k}^{[B,\zeta=2]} = b_{k}^{\dagger}$, and $O_{k+1}^{[B,\zeta=2]} = b_{k+1}$.
The last term is a two-site interaction at arbitrary distance governed by
a coupling $c_{k',k} = c(k' - k)$ depending only on the distance between
the sites. This term extends many models where the
nearest-neighbor interaction is an approximation, and a more accurate Hamiltonian
includes further terms. The extended Bose-Hubbard model with long-range tunneling
is one example. In the following, we refer to the three terms as site rules,
bond rules, and infinite function rules. Appendix~\ref{ed:app:rules} contains a
complete description of more possible terms in a Hamiltonian.
We need to build the Hamiltonian on the complete Hilbert space to determine
the statics. The same applies to the dynamics if using the complete matrix
for taking the exponential or for the Krylov method. For systems
without symmetry, this construction can be done purely using the Kronecker product. Using \py{}
for our implementation, this Kronecker product is either done with \texttt{numpy.kron}
when building a dense Hamiltonian or with \texttt{scipy.sparse.kron}
aiming for a sparse Hamiltonian. Dense methods allow one to calculate the complete
set of eigenvalues and vectors while sparse methods calculate only the
eigenvalues and eigenvectors close to the ground state for the static case.
Considering the entries in the final matrix in case of a site rule presented
in Eq.~\eqref{ed:eq:genericH}, the computational scaling $\cc$ is then
\begin{eqnarray}                                                                \label{ed:eq:OH}
  \cc_{\mathrm{site,dense}} &=& \mathcal{O}(d^{2L}) \, , \qquad \qquad
  \cc_{\mathrm{site,sparse}} = \mathcal{O}(d^{L + 1}) \, ,
\end{eqnarray}
where $\cc$ is the number of floating point operations.
We introduced $d$ as the local dimension of the Hilbert space of a single
site, e.g., for the Ising model $d=2$. The scaling in Eq.~\eqref{ed:eq:OH}
includes the number of multiplications, but not any overhead keeping
track of the sparse matrices. Similar calculations hold for the remaining
rules defined in Eq.~\eqref{ed:eq:genericH}.

The picture gets more complicated when introducing symmetries. We support
$\mathcal{U}(1)$ and $\mathbb{Z}_2$ symmetries defined by their diagonal
generator defined on the local Hilbert space. For example, the Bose-Hubbard
model conserves the number of particles $-$ a $\mathcal{U}(1)$ symmetry $-$
and the simulation can be executed in the symmetry sector with $N$ bosons.
The eigenstates of the basis in the symmetry sector can either be passed as an
argument, or the algorithm iterates once over all $D = d^L$ states checking
if the state obeys the symmetry, which is possibly slow. Either way, we have
a reduced basis of $D_S$ basis states
\begin{eqnarray}
  | i \rangle = | i_1, i_2, \ldots, i_L \rangle \, , \qquad
  i \in \left\{ 1, \ldots, D_s \right\} \, ,
\end{eqnarray}
where $i_{k}$ are the corresponding indices in the local Hilbert space.
The complete wave function is written as $\ket{\psi} = \sum_{i} C_i \ket{i}$
with coefficients $C_i$ fulfilling the normalization constraint of the
wave function.
Neither can the local basis $| i_k \rangle$ be deduced easily from
$| i \rangle$ nor $| i \rangle$ from a set of $| i_k \rangle$, because
the complete Hilbert space is not a tensor product
of the local Hilbert spaces. Therefore, a \py{} class \texttt{SymmetrySector}
provides two key functions:
\begin{itemize}
\item{A dictionary returns the index $| i \rangle$ providing the tuple of all
  local dimensions as key. Since dictionaries in \py{} have an underlying hash
  table, the lookup is fast. Henceforward we use the class method
  \texttt{\_\_getitem\_\_} for getting dictionary entries in \py{}, e.g.,
  \texttt{SymmSec[(0, 0, 1)]}.}
\item{\texttt{SymmetrySector} provides the local indices for the state
  $| i \rangle$. This implementation is achieved with a simple matrix
  $\mathcal{B}$. The
  $i^{\mathrm{th}}$ row contains the local indices for state $| i \rangle$.
  The vector of indices is obtained via the \texttt{\_\_call\_\_} attribute
  as \texttt{SymmSec(i)}.}
\end{itemize}
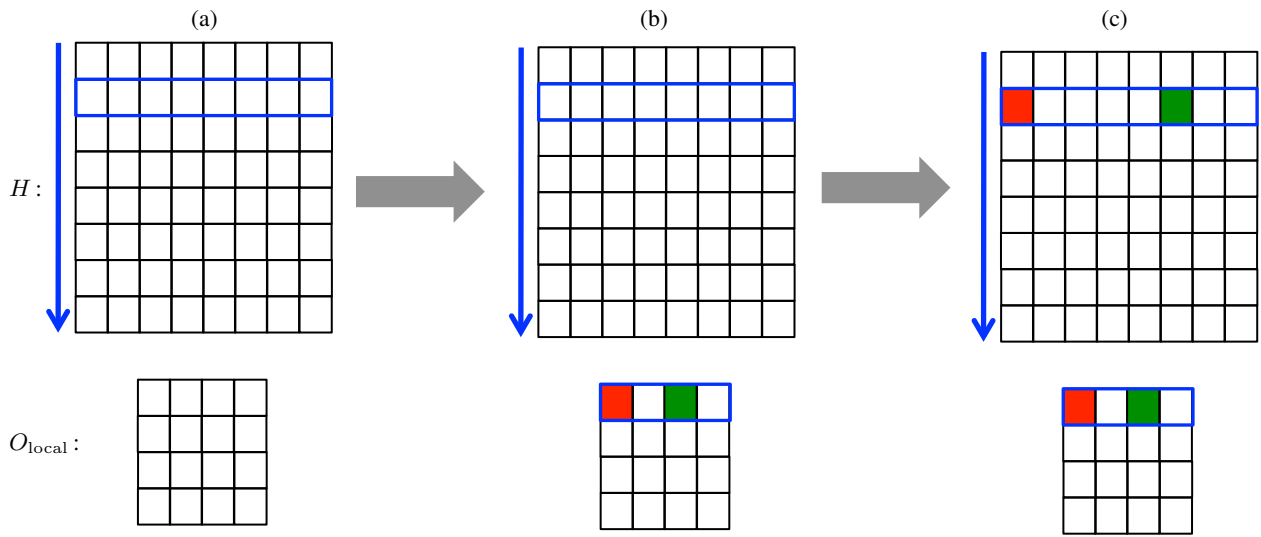
\begin{figure}[t]
  \begin{center}
    \vspace{1.0cm}
    \begin{overpic}[width=0.9 \columnwidth, unit=1mm]{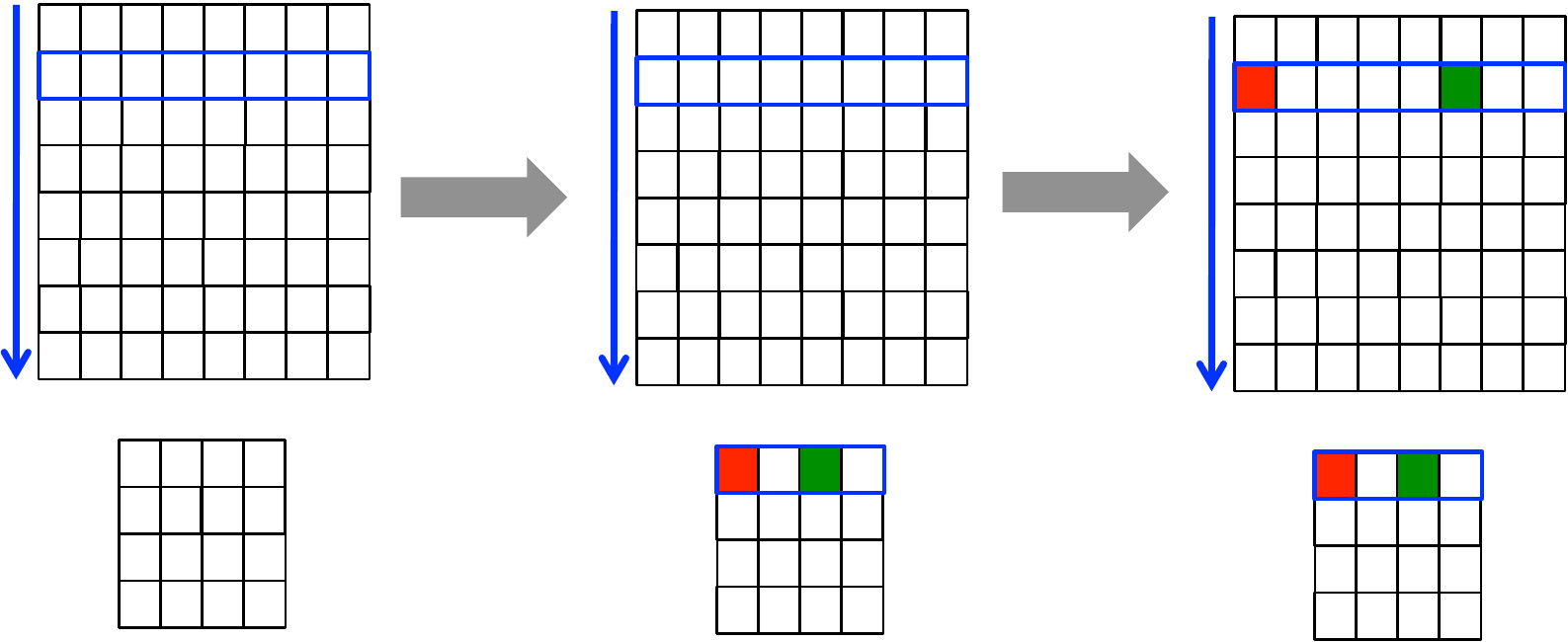}
      \put( 12, 42){(a)}
      \put( 49, 42){(b)}
      \put( 87, 42){(c)}
      \put( -3, 28){$H \,$:}
      \put( -3,  7){$O_{\mathrm{local}} \,$:}
    \end{overpic}
    \caption[Building the Hamiltonian with Symmetries.]
    {\emph{Building the Hamiltonian with Symmetries.} The process of
      obtaining the Hamiltonian $H$ with symmetry for a site rule can be split
      into three steps.
      (a) To build the Hamiltonian, we loop over the rows of $H$.
      (b) For each row, we find the corresponding row in the local matrix
      $O_{\mathrm{local}}$ defining the term in the rule set. We mark the
      non-zero entries in this row. (c) Starting from the non-zero entries
      marked in step~(b), we find the corresponding columns in $H$ and
      update the matrix element.
                                                                                \label{ed:fig:BuildSymm}}
  \end{center}
\end{figure}
Knowing the basic structure for the symmetries, we continue with the
construction of the Hamiltonian in the example of a site rule. To
set the matrix elements for the local operator $O$ on site $k$, we iterate over
the $D_S$ basis states in the Hamiltonian. We choose to iterate over the rows
$| i \rangle$ and obtain the corresponding local basis states
$| i_1, i_2, \ldots, i_L \rangle$ via the dictionary. We need the index for the
columns identified via $| i' \rangle$ and the local indices
$| i_1', i_2', \ldots, i_L' \rangle$. Considering the structure
$\1 \otimes O_k \otimes \1$, we know that all indices except $k$ stay equal:
$i_{k'}' = i_{k'}, \, \forall k' \neq k$. The index $i_k'$ can be determined
by finding the columns with non-zero entries of row $i_k$ of $O_k$.
Therefore, we loop next over the local dimension $i_k'$ of site $k$ where
$| i_1, i_2, \ldots, i_k', \ldots, i_L \rangle$ corresponds to the entry
in the column. We depict this procedure in Fig.~\ref{ed:fig:BuildSymm}. The
computational time for building a site rule in the complete Hamiltonian
$H$ then scales as
\begin{eqnarray}
  \cc_{\mathrm{Build~}H\mathrm{,site}} &=& \mathcal{O}(L D_S d) \, .
\end{eqnarray}
This scaling can be further improved when iterating only over non-zero elements
in the columns of the local operator $O_k$ instead of all the local dimension
$d$.
The fact of building a sparse or a dense Hamiltonian does not influence
the algorithm apart from setting the matrix elements in a sparse or dense
matrix. The interface for addressing a matrix entry in \py{} is equivalent
for both approaches.

The ground and low-lying excited states can now be calculated via
the corresponding numpy and scipy functions. For dense matrices, we obtain the
full spectrum via \texttt{numpy.linalg.eigh} containing all eigenvalues, while
in the sparse case only the lowest-lying eigenvalues and corresponding
eigenvectors are calculated in \texttt{scipy.sparse.linalg.eigsh}. The ground
state serves as the default initial state for the dynamics if no other state
is specified. At present, the ground state is calculated from the complete
Hamiltonian and therefore large systems are limited through the memory when
representing the corresponding matrix.

\section{Methods for Time Evolution of Closed Systems                          \label{ed:sec:dynamics}}

Having explained the general setup of the system and how to obtain the ground
state, we consider the more important question of time evolution. The first
approach for the time evolution of a quantum state is (1) to calculate the
propagator as a matrix exponential of the Hamiltonian on the complete Hilbert
space. This method requires few modifications with regards to the ground state
where we already have functions to construct the Hamiltonian. For
time-dependent Hamiltonians, the procedure is especially expensive since the
Hamiltonian and its propagator have
to be calculated for every time step. Also, the Hilbert space grows
exponentially for increasing system size $L$. Therefore, we consider two additional
approximations: (2) the Suzuki-Trotter decomposition applies local propagators to the quantum state, and
(3) the Krylov subspace approximates the state directly after applying the
propagator. After a brief overview of two test models and their computational
scaling in Sec.~\ref{ed:sec:modelscaling}, we proceed to describe the three
time evolution methods in Sec.~\ref{ed:sec:expm} to \ref{ed:sec:krylov}.

\subsection{Computational Scaling in the Quantum Ising and Bose-Hubbard Model  \label{ed:sec:modelscaling}}

We introduce the first two models which we use to study the
scaling of CPU time $\Tcpu$ for the different time evolution methods. First,
we consider the 1D quantum Ising model \cite{Ising1925,SachdevQPT} defined as
\begin{eqnarray}                                                                \label{ed:eq:HQI}
  H_{\mathrm{QI}} &=&
  - \sum_{k=1}^{L-1} \sigma_{k}^{z} \sigma_{k+1}^{z}
  - h \sum_{k=1}^{L} \sigma_{k}^{x} \, ,
\end{eqnarray}
which obeys $\mathbb{Z}_{2}$ symmetry. The operators $\sigma_{k}^{x}$ and
$\sigma_{k}^{z}$ are the Pauli matrices acting on site $k$. The interaction
energy is normalized to $1$, and the value of the external field is $h$.
$L$ is the number of sites in the system. Second, we study the Bose-Hubbard
Hamiltonian introduced in Eq.~\eqref{ed:eq:HBH}, where the chemical
potential $\mu$ can be disregarded when we use the $\mathcal{U}(1)$ symmetry
leading to a fixed number of bosons $N = \sum_{k=1}^{L} n_{k}$.

\begin{figure}[t]
  \begin{center}
    \vspace{0.8cm}\begin{minipage}{0.47\linewidth}
      \begin{overpic}[width=1.0 \columnwidth,unit=1mm]{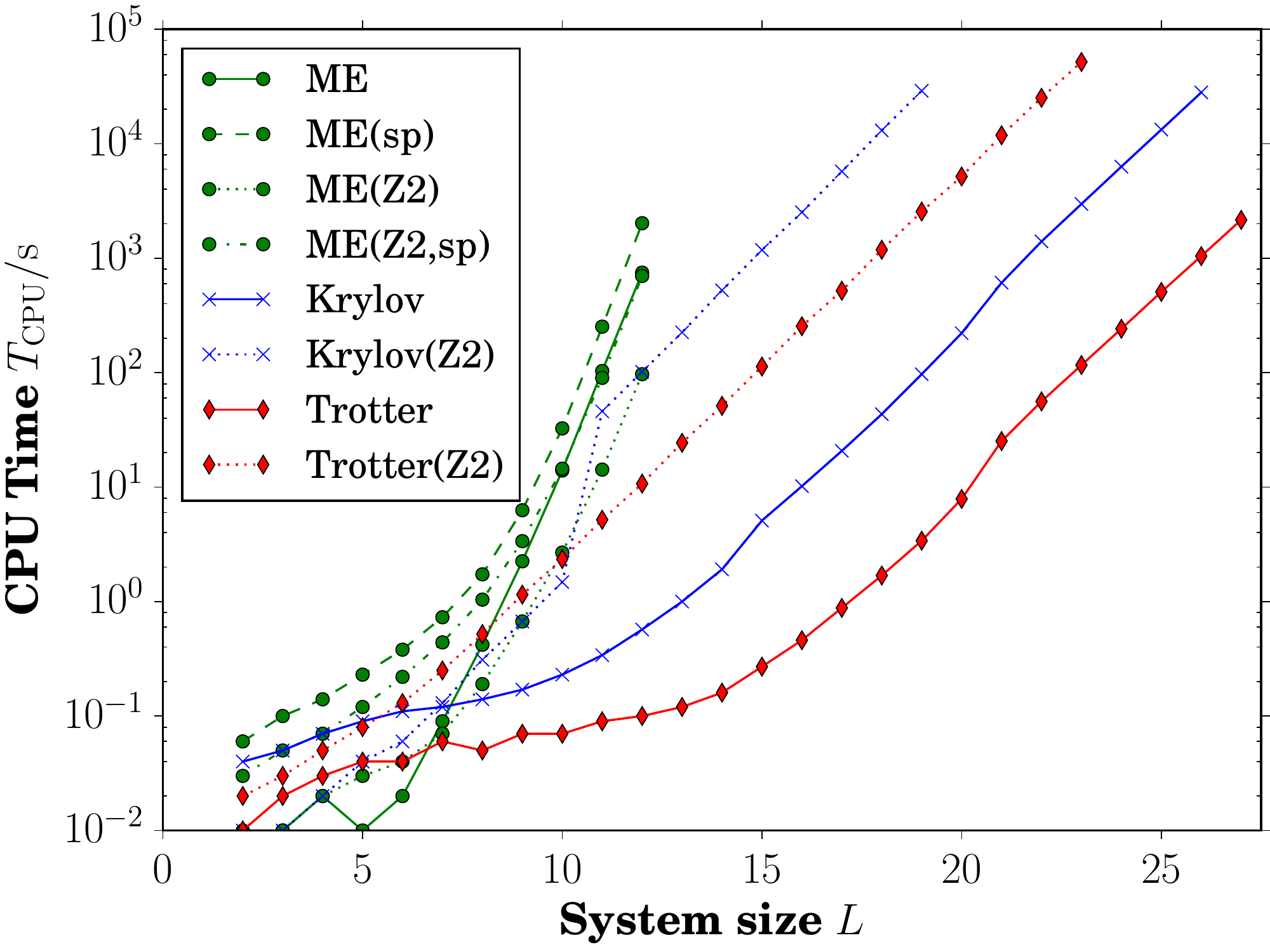}
        \put( 0,80){(a)}
      \end{overpic}
    \end{minipage}\hfill
    \begin{minipage}{0.47\linewidth}
      \begin{overpic}[width=1.0 \columnwidth,unit=1mm]{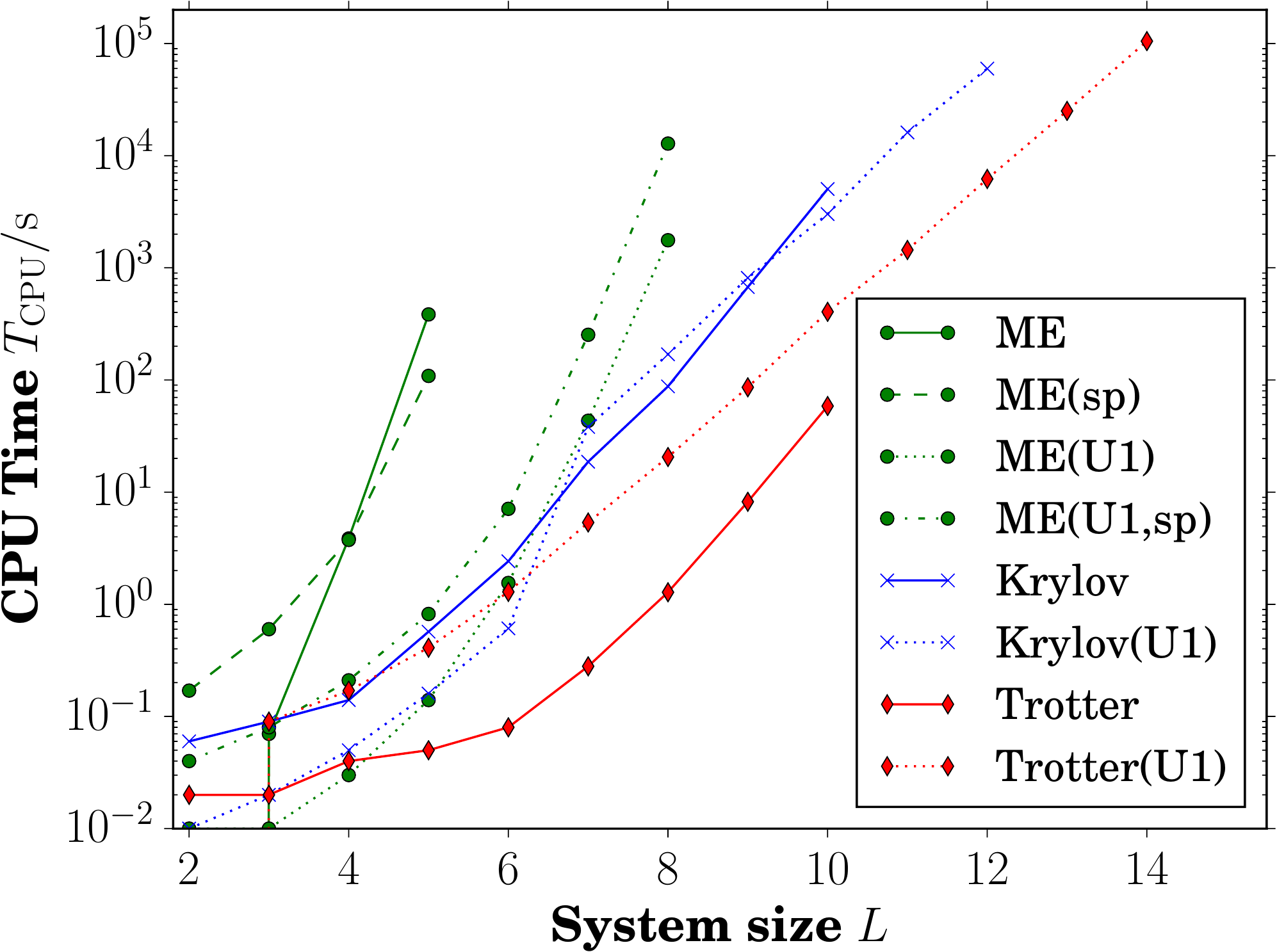}
        \put( 0,80){(b)}
      \end{overpic}
    \end{minipage}
    \caption[Scaling for Time Evolution Methods.]
    {\emph{Scaling for Time Evolution Methods.} We profile five time
      steps for different system sizes $L$ comparing the matrix exponential
      (ME) with sparse matrices (sp) and with dense matrices to the Trotter decomposition
      and to the default Krylov method. The computation time in seconds is plotted on a
      lin-log-scale and all times are taken on a \mionineteenx{}.
      (a)~The nearest neighbor quantum Ising model has a $\mathbb{Z}_{2}$
      symmetry, which we use in comparison to the methods without symmetry.
      The main conclusion is that the $\mathbb{Z}_{2}$ does not accelerate
      simulations except for the matrix exponential where the size of the
      Hamiltonian to be exponentiated is crucial, which is reduced by a
      factor of 2. For the Krylov or Trotter
      method, it is faster to use no symmetries.
      (b)~The Bose-Hubbard model is considered with a local dimension of
      $d = 5$, and the $\mathcal{U}(1)$ symmetry is used with unit filling
      in the simulations marked with $U1$. In contrast to the quantum
      Ising model, use of the $\mathcal{U}(1)$ symmetry leads to a much smaller
      fraction of the total Hilbert space and can accelerate simulations and
      allow for larger system sizes.
                                                                                \label{ed:fig:ising_scal_timestep}}
  \end{center}
\end{figure}

We present the scaling of
the CPU time with increasing system size $L$ for the quantum Ising model in
Fig.~\ref{ed:fig:ising_scal_timestep}~(a). We see that the matrix exponential (ME) is
limited first, where we simulated up to twelve qubits. Within the different
algorithms using the matrix exponential on the complete Hilbert space, the
choice of using dense matrices and $\mathbb{Z}_{2}$ symmetry is optimal.
Sparse methods are labeled with \emph{sp}.
The symmetry reduces the size of the matrix to be exponentiated in the case
of the quantum Ising model by a factor of $2$ and therefore reduces the time
significantly. The different memory requirements themselves are not the restricting
limit. Other simulations have higher memory needs.
For the Suzuki-Trotter or Krylov subspace method the use of
symmetries is slower in case of the Ising model due to matrix-vector
operations, which are needed to decode the basis. In the Krylov method with
$\mathbb{Z}_{2}$ symmetry, we observe the step when switching to a memory-optimized
but slower Krylov method for the 10$^{\mathrm{th}}$ qubit. This memory-optimized
method is necessary for larger systems as the set of Krylov vectors has to be
stored, where each Krylov vector is of the same size as the state vector. The slow
down for the memory-optimized version originates in reading and writing vectors
to and from the hard disk.
The Bose-Hubbard
model with a local dimension of $d = 5$, corresponding to maximally four bosons
per site, is shown in Fig.~\ref{ed:fig:ising_scal_timestep}~(b). Simulations with
$\mathcal{U}(1)$ are at unit filling. Due to the large local dimension
simulations are limited to smaller system sizes. The benefits of using the
symmetry become relevant for the Krylov and Trotter method when going to
large system sizes. The details on the implementation leading to those results
for each method are described in the following three subsections.

\subsection{Time Evolution Method 1: The Propagator as a Matrix Exponential    \label{ed:sec:expm}}

The approach for the time evolution without making any approximations apart
from the time-dependence of the Hamiltonian $H(t)$ and the base numerical
resolution inherent in any floating point computational scheme is to calculate
the propagator as
\begin{eqnarray}
  U(t \to t + dt) = \exp\left(-\i H\left(t + dt / 2 \right) dt \right) \, .
\end{eqnarray}
We take the corresponding functions \texttt{scipy.sparse.linalg.expm}
in the sparse case and \texttt{scipy.linalg.expm} in the dense case. The
construction of $H$ is identical to the static case except that
time-dependent coupling may be introduced via the \OSMPS{} interfaces. We
point out that the coupling constants are estimated for all methods at
mid-time step $t + \frac{dt}{2}$, which corresponds to a time-ordering of
$\mathcal{O}(dt^2)$.
Finally, the state is propagated with a matrix-vector multiplication. The main
problem is that taking the matrix exponential on the complete Hilbert space is
expensive in memory and time. The scaling in memory is quadratic in $D$;
the calculation time of the matrix exponential scales cubically with $D$.
Therefore, computation time limits the system size before memory problems.
We recall the dimension of the full Hilbert space $D$ scales exponentially with
the number of sites $L$. In the following two subsections we explain how to
simplify this exhaustive computational problem with approximate methods.

\subsection{Time Evolution Method 2: Trotter Decomposition for
  Nearest-Neighbor Models                                                      \label{ed:sec:trotter}}

The main bottleneck so far is taking the exponential of a matrix due to CPU time, which
grows exponentially with system size. In its original version, the Trotter
decomposition allows us to approximate the exponential of the Hamiltonian on
the complete Hilbert space with local exponentials on two sites. We
present in the following an efficient scheme for applying these two-site
propagators to the state, starting without symmetry. We point out that the
method in this form is only applicable to Hamiltonians built of site and bond
rules \cite{Wall2012}. The Suzuki-Trotter decomposition \cite{NielsenChuang}
in the first order separates the Hamiltonian into
\begin{eqnarray}                                                                \label{ed:eq:trotter1}
  \exp(- \i H dt)
  &\approx& \exp\left(- \i dt \sum_{k} H_{2k - 1, 2k}\right)
            \exp\left(- \i dt \sum_{k} H_{2k, 2k + 1}\right)                    \nonumber \\
  &=& \exp(- \i dt H_{o}) \exp(- \i dt H_{e}) \, .
\end{eqnarray}
$H_{o}$ contains terms acting on odd sites and their right nearest neighbors,
and likewise in $H_{e}$, except for even sites, with $H = H_{o} + H_{e}$.
We leave the upper limit for the summation intentionally open since it depends
on open versus closed boundary conditions and even vs. odd system sizes
$L$. This decomposition in Eq.~\eqref{ed:eq:trotter1} neglects the second order
terms of $\mathcal{O}(dt^2)$ according to the Baker-Campbell-Hausdorff
equation \cite{NielsenChuang}. We use higher-order Trotter approximations,
see \cite{Schollwoeck2011}, in the implementations defined as
\begin{eqnarray}                                                                \label{ed:eq:trotter2}
  \exp(- \i dt H)
  &\approx& \exp \left(- \i \frac{dt}{2} H_{e} \right)
          \exp(- \i dt H_{o}) \exp \left(- \i \frac{dt}{2} H_{e} \right) \, ,
\end{eqnarray}
for the second order, and
\begin{eqnarray}                                                                \label{ed:eq:trotter4}
  \exp(- \i dt H)
  &\approx& \exp(- \i \tau_3 dt H_{e})
            \exp(- \i \tau_4 dt H_{o})
            \exp(- \i \tau_4 dt H_{e})
            \exp(- \i \tau_4 dt H_{o})                                          \nonumber \\
  &&        \times \exp(- \i \tau_2 dt H_{e})
            \exp(- \i \tau_1 dt H_{o})
            \exp(- \i \tau_2 dt H_{e})                                          \nonumber \\
  &&        \times \exp(- \i \tau_4 dt H_{o})
            \exp(- \i \tau_4 dt H_{e})
            \exp(- \i \tau_4 dt H_{o})
            \exp(- \i \tau_3 dt H_{e})                                          \\
  \tau_3 &=& \frac{1}{2 (4 - 4^{\frac{1}{3}})} \, , \tau_1 = 1 - 8 \tau_3 \, ,
  \tau_2 = \frac{1}{2} (1 - 6 \tau_3) \, , \tau_4 = 2 \tau_3 \, ,               \label{ed:eq:trotter4c}
\end{eqnarray}
for the fourth order, respectively. Other formulations of the fourth order
approximation exist.
We remark that the two-site terms either in $H_{e}$ or $H_{o}$ act on different
sites and therefore commute. Their exponentials can thus be taken separately.
Building the local exponentials, we then distribute the site rules into
$H_{o}$ and $H_{e}$, where we choose to weight them with $1/2$ in each layer.
This weighting applies to periodic boundary conditions, while in systems with open
boundary conditions the first and last site receive a weight of $1$.

\begin{figure}[t]
  \begin{center}
    \vspace{1.0cm}
    \begin{overpic}[width=0.95 \columnwidth,unit=1mm]{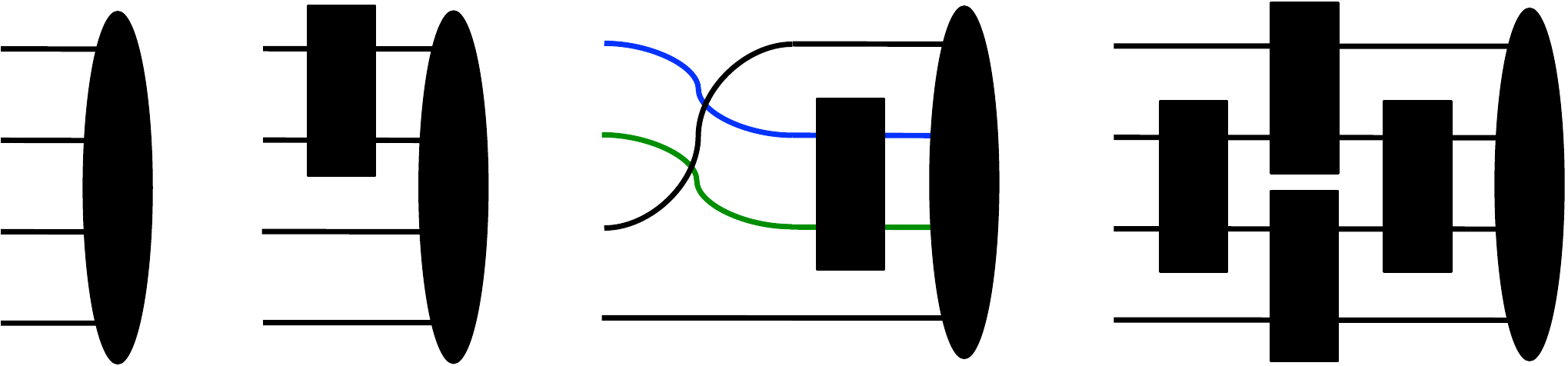}
      \put( 0,25){(a) $\ket{\psi}$}
      \put(16,25){(b) $(P_{1,2} \otimes \1) \ket{\psi}$}
      \put(37,25){(c) \texttt{tensordot(Op, Psi)}}
      \put(71,25){(d) $\ket{\psi(t + dt)}$}
    \end{overpic}
    \caption[Tensor Operations on a State Vector without Symmetries.]
    {\emph{Tensor Operations on a State Vector without Symmetries.}
      (a) The Hilbert space is a tensor product of local Hilbert spaces, and
      therefore we can transform our state vector $| \psi \rangle$ into a
      rank-$L$ tensor.
      (b) The application of a two-site propagator is reduced to a contraction
      with the two corresponding indices at a cost of $\cc = \mathcal{O}(d^{L+2})$.
      (c) The implementation of the \texttt{tensordot} in numpy permutes the
      indices which we have to take into account in the algorithm.
      (d) One step of the second order Trotter decomposition for four sites
      contains the contraction of the propagators, depicted as rectangles.
                                                                                \label{ed:fig:TrotterState}}
  \end{center}
\end{figure}

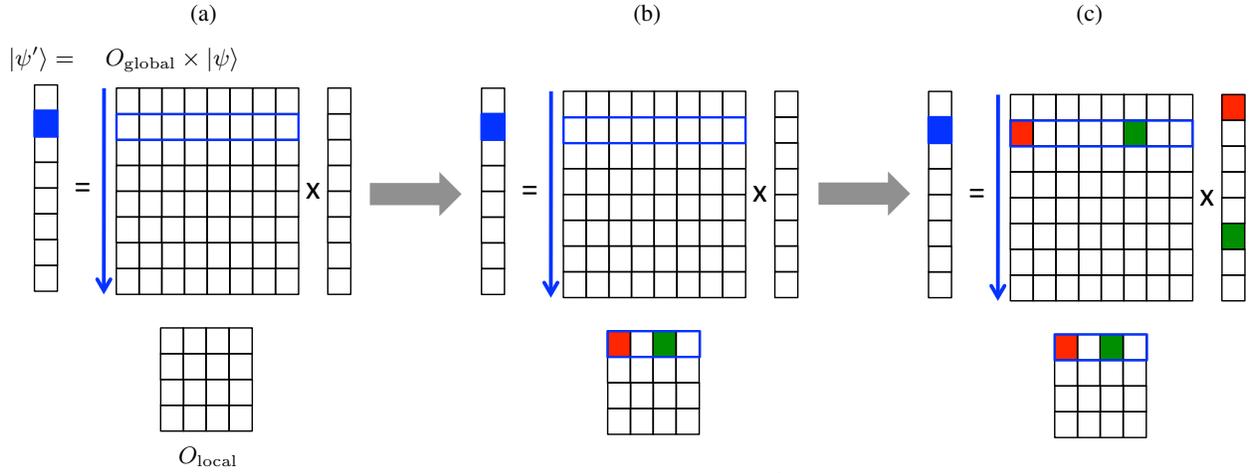
\begin{figure}[t]
  \begin{center}
    \vspace{1.0cm}
    \begin{overpic}[width=0.9 \columnwidth, unit=1mm]{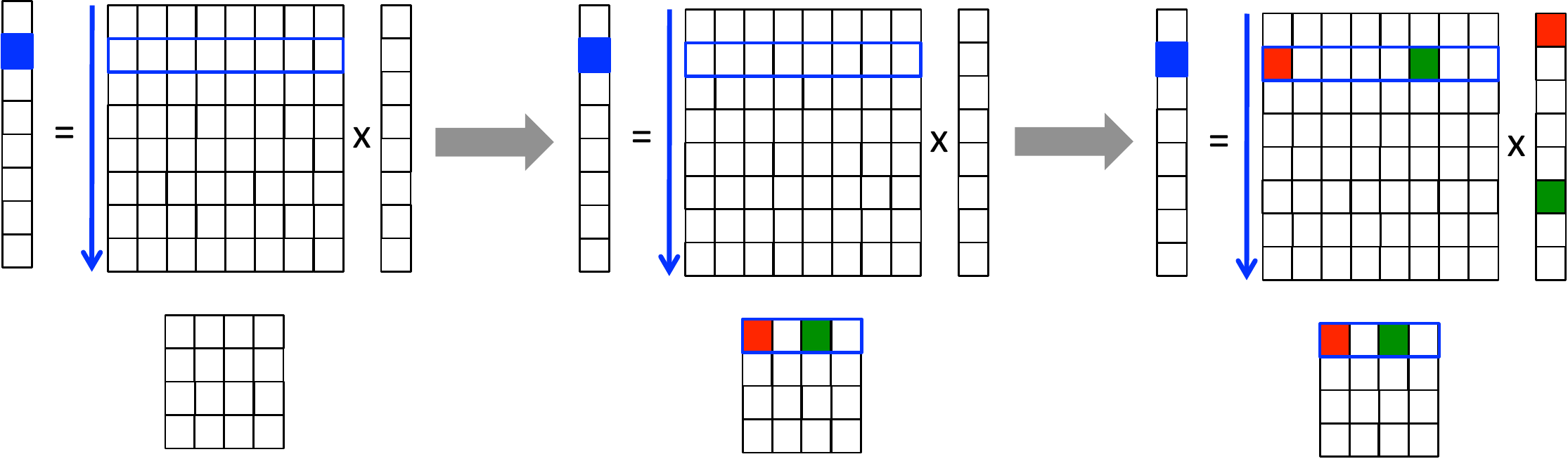}
      \put(  13, 34.5){(a)}
      \put(49.5, 34.5){(b)}
      \put(  86, 34.5){(c)}
      \put(  6,  30.8){$O_{\mathrm{global}} \times \ket{\psi}$}
      \put( 12, -2){$O_{\mathrm{local}} \,$}
      \put(-2,  30.8){$\ket{\psi'} = $}
    \end{overpic}
    \caption[Multiplying the Wave Function by a Local Matrix.]
    {\emph{Multiplying the Wave Function by a Local Matrix.} The
      multiplication $\ket{\psi'} = O_{\mathrm{global}} \ket{\psi}$ is
      carried out in multiple steps, where $O_{\mathrm{global}}$ is defined
      on local Hilbert spaces as operator $O_{\mathrm{local}}$ and padded
      with identities on all other sites. (a) We loop over the rows of the
      matrix $O_{\mathrm{global}}$ and final state $\ket{\psi'}$ and select
      the corresponding row  marking the selected row in blue. (b) In the
      next step, we find the corresponding row in the matrix
      $O_{\mathrm{local}}$ defined on the local Hilbert space and mark the
      non-zero elements (red, green) in this row. (c) In the final, third
      step we find the corresponding column index for each non-zero element
      in $O_{\mathrm{global}}$, which is identical to the row-index of the
      wave function $\ket{\psi}$ to be multiplied. Therefore, we can update
      the blue entry in the new vector $\ket{\psi'}$.
                                                                                \label{ed:fig:MultSymm}}
  \end{center}
\end{figure}

Now we search for an efficient application scheme of these local propagators
for systems without symmetries. A transformation of the state vector into a
rank-$L$ tensor allows one to contract the rank-4 tensor of the local propagator
with the corresponding sites. This approach avoids building up the propagator on the
complete Hilbert space. The function \texttt{numpy.tensordot}, depicted in
Fig.~\ref{ed:fig:TrotterState}~(b) and (c), has a computational scaling
\begin{eqnarray}
  \cc_{\mathrm{contract~2-site~operator}} &=& \mathcal{O}(d^{L+2}) \, .
\end{eqnarray}
In addition, we have to permute the indices of the state vector back into
their original position, which is done once in case of the pure state at
the end of one layer. The sketch for the second order decomposition can be
found in Fig~\ref{ed:fig:TrotterState}~(d), where the different steps described
previously are shown. Considering the second order decomposition, the number
of terms can be approximated by $3L / 2$ leading to an overall scaling of
the method for one time step as
\begin{eqnarray}
  \cc_{\mathrm{Trotter-2~step}} &=& \frac{3 L}{2} \mathcal{O}(d^{L+2}) \, ,
  \qquad
  \cc_{\mathrm{Trotter-4~step}} = \frac{11 L}{2} \mathcal{O}(d^{L+2}) \, .
\end{eqnarray}
A similar result counting the number of layers is shown for the
4$^{\mathrm{th}}$ order. Equation~\eqref{ed:eq:trotter4} shows that we
have eleven layers in comparison to three layers in the second order
represented in Eq.~\eqref{ed:eq:trotter2}. In comparison, systems with
symmetries scale as
\begin{eqnarray}
  \cc_{\mathrm{contract~2-site~operator~symm}} &=& \mathcal{O}(D_S d^2) \, , \quad
  \cc_{\mathrm{Trotter-2~step~symm}} = \frac{3 L}{2} \mathcal{O}(D_S d^{2}) \, .
\end{eqnarray}

The trick of rewriting the wave function as a rank-$L$ tensor does not work with
symmetries since the basis is not a tensor product of local Hilbert spaces.
Here, we take the same approach as when building the Hamiltonian and show a
visualization of the algorithm in Fig.~\ref{ed:fig:MultSymm}. First, we obtain the
two-site Hamiltonian and build the subblocks, which conserve the symmetry.
This reduction to subblocks is a minor improvement on the performance, taking exponentials on a
smaller space and searching for non-zero elements only for the subblocks forming
the propagator. In the matrix-vector multiplication, we loop first over the
rows of the matrix consisting of the $D_S$ basis states. The local propagator
$U_{k, k+1}$ is padded with identity matrices to the left and right, which
do not modify the local basis in $| i' \rangle = | i_1', i_2', \ldots, i_L'
\rangle$ on any site $k' \notin \{k, k+1 \}$. Looping over the non-zero
entries in the columns of the propagator $U_{k, k+1}$ yields the corresponding
local indices for $i_{k}, i_{k+1}$. Then the basis state $|i \rangle$ corresponding
to the column in the full propagator can be easily obtained. The value
is multiplied by the coefficient of $C_{i}$ of $\ket{\psi(t)}$
and added to $C_{i'}$ of the new wave function
$ \ket{\psi'}  = \ket{\psi(t + dt)}$. Therefore, we have two copies
of a wave function $\ket{\psi}$ in memory at the same time.

\subsection{Time Evolution Method 3: Krylov Approximation                      \label{ed:sec:krylov}}

The approach via the Trotter decomposition in Sec.~\ref{ed:sec:trotter} allows one to avoid the matrix
exponential of the complete Hilbert space as long as there are no long-range
interactions. As mentioned before in the introduction in Sec.~\ref{ed:sec:intro}, this is not the case for the exponential
rule, infinite function rule, and the MBString rule. An
alternative is to approximate the exponential of the complete Hilbert
space on a subspace, where the Krylov subspace approach \cite{Moler2003}
seems to be convenient
for this purpose. In detail, it does not approximate the exponential, but
instead directly the propagated state
\begin{eqnarray}
  | \psi(t + dt) \rangle &\approx& \exp(- \i H dt) | \psi(t) \rangle \, .
\end{eqnarray}
The new state is written in terms of the basis vectors $v_{i}$ of the Krylov
subspace
\begin{eqnarray}
  | \psi(t + dt) \rangle \approx \sum_{i=1}^{m} \xi_{i} v_{i} \, ,
\end{eqnarray}
where $v_i$ is $H^{i-1} v_i$ orthogonalized against all previous eigenvectors
$v_j, j < i$, and $\xi_{i}$ is the matrix entry $U_{i,1}^{[\mathrm{Krylov}]}$
in the exponential transformed to the Krylov basis. This approach only
saves computational effort when the number of Krylov vectors $m$ is much
smaller than the dimension of the Hilbert space. The algorithm can be
controlled through convergence parameters. We provide four different modes:
(0) the built-in scipy method based on the complete Hamiltonian is used for
\texttt{edlibmode=0}. (1-3) These modes use the Krylov method of
\OSMPS{}, which can be tuned by the following convergence parameters. On the
one hand, we can set a maximal number of Krylov vectors $m$ building the
basis. On the other hand, the algorithm can check a tolerance
\texttt{lanczos\_tol}, which stops the iterations before reaching $m$.
The modes are distinguished as follows:
\begin{description}
\item[(0)]{Built-in scipy method based on the complete Hamiltonian. The complete
  Hamiltonian is built from the rule sets.}
\item[(1)]{The Hamiltonian is built as a matrix and multiplied with the state vector
  to construct the Krylov method. Being similar to the scipy method, it cannot
  achieve the same performance as the scipy implementation in (0).}
\item[(2)]{The Krylov vectors are constructed from a multiplication of the rule set
  with the state vector and therefore does not need to allocate the memory
  to represent the matrix $H$.}
\item[(3)]{The Krylov vectors are constructed from a multiplication of the rule set
  with the state vector. Instead of keeping all Krylov vectors in memory, this
  mode saves them temporarily on the hard disk to save memory and reads them on
  demand.}
\end{description}
\begin{figure}[t]
  \begin{center}
    \vspace{0.5cm}\begin{minipage}{0.47\linewidth}
      \begin{overpic}[width=1.0 \columnwidth,unit=1mm]{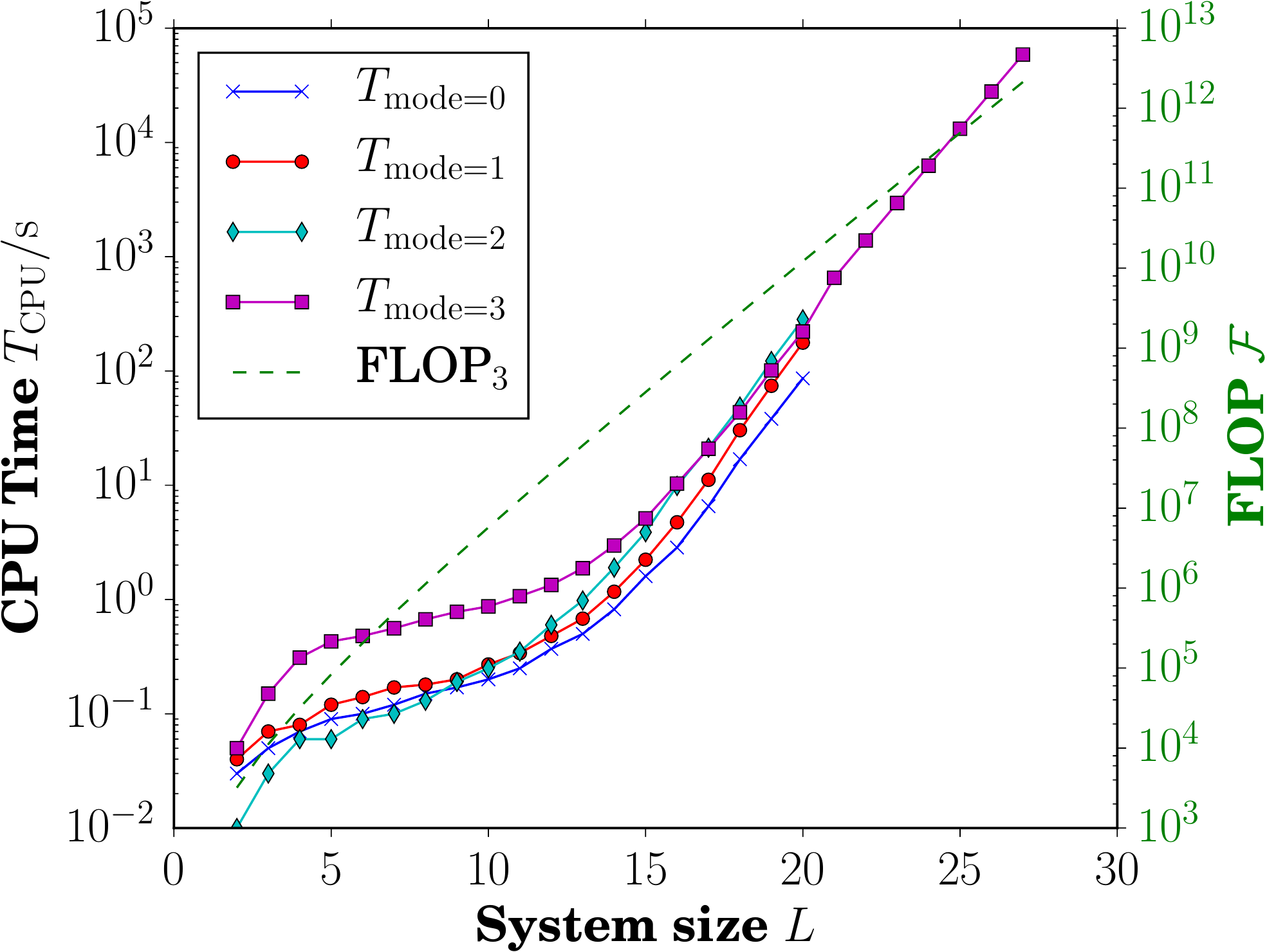}
        \put( 0,80){(a)}
      \end{overpic}
    \end{minipage}\hfill
    \begin{minipage}{0.47\linewidth}
      \begin{overpic}[width=1.0 \columnwidth,unit=1mm]{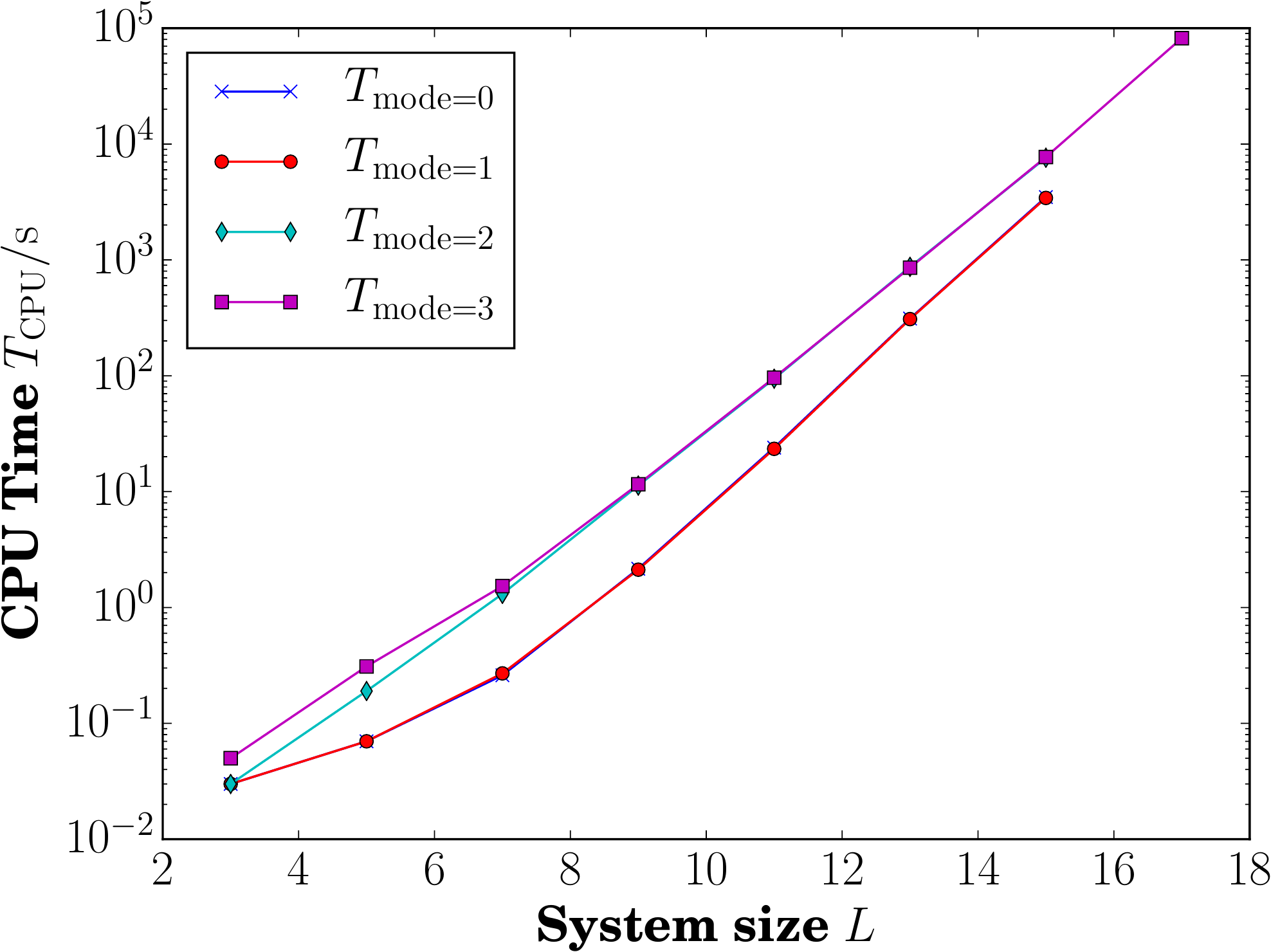}
        \put( 0,80){(b)}
      \end{overpic}
    \end{minipage}\vspace{0.5cm}
    \caption[Scaling of the different Krylov Modes.]
    {\emph{Scaling of the different Krylov Modes.}
      (a) The Ising model without using its $\mathbb{Z}_{2}$ symmetry shows
      that the built-in scipy method with \texttt{edlibmode=0} and
      $T_{\mathrm{mode=0}}$ is fastest except for small systems.
      Large systems are only considered for mode \texttt{edlibmode=3} and we
      show the corresponding scaling of the floating point operations for
      this mode in comparison.  Times are determined on a
      \mioseventeenx{}.
      (b) The Bose-Hubbard model for the double-well problem employs the
      $\mathcal{U}(1)$ symmetry, and we choose a local dimension $d=5$ and
      a filling of $N = (L - 1) / 2$ bosons. The implementations using the
      rule set for the matrix-vector multiplication (modes 2 and 3) are
      slower than building the Hamiltonian and calculating the matrix-vector
      multiplication (modes 0 and 1). Times are determined on a
      \mionineteenx{}.
                                                                                \label{ed:fig:timeskrylov}}
  \end{center}
\end{figure}

Figure~\ref{ed:fig:timeskrylov} shows the CPU times for the different
modes for five time-steps, where we introduce the double-well problem
based on the Bose-Hubbard Hamiltonian introduced in Eq.~\eqref{ed:eq:HBH}. The
potential at the center site splits the system into the two wells,
\begin{eqnarray}                                                                \label{ed:eq:HDW}
  H_{\mathrm{DW}} &=&
  H_{\mathrm{BH}} + V \cdot n_{(L - 1) / 2} \, ,
\end{eqnarray}
which is always considered at a filling with $N = (L - 1) / 2$ particles, $L$ odd. The
fraction of Hilbert space needed in comparison to unit filling is smaller
and allows us to simulate more sites than with unit filling. Mode 2 is preferable over 0, 1
and 3 for small systems from the viewpoint of CPU times in the case of the
quantum Ising Hamiltonian $H_{\mathrm{QI}}$ from Eq.~\eqref{ed:eq:HQI} shown
in Fig.~\ref{ed:fig:timeskrylov}(a). For larger systems, the single matrix-vector
multiplication in mode 0
may have a favorable scaling over the multiple contractions involved for the
rule set in mode 2. For the Bose-Hubbard double-well problem shown in (b),
mode 0 is always preferable. Mode 3 is intended for large systems
and can be scaled up to 27 qubits without encountering memory limitations,
derived below in Eq.~\eqref{ed:eq:OKrylov}.

Apart from the implementation of the Krylov algorithm, we need an efficient
multiplication $H | \psi \rangle$. Seeking limits beyond twenty qubits,
building the Hamiltonian in the Hilbert space is not appropriate. Therefore,
we implement directly the matrix-vector product of a rule set with a vector.
In the case without symmetries, we transform $| \psi \rangle$ into a rank-$L$
tensor and then use \texttt{np.tensordot} in addition to a permutation to obtain the
results similar to the procedure in the Trotter decomposition, which requires
nested loops over every rule and site. The same applies to the case with
symmetries: instead of multiplying the propagator with the state, we apply
the multiplication with the matrices of the Hamiltonian rule sets looping
over all rules sets and sites. Considering the case with $n_s$ site rules and
$n_b$ bond rules without using symmetries, the scaling for $H \ket{\psi}$
is equal to
\begin{eqnarray}                                                                \label{ed:eq:OKrylov}
  \cc_{\mathrm{Rule~set}\times \ket{\psi}} &=&
  L n_s \mathcal{O}(d^{L + 1}) + 2 (L - 1) n_b \mathcal{O}(d^{L + 1})
  \approx L (2 n_b + n_s) \mathcal{O}(d^{L + 1}) \, .
\end{eqnarray}
For one time step, we have to build $m$ Krylov vectors. The leading term in
the computational scaling should be the matrix-vector multiplication;
therefore, we have
\begin{eqnarray}
  \cc_{\mathrm{Krylov~step}} &=& m L (2 n_b + n_s) \mathcal{O}(d^{L + 1}) \, .
\end{eqnarray}
We introduce the number of infinite function rules as $n_i$.
Replacing the $(L - 1)$ interactions of a bond rule with the $L (L - 1) / 2$
interactions of an infinite function rule ($n_i = n_b$), the scaling for one
time steps turns into
\begin{eqnarray}
  \cc_{\mathrm{Krylov~step}}' &=& m L \left[n_s + (L - 1) n_i \right] \mathcal{O}(d^{L + 1}) \, ,
\end{eqnarray}
assuming the phase term is the identity and not contracted as done in the
implementation. For fermionic systems with phase operators, the number of
floating point operations would further increase.


If we compare the Krylov method to the 4$^{\mathrm{th}}$ order Trotter
decomposition for the nearest neighbor case of the quantum Ising model, we
estimate that the Krylov takes longer than the Trotter decomposition by a
factor of
\begin{eqnarray}                                                                \label{ed:eq:Krylov_vs_Trotter}
  \frac{\cc_{\mathrm{Krylov}}}{\cc_{\mathrm{Trotter}}}
 &=& \frac{2 d}{11} m \left(2 n_{b} + n_{s} \right) \mathcal{O}(m n_b) \, .
\end{eqnarray}
Plugging in the actual numbers with $n_{b} = n_{s} = 1$ and $d = 2$, we obtain
finally a factor of $3 m / 11$. We can take the calculation times of
Fig.~\ref{ed:fig:ising_scal_timestep} without $\mathbb{Z}_{2}$ symmetry and
obtain a factor of approximately $30$ between the two methods for large
systems. That would yield $m \approx 110$, which is above the maximal number
specified in the convergence parameters as $100$.

In conclusion, these scaling considerations can only serve
as estimates. But we point out why long-range simulations are so
expensive, according to Eq.~\eqref{ed:eq:Krylov_vs_Trotter}. If we assume that we replace the
nearest-neighbor interactions in the quantum Ising model with long-range
interactions, i.e., we replace the bond rule by one infinite function rule
without phase operators, the resulting long-range model has the following factor
in comparison to the nearest-neighbor Trotter decomposition:
\begin{eqnarray}
  \frac{\cc_{\mathrm{Krylov}}^{\mathrm{exp}}}
       {\cc_{\mathrm{Trotter-4}}^{\mathrm{NN}}}
  &=& \frac{2m}{11d} \left[n_s + (L - 1) n_i \right]
   \approx \frac{2m}{11d} (L - 1) n_i \, ,
\end{eqnarray}
where the approximation in the last steps assumes that the site rule does
not contribute the major part to the total computation. The quantum Ising
model has one non-local rule $n_i = 1$ and a local dimension $d = 2$,
which leads to $m (L - 1) / 11$. Taking, for
example, $L = 18$ and $m = 50$ for the long-range Ising model, one could
calculate in the same compute time the nearest-neighbor Ising model with a Trotter
decomposition for $L=24$, since $m (L - 1) / 11 \approx 77.3 \approx 2^6$.

\section{Efficient Measurements of Pure States                                 \label{ed:sec:meas}}

\begin{figure}[t]
  \begin{center}
    \vspace{0.5cm}\begin{minipage}{0.47\linewidth}
      \begin{overpic}[width=0.98 \columnwidth,unit=1mm]{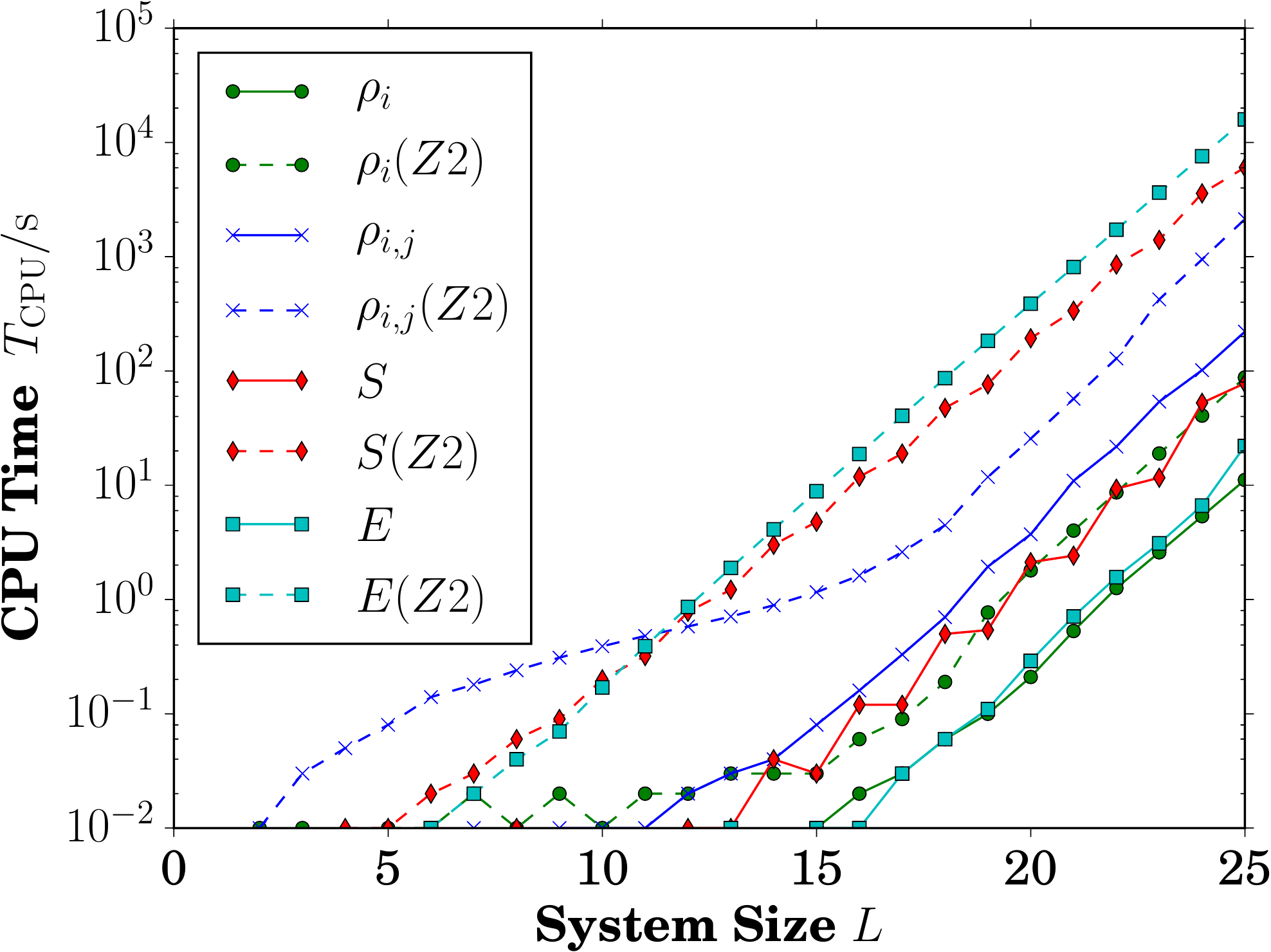}
        \put( 0,80){(a)}
      \end{overpic}
    \end{minipage}\hfill
    \begin{minipage}{0.47\linewidth}
      \begin{overpic}[width=0.98 \columnwidth,unit=1mm]{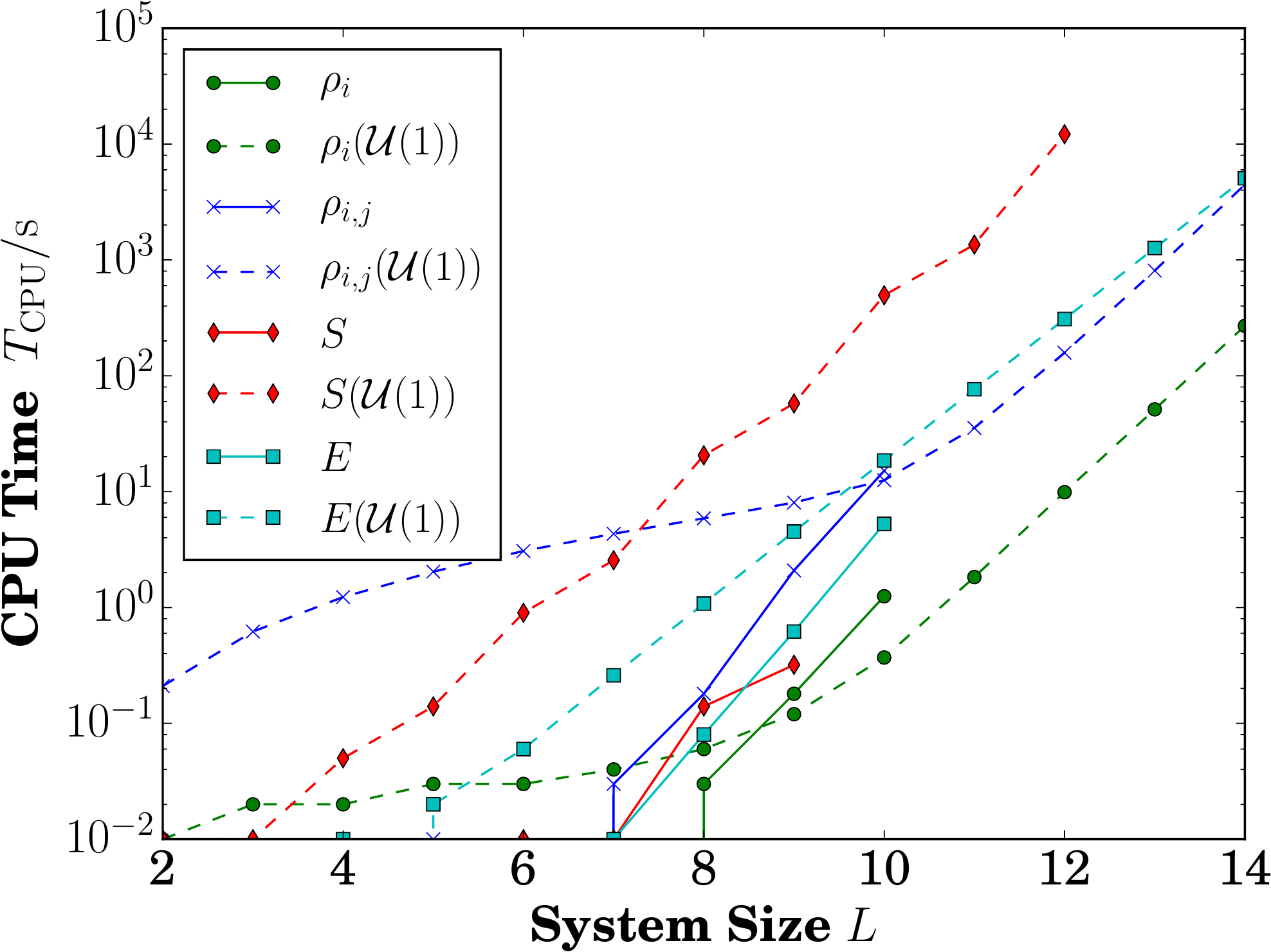}
        \put( 0,80){(b)}
      \end{overpic}
    \end{minipage}\vspace{0.5cm}
    \caption[Scaling for Measurements.]{\emph{Scaling for Measurements.}
      We show CPU times for the construction of local and two-site
      density matrices $\rho_k$ and $\rho_{k,j}$, the bond entropy $S$, and the energy measurement $E$.
      All times are determined on a \mioseventeenx{}.
      (a) We observe from the scaling for the quantum Ising model with and
      without $\mathbb{Z}_{2}$ symmetry that the measurements with symmetry are
      slowed down. The jagged behavior for every second data point in the
      bond entropy originates in the fact that every even system size
      and the next bigger odd system size have their eigenvalue
      decomposition on a matrix of the same dimension, although the odd
      system size has two of them. But the dimension increases again
      for the next even system size. 
      (b) For the Bose-Hubbard model, we show the scaling without symmetry
      and with $\mathcal{U}(1)$ symmetry at unit filling. Due to the local
      dimension of $d = 5$, system sizes are especially limited if the
      $\mathcal{U}(1)$ symmetry is not used.
                                                                                \label{ed:fig:ising_scal_meas}}
  \end{center}
\end{figure}

As our focus is on many-body systems, we draw our attention to the corresponding
measurement procedures. In addition to local observables and correlation
measurements, we are especially interested in the entanglement between two
subsystems known in MPS methods as \emph{bond entropy}. For the local measurement,
we use the reduced density matrix $\rho_k$ for site $k$, tracing out all other sites.
This operation scales as $\mathcal{O}(d^{L+1})$. The actual measurement
$\mathrm{Tr}(O_k \rho_k)$ is negligible with $\mathcal{O}(d^3)$. The first
alternative method is to use the approach of the Trotter decomposition, i.e., contracting
a local operator with the state vector leading to $\ket{\psi'} = (O_k, | \psi \rangle)$.
Then, we calculate the measurement outcome $\braket{\psi}{\psi'}$ at a cost
of $\mathcal{O}(d^{L+1}) + \mathcal{O}(d^L)$. Using the observable $O_k$ padded
with identities in the complete Hilbert space has a leading scaling of
$\mathcal{O}(d^{2L})$ for the measurement $\langle \psi | O_k | \psi \rangle$,
let alone the tensor products to obtain $O_k$. These considerations are
valid for non-symmetry conserving systems. To summarize, local
and two-site density matrices are an efficient approach to measurements. In
addition, we need the bond entropies. The scaling with system size $L$ for
the complete sets of single-site, two-site density matrices, the energy, and
bond entropies are shown in Fig.~\ref{ed:fig:ising_scal_meas}. The CPU times show
that the symmetry-adapted algorithms are slower for all four different
measurements in the case of the quantum Ising model and its $\mathbb{Z}_{2}$
symmetry. The Bose-Hubbard model shows a more diverse picture. In the following,
we explain the details of the algorithms.

%
%
%
%
%
%
%

The simplest way to build the single-site density matrix is to use the function
\texttt{np.tensordot} to contract $\ket{\psi}$ and $\bra{\psi}$ over all indices
not appearing in the reduced density matrix after transforming the vectors into
rank-$L$ tensors. This function is helpful for systems without symmetries,
but we lay out another path to lead to the calculation of reduced density
matrices for systems with conserved quantities. Assuming we calculate the
density matrix of the first site in \py{} starting from a state vector
$| \psi \rangle$, we have the following order in the row-major memory\footnote{
The memory in a computer is linear, but matrices have two dimensions. Row-major
memory stores matrices row-by-row. Column-major memory stores matrices
column-by-column. Each memory order is generalized to higher rank tensors.
Python uses row-major memory order; therefore, we use this order for our
example.}
\begin{eqnarray}
  | \psi \rangle = [ |0\rangle_{1} \otimes | \ldots \rangle_{2, \ldots, L} ]
                   [ |1\rangle_{1} \otimes | \ldots \rangle_{2, \ldots, L} ]
                 = [ A ] [ B ] \, ,
\end{eqnarray}
where we focus on qubits in this example and refer to the blocks as
$[ A ] [ B ]$. The reduced density matrix of the first site is then
\begin{eqnarray}
  \rho_{1} = \begin{pmatrix}
    \sum A A^{\ast} & \sum A B^{\ast} \\ \sum B A^{\ast} & \sum B B^{\ast}
  \end{pmatrix} \, .
\end{eqnarray}
The $^{\ast}$ takes the complex conjugate of the elements.
Each of the entries in $\rho_{1}$ can be calculated efficiently using
the \texttt{numpy.sum} and the
element-wise multiplication. In fact, it is sufficient to calculate the
upper (lower) triangular part of $\rho$ and use the property
$\rho = \rho^{\dagger}$ to fill the lower (upper) triangular part of $\rho$.
Reusing the previous method for $k$ with $k \neq 1$ we choose the following
path to avoid writing specific summations for different sites $k$:
\begin{enumerate}
\item{Reshape the vector $| \psi \rangle$ into a rank-$L$ tensor with
  the local dimension (\texttt{numpy.reshape}).}
\item{Permute the index or indices, which are not traced out, to the front
  (\texttt{numpy.transpose}).}
\item{The reduced density matrix can now be calculated as explained
  for the first site. We point out that BLAS subroutines
  \texttt{scipy.blas.dsyrk} and \texttt{scipy.blas.zherk} allow us to
  calculate symmetric matrices as well. These functions only build the
  upper or lower triangular matrix; they are not used right now, but may
  enhance the computation time for systems without symmetry.}
\end{enumerate}


The previous method does not work directly for the reduced density matrices of
states in a specific symmetry sector. The first possibility is to map the
state vector from the Hilbert space with symmetries back to the complete Hilbert
space without symmetries. This method has the disadvantage that the memory needs
might increase drastically in addition to the fact of multiplying a lot of zeros.
For systems where only a small fraction of the complete Hilbert space is needed,
this  increase in memory imposes a problem. For example, the Bose-Hubbard model at unit filling for
a local dimension of $d = 5$ and fourteen sites corresponds to approximately
$23.67$ qubits. The complete Hilbert space has a dimension of $5^{14}$, which
corresponds to approximately $32.51$ qubits. We can overcome this problem in
two ways. We may use booleans to index the complete Hilbert space into indices
being part of the symmetric Hilbert space (True) and not belonging to the
symmetry sector specified (False).
Thus, we use less memory: a factor of 16 (8) in comparison to complex-valued
(real-valued) states knowing that numpy stores each boolean in a byte, not a
bit. Assuming this approach is not always
enough gain, we use a sparse structure. There are no sparse structures for
tensors or vectors in \emph{scipy}, so we save the state vector in a sparse
matrix. Therefore, the permutation has to be accounted for while writing the
sparse structure. We propose, again in the simple example of qubits and a
single-site density matrix for site $k$, the following steps focusing solely
on the method with the sparse matrix:
\begin{enumerate}
\item{For each basis state $\ket{i}$ we need the index in the Hilbert space
  only considering the traced out or retained sites. For the new Hilbert space
  of the retained sites, we create the vector
  \begin{eqnarray}
    v_{\mathrm{retain}} &=& (0, \ldots, 2^0, \ldots, 0) \, ,
  \end{eqnarray}
  where the entry $2^0$ is at the position $k$. Retaining more than one index,
  the entries for the $l$ retained sites would be $2^{l-1}, \ldots, 2^0$. In
  contrast, the vector for the traced out part of the Hilbert space is
  \begin{eqnarray}
    v_{\mathrm{trace}} &=& (2^{L-2}, 2^{L-3}, \ldots, 2^{L-k}, 0, 2^{L-k-1}, \ldots, 2^0) \, ,
  \end{eqnarray}
  where the entry $0$ is at the position $k$. The indices in the corresponding
  complete subspace disregarding the symmetry can be obtained with the matrix
  $\mathcal{B}$, which contains the basis, via a matrix-vector multiplication:
  \begin{eqnarray}
    i_{1} = \mathcal{B} \cdot v_{\mathrm{retain}} \, , \qquad
    i_{2} = \mathcal{B} \cdot v_{\mathrm{trace}} \, .
  \end{eqnarray}
  $i_{1}$ and $i_{2}$ are vectors. Building the vectors is cheap and scales
  linearly with the number of sites. The expensive part is the two
  matrix-vector multiplications at $\mathcal{O}(D_S L)$.
}
\item{Looping over all basis states $i$ in the reduced density matrix, we create
  for each state $i$ a sparse vector. The indices in the present (symmetry-adapted)
  Hilbert space $i_{\mathrm{match}}$ are identified via the comparison $i_{1} = i$.
  The sparse vector contains the entries of $\psi_{i_{\mathrm{match}}}$ at the
  corresponding indices in the traced out Hilbert space $i_{2}(i_{\mathrm{match}})$.
  If we retain $l$ sites, they approximately have a scaling of
  $\mathcal{O}(d^l D_S)$, where $d^l$ is due to the loop over the basis
  states and $D_S$ corresponds to the comparisons $i_{1} = k$.
}
\item{Looping over all upper triangular entries $\rho_{i,i'}$ of the density
  matrix, each entry is calculated via the dot product of the sparse vectors
  with indices $i$ and $i'$, where we have to take the complex conjugated vector
  for $i'$. Although we loop only over the upper triangular part, the leading
  scaling for the loops is still $\mathcal{O}(d^{2l})$. For the vector-vector
  multiplication, we can only give an upper limit of $\mathcal{O}(D_S)$
  valid for the limit $l=0$ leading overall to $\mathcal{O}(d^{2l} D_S)$.
}
\end{enumerate}
So the overall scaling is $\mathcal{O}(D_S L) + \mathcal{O}(d^{2l} D_S)$. The
sparse structure in scipy is limited to a dimension of $2^{31} - 1$, which
becomes an issue if the complete Hilbert space of the sites traced over exceeds
this value. The example above with fourteen sites with a local dimension of
$d = 5$ exceeds this limit. Instead of using the indices of the complete
Hilbert space, we build a symmetry-adapted basis for the reduced density
matrix and use its indices, which can never exceed the number of basis states
of the original quantum state. A similar algorithm can be formulated if the
resulting density matrix should keep the symmetry-adapted space to the extent
possible.


For the calculation of the bond entropy we employ the following algorithm; we do not
use the characteristic $\rho = \rho^{\dagger}$ so far. Starting in the middle
of the system we create the density matrix for the left bipartition
$1, \ldots, l$ as matrix-matrix multiplication
\begin{eqnarray}
  \rho_{l} = M M^{\dagger} \, ,
\end{eqnarray}
where $M$ is the vector $| \psi \rangle$ reshaped into the dimension
$d^l, d^{L-l}$. In the case of symmetries present, we create $M$ as a sparse
matrix. An eigenvalue decomposition yields the bond entropy. The next reduced
density matrix is obtained via a partial trace
\begin{eqnarray}
  \mathrm{Tr}_{l+1} \rho_{l+1} = \rho_{l} \, .
\end{eqnarray}
This approach should have a favorable scaling with $d^{2l + 1}$ additions
in addition to a permutation to form blocks of the memory along
\texttt{np.transpose((0, 2, 1, 3))}) in regard to calculating the reduced
density matrix for every site with a matrix multiplication scaling with
$d^{L+l}$. The summation for each of the $d^{2l}$ entries runs only over
the $d$ diagonal entries of the memory block. In the case with symmetries,
the scaling for the matrix multiplication does not hold.
%
%
%
%
%
The second step takes care of the right half of the system building an initial,
reduced density matrix for $l + 2, \ldots, L$, that is,
\begin{eqnarray}
  \rho_{l'} = M^{\dagger} M \, .
\end{eqnarray}
In the following we trace out the first site in the reduced density matrix
instead of the last. The permutation with \texttt{np.transpose((1, 3, 0, 2))}
delivers the entries to be summed over in blocks of the memory, although we
recall that only the diagonal elements are summed over. Overall the bond
entropy can be obtained equally well with singular value decomposition of the
matrix $M$ build for each splitting. In the tensor network section of
\OSMPS{}, eigendecompositions are preferred over SVD, so we use them in this
case as well.

\section{Convergence of Time Evolution Methods                                 \label{ed:sec:conv}}

Finally, we are interested in the convergence of those methods, especially in
the error of the Trotter decomposition and Krylov evolution in comparison to
the matrix exponential on the complete Hilbert space (ME). For the convergence
studies we consider the following four kinds of errors taking into account
local measurement, correlation measurements between operators on two sites,
the energy $E$ and the entanglement specified through the bond entropy or Schmidt
entropy $S$. These error measurements are defined as:
\begin{eqnarray}
  \erhoi &=& \max_{k} \mathcal{D}(\rho_k, \rho_k^{\mathrm{ME}}),
  k = 1, \ldots L                                                               \label{ed:eq:erhoi} \\
  \erhoij &=& \max_{(k, j)} \mathcal{D}(\rho_{k,j}, \rho_{k,j}^{\mathrm{ME}}),
  k,j = 1, \ldots L, k < j                                                      \label{ed:eq:erhoij} \\
  \eener &=& | E - E^{\mathrm{ME}} |                                            \label{ed:eq:eener} \\
  \eentr &=& | S(L / 2) - S^{\mathrm{ME}}(L / 2) | \, ,                         \label{ed:eq:eentr}
\end{eqnarray}
where the superscript $\mathrm{ME}$ specifies that the matrix
exponential was used as reference method. The trace distance $\mathcal{D}$
and the bond entropy $S$ are defined as
\begin{eqnarray}
  \mathcal{D} &=& \frac{1}{2} | \rho - \rho' |, \quad |\mathcal{A}| = \sqrt{\mathcal{A}^{\dagger} \mathcal{A}} \\
  S(l) &=& - \sum_j \Lambda_j(l) \log(\Lambda_j(l)) \, .
\end{eqnarray}
$\mathcal{A}$ is a matrix used to define the norm and is replaced in this case with the
difference of the two density matrices $\rho - \rho'$.
$\Lambda_j(l)$ are the eigenvalues of the reduced density matrix of the subsystem
reaching from site $1$ to $l$. An alternative distance measure would be the
infidelity
$\mathcal{I} = 1 - \mathrm{Tr} \sqrt{ \sqrt{\rho_{A}} \rho_{B} \sqrt{\rho_{A}}}$,
but seems to introduce an additional instability taking the square root of a
matrix twice. For the convergence rates, we use
%
\begin{eqnarray}                                                                \label{ed:eq:convrate}
  r &=& \frac{1}{\ln(\alpha)}
        \ln \left( \frac{\epsilon(\alpha dt)}{\epsilon(dt)} \right) \, ,
\end{eqnarray}
where $\alpha$ is the proportion between the size of the time steps for the
two evolutions carried out with time steps $dt$ and $\alpha \cdot dt$. $\epsilon$
is the error, e.g., one of the measures introduced in Eqs.~\eqref{ed:eq:erhoi}
to \eqref{ed:eq:eentr}. We consider the error with regards to the simulations
with the matrix exponential and the smallest time step.
We consider two different models and two scenarios for the dynamics. We
recall that we introduced the definitions of the Ising Hamiltonian
$H_{\mathrm{QI}}$ in Eq.~\eqref{ed:eq:HQI} and the Bose-Hubbard model with
$H_{\mathrm{BH}}$ in Eq.~\eqref{ed:eq:HBH}. The two scenarios considered are
as follows.
\begin{enumerate}
\item{A linear quench in the paramagnetic (Mott insulator) phase for the
  Ising (Bose-Hubbard) model starting at $h=5.0$ ($U=10$)
  and reaching after $0.5$ time units $h=4.5$ ($U=8$). The time is in
  units of the interaction $\sigma^z \sigma^z$ (tunneling strength $J$).}
\item{A sudden quench in the paramagnetic (Mott insulator) phase evolving the
  ground state of the Ising (Bose-Hubbard) model $h=5.0$
  ($U=10$) with a value of $h=4.5$ ($U=8$) for $0.5$ time units.}
\end{enumerate}

\begin{figure}[t]
  \begin{center}
    \vspace{0.8cm}\begin{minipage}{0.47\linewidth}
      \begin{overpic}[width=1.0 \columnwidth,unit=1mm]{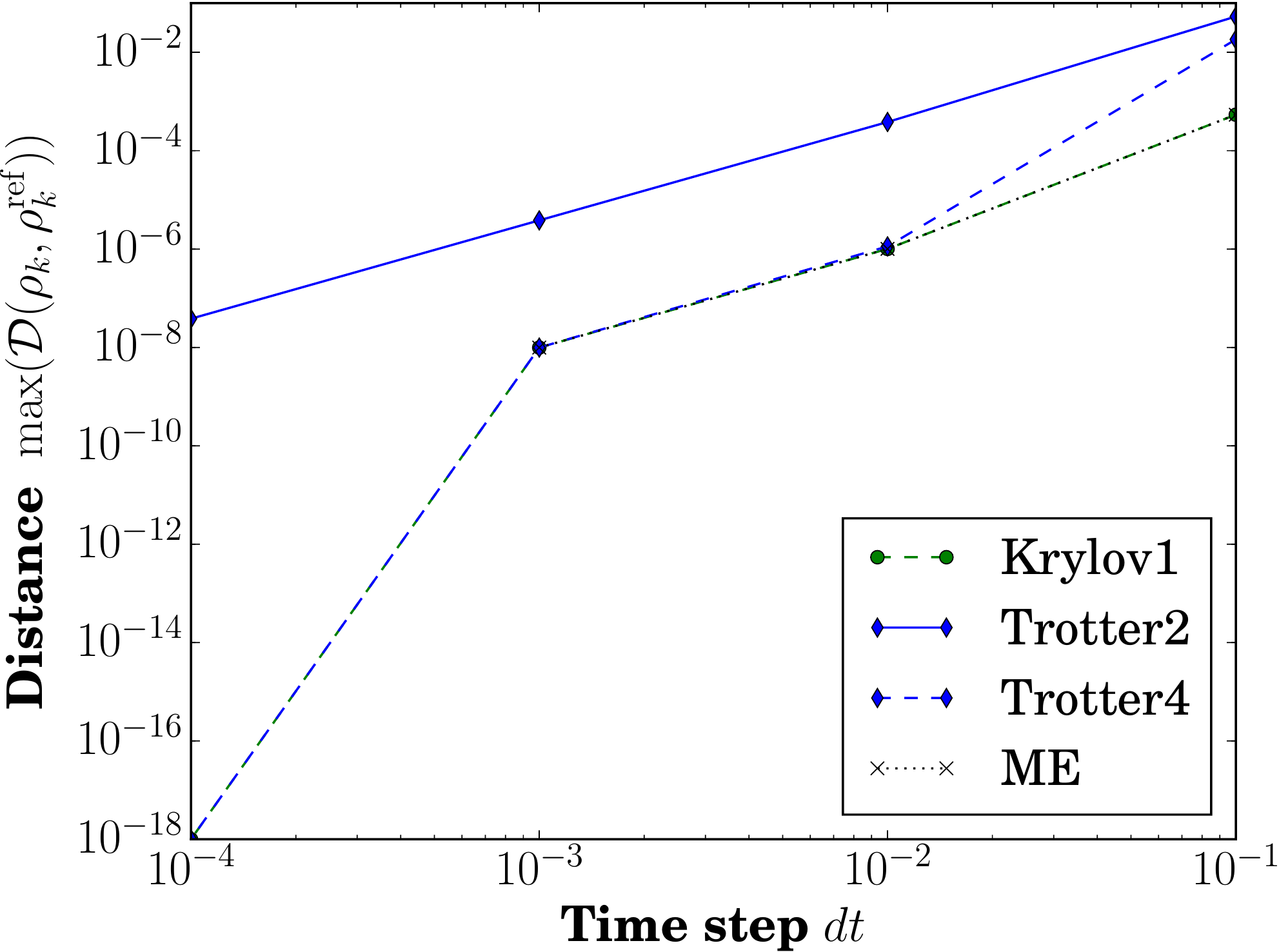}
        \put( 0,78){(a)}
      \end{overpic}
    \end{minipage}\hfill
    \begin{minipage}{0.47\linewidth}
      \begin{overpic}[width=1.0 \columnwidth,unit=1mm]{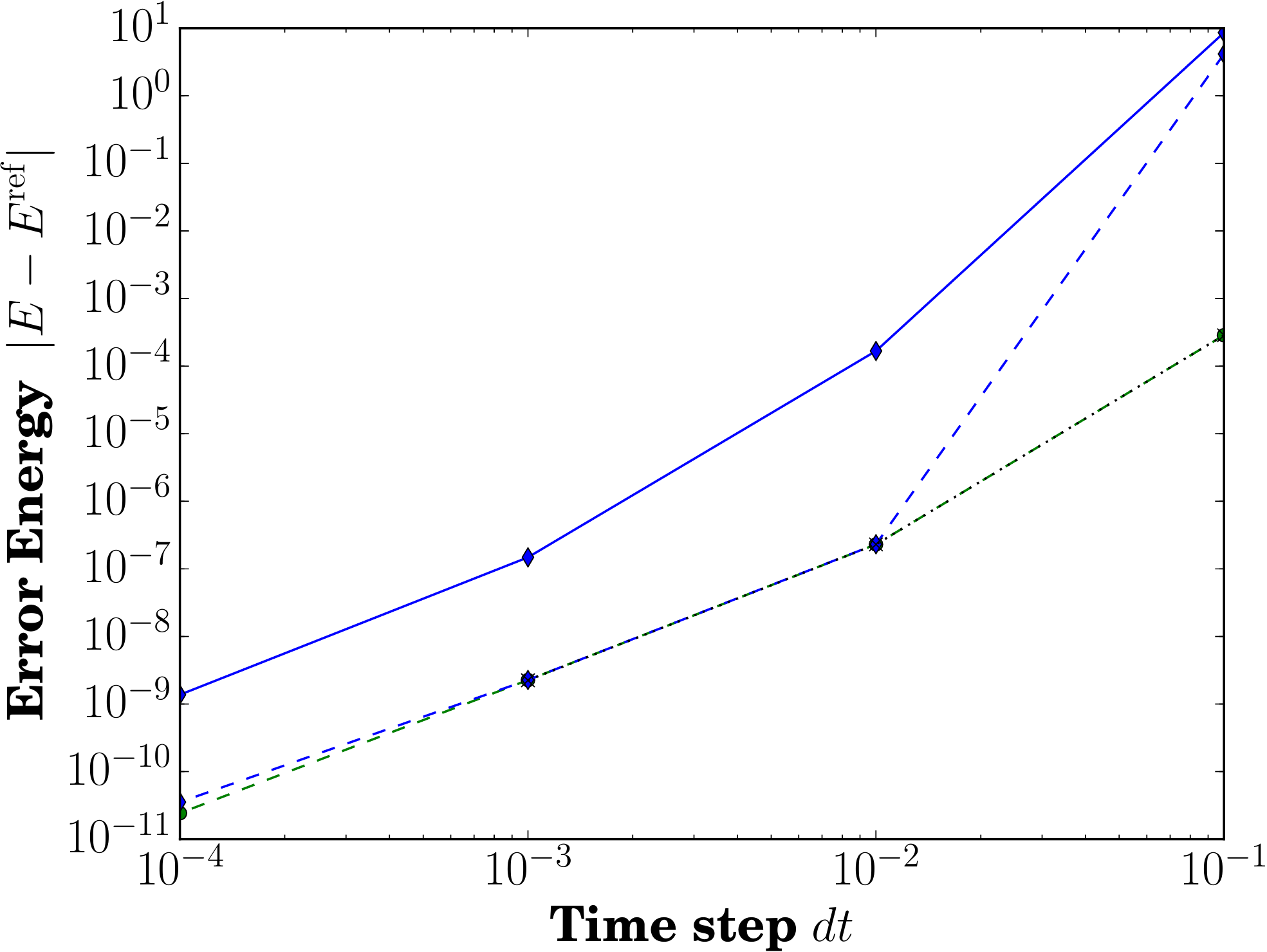}
        \put( 0,78){(c)}
      \end{overpic}
    \end{minipage}\vspace{0.4cm}
    
    \begin{minipage}{0.47\linewidth}
      \begin{overpic}[width=1.0 \columnwidth,unit=1mm]{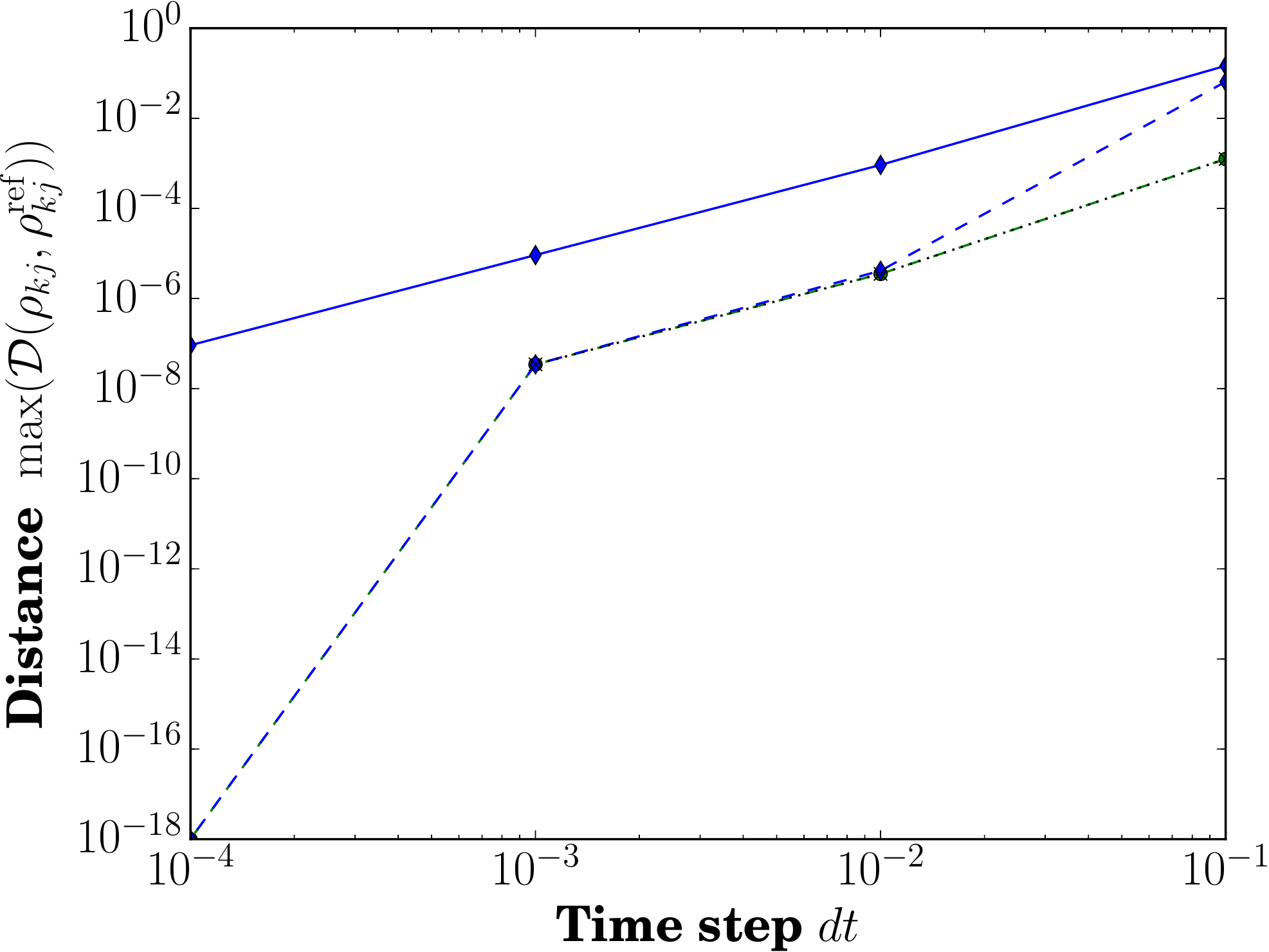}
        \put( 0,78){(b)}
      \end{overpic}
    \end{minipage}\hfill
    \begin{minipage}{0.47\linewidth}
      \begin{overpic}[width=1.0 \columnwidth,unit=1mm]{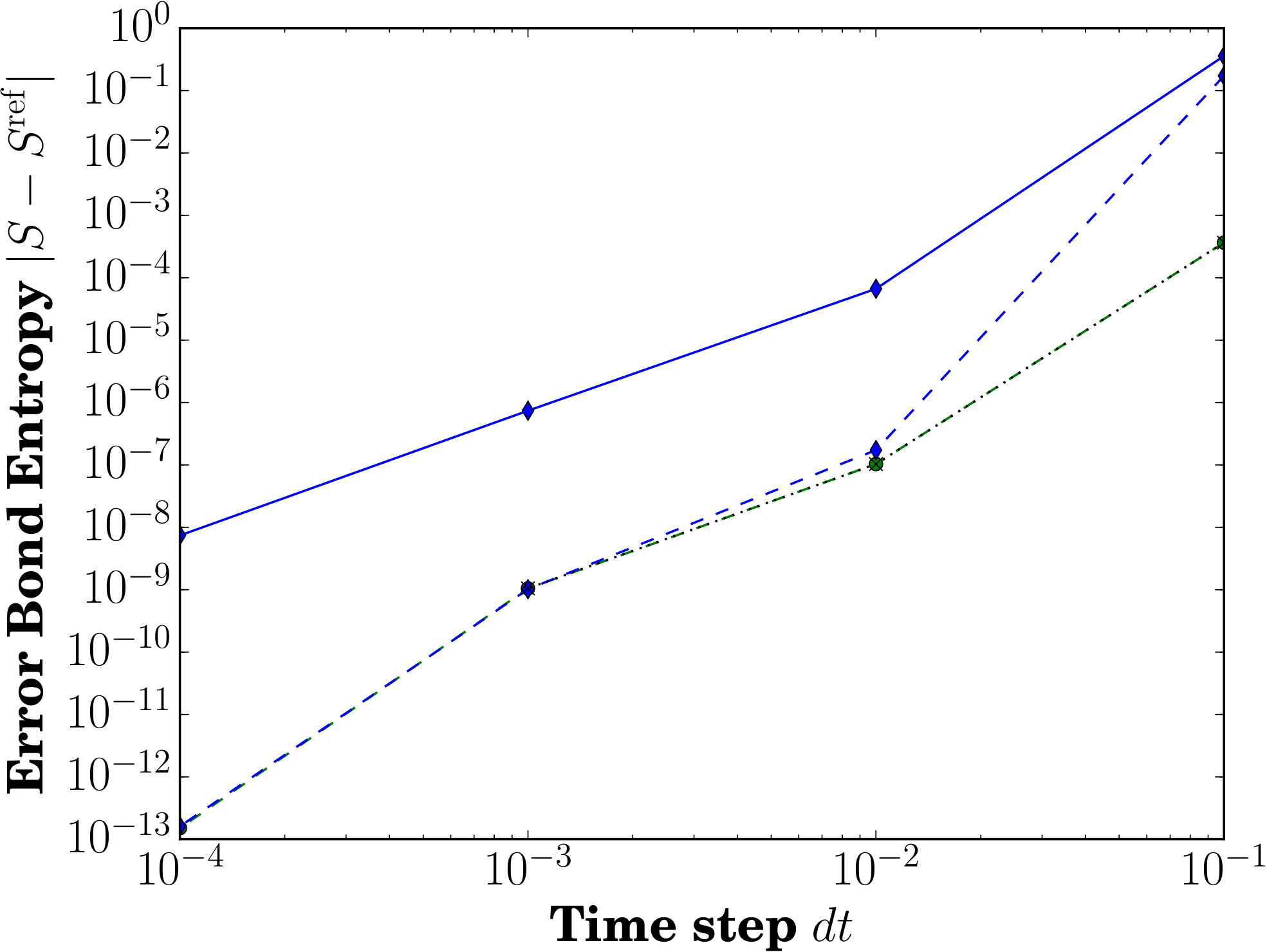}
        \put( 0,78){(d)}
      \end{overpic}
    \end{minipage}
    \caption[Convergence of Methods for the Bose-Hubbard Model.]
    {\emph{Convergence of Methods for the Bose-Hubbard Model.}
      We compare the 2$^{\mathrm{nd}}$
      and 4$^{\mathrm{th}}$ order Trotter time evolution and the Krylov method
      with mode $1$ against the reference (ref) taken as the matrix exponential
      (ME) with
      $dt = 0.0001$ for various measures. We consider the Bose-Hubbard model
      in a linear quench from on-site interaction strength $U(t=0) = 10.0$ to
      $U(t=0.5) = 8.0$ for a system size of $L = 6$ and unit filling.
      The minimal error is set to $10^{-18}$ indicating machine
      precision. The measures are (a) the minimal distance over all
      single-site reduced density matrices, (b) the minimal distance over all
      two-site density matrices, (c) the error in energy, and (d) the error
      in the bond entropy for the splitting in the middle of the system.
                                                                                \label{ed:fig:bose_qsu}}
  \end{center}
\end{figure}

The results for the scenario \emph{(1)} in case of the Bose-Hubbard model can
be found in Fig.~\ref{ed:fig:bose_qsu}. We point out that the Krylov method has
a similar error to taking the complete matrix exponential. Greater errors in
the Trotter decomposition are due to non-commutating terms in the decomposition
originating from Eq.~\eqref{ed:eq:trotter2}. For the example in
Fig.~\ref{ed:fig:bose_qsu}, we obtain a convergence rate of $dt^2$ for
the second order Trotter decomposition ($2.15$, $2.0$), the fourth order
Trotter decomposition ($4.21$, $2.05$), and the Krylov methods
($2.73$, $2.0$), where we consider the pairs $(dt = 0.01, dt' = 0.1)$
and $(dt = 0.001, dt' = 0.01)$ to evaluate the rate of convergence
according the distance of the single site density matrices. The limitation
to the second order convergence rate for all methods can be explained
since the time-ordering is correct up to second order for the time-dependent
Hamiltonian. Figure~\ref{ed:fig:bose_ssu} in the appendix shows that the
convergence rates differ for a time-independent Hamiltonian. The Krylov
method is then exact independent of the time step, and the fourth-order Trotter
convergence rate is $4.59$ for $(dt = 0.01, dt' = 0.1)$ before reaching
machine precision.
%
%
Additional plots for the Ising model and other scenarios can be found in
App.~\ref{ed:app:addons}.

\section{Open Systems in Untruncated Space                                     \label{ed:sec:open}}

The methods so far described are applicable for closed systems described by
pure states. These methods ignore the fact that any quantum system is coupled to some extent
to its environment. The Lindblad master equation is one way to describe an
open quantum system through the evolution of its density matrix $\rho$. The
assumptions are that the system is never entangled with its environment,
the correlation time of the environment is small in comparison to the
evolution time, and the evolution time itself is small in comparison to the
equilibration time of the system. These approximations apply if the environment is large
and has no memory. They are referred to as Born-Markov approximation and
secular approximation. In addition to Hamiltonian dynamics described via the
von Neumann equation, the evolution is described via $\eta$ Lindblad channels
acting on different sites $k$:
\begin{eqnarray}                                                                \label{ed:eq:lindbladme}
  \dot{\rho}
 &=& - \frac{\i}{\hbar} \left[ H, \rho \right]
     + \sum_{\eta, k} L_{\eta, k} \rho L_{\eta, k}^{\dagger}
     - \frac{1}{2} \left \{ L_{\eta, k}^{\dagger} L_{\eta, k}, \rho \right \}
  =  - \frac{\i}{\hbar} \left[ H, \rho \right]
     + \sum_{\nu} L_{\nu} \rho L_{\nu}^{\dagger}
     - \frac{1}{2} \left \{ L_{\nu}^{\dagger} L_{\nu}, \rho \right \}
     \, ,                                                                       \nonumber \\
\end{eqnarray}
where the indices $\eta$ and $k$ can be combined in another index
$\nu = (\eta, k)$. One of our additional rule sets for the open
system is a local Lindblad operator
\begin{eqnarray}
  \textrm{Lind1 rule}&:& \sum_{k=1}^{L} w \cdot c_i \cdot L_i \, ,
\end{eqnarray}
which would correspond to one channel $\eta$. We can always choose zero Lindblad
channels and have an evolution of a possibly initially mixed system under the von
Neumann equation
\begin{eqnarray}                                                                \label{ed:eq:rhovonneumann}
  \dot{\rho} &=& - \frac{\i}{\hbar} \left[ H, \rho \right] \, . 
\end{eqnarray}
In the following, we describe two approaches to simulate the Lindblad master
equation. On the one hand, we can simulate the whole density matrix $\rho$ as
represented in Eq.~\eqref{ed:eq:lindbladme}. Therefore, $\rho$ is represented as
superket vector $\Lket{\rho}$ in the Liouville space obtaining a Schr\"odinger-like
equation. The superket vector is the density matrix written as a vector.
A $D \times D$ density matrix turns into a vector with $D^2$ entries.
But first, we lay out the quantum trajectories, which average over
many simulations. Each simulation describes a single realization of what
possibly could happen in an experiment.

\subsection{Quantum Trajectories                                               \label{ed:sec:QT}}

The method of quantum trajectories has been proposed to avoid simulation
of the complete density matrix at the cost of sampling over many realizations
of a simulation with pure states. The algorithm used can be described by
the following steps \cite{Dalibard1992,Dum1992,Daley2014}:
\begin{enumerate}
\item{Throw a random number $r$ and evolve under the effective Hamiltonian
  $H_{\mathrm{eff}}$ until the norm drops below $r$,
  \begin{eqnarray}
    H_{\mathrm{eff}} &=& H - \frac{1}{2} \sum_{\nu}  L_{\nu}^{\dagger} L_{\nu} \, .
  \end{eqnarray}
}
\item{Throw another random number $r_\kappa$ and calculate the probabilities
  $p_{\nu}$ to apply any possible Lindblad operator and the cumulative
  probability $P_{\nu}$,
  \begin{eqnarray}
    p_{\nu} &=& \bra{\psi} L_{\nu}^{\dagger} L_{\nu} \ket{\psi} \, , \qquad
    P_{\nu} = \frac{\sum_{j=1}^{\nu} p_j}{\sum_{j=1}^{\nu_{\max}} p_j} \, .
  \end{eqnarray}
}
\item{Apply Lindblad operator $\kappa$ with $P_{\kappa-1} < r_{\kappa} \le P_{\kappa}$ and
  renormalize $\ket{\psi'} = L_{\kappa} \ket{\psi}$. Restart with step \emph{(1)}.}
\end{enumerate}
This approach can be used with the matrix exponential on the complete Hilbert
space, the Trotter and Krylov evolution.


\subsection{Liouville Space                                               \label{ed:sec:Liouville}}

\begin{figure}[t]
  \begin{center}
    \vspace{1.0cm}
    \begin{overpic}[width=0.45 \columnwidth,unit=1mm]{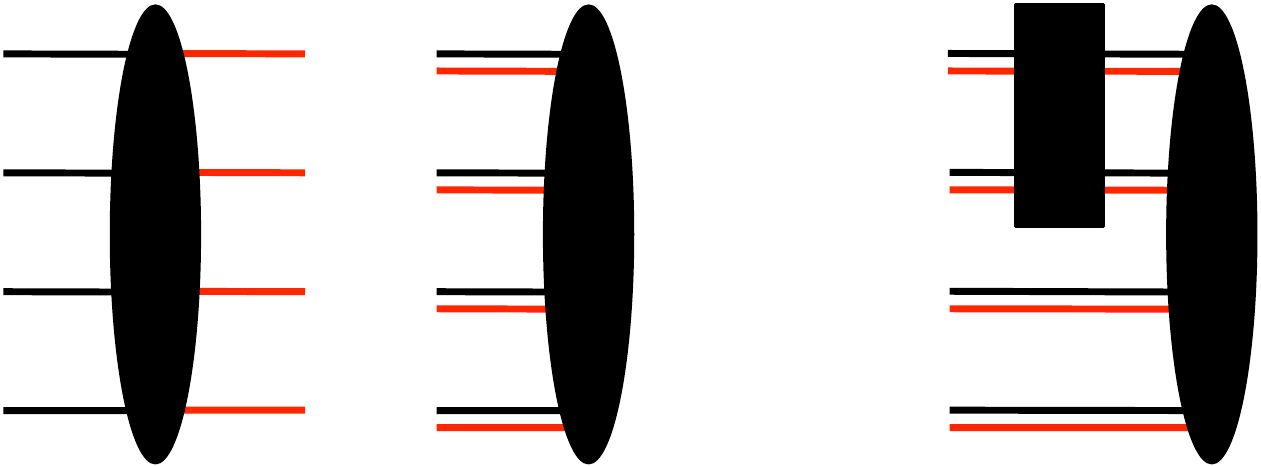}
      \put( 0,42){(a) $\rho$ and $\Lket{\rho}$}
      \put(71,42){(b) $(P_{1,2} \otimes \1) \Lket{\rho}$}
      \put(28,18){$\Rightarrow$}
    \end{overpic}
    \caption[Tensor Operations for the Density Matrix without Symmetries.]
    {\emph{Tensor Operations for the Density Matrix without Symmetries.}
      (a) The density matrix is rewritten as a rank-$(2L)$ tensor where the first
      $L$ indices represent the rows of the density matrix (black lines) and
      the second $L$ indices run over the columns of the density matrix (red)
      for both $\rho$ and $\Lket{\rho}$.
      (b) The application of a two-site operator to the density matrix
      corresponds to the contraction over the four indices of the two sites
      $k$ and $k+1$. This operation comes at a cost of $\mathcal{O}(d^{2L + 4})$.
                                                                                \label{ed:fig:TrotterMatrix}}
  \end{center}
\end{figure}

The Liouville space is a mapping from the Hilbert space defined over
\begin{eqnarray}                                                                \label{ed:eq:lioutrafo}
  O' \rho \; O \rightarrow O' \otimes O^T \Lket{\rho} \, ,
\end{eqnarray}
where the original density matrix $\rho$ is now represented as a vector, the
super-ket $\Lket{\rho}$. The superscript~$^{T}$ is the transpose of the
matrix. A quick
derivation of the transformation is given in App.~\ref{ed:app:openaddons}. The
corresponding equation governing the evolution of the density matrix is
Schr\"odinger-like,
\begin{eqnarray}                                                                \label{ed:eq:liousuperket}
  \frac{\partial}{\partial t} \Lket{\rho}
  &=& \left[- \frac{\i}{\hbar} \left( H \otimes \1 - 1 \otimes H^T \right)
      + \sum L_{k} \otimes L_{k}^{\ast}
      - \frac{1}{2} \left( L_{k} L_{k}^{\dagger} \otimes \1
      + \1 \otimes L_{k}^{\ast} L_{k}^{T} \right) \right] \Lket{\rho} \, .  
\end{eqnarray}
The short-hand notation is $\partial / \partial t \Lket{\rho} = \mathcal{L}
\Lket{\rho}$, where $\mathcal{L}$ represents Hamiltonian part and dissipative
part of the time evolution in Liouville space.
Thanks to the structure we can reuse two of the previous methods with small
modifications. In the first method presented, we obtain the propagator via
the matrix exponential. Calling appropriate methods for non-hermitian
matrices $-$ the matrix in Eq.~\eqref{ed:eq:liousuperket} is not necessarily
hermitian due to the term $L_{\nu} \otimes L_{\nu}^{\ast}$ $-$ we can reuse the
method. For the Trotter decomposition, the same strategy is necessary for
the local two-site propagators. The matrix exponential for the two site
propagator is now of dimension $d^4 \times d^4$ instead of $d^2 \times d^2$
for state vectors. Figure~\ref{ed:fig:TrotterMatrix} shows how to apply the two-site
operators acting on the density matrix formatted as a rank-$(2L)$ tensor. Based on
the contraction over four indices for two sites, the application scheme for
the Trotter decomposition can be adapted. In contrast to the pure system,
we permute the tensor after every contraction back to its original order.
The Krylov method used previously has to be modified for
non-hermitian matrices since the exponential of the matrix in the
Krylov subspace is no longer tridiagonal, but of upper Hessenberg form.

\subsection{Error Analysis in Case Study 1: Coupled Cavities                   \label{ed:sec:opencavities}}

To check the convergence of the code, we analyze two coupled cavities,
or photon Josephson junctions (PJJ), described through the Jaynes-Cummings
model with loss \cite{Schmidt2010}. The Hamiltonian of the system without a
loss is defined as
\begin{eqnarray}
  H_{\mathrm{PJJ}}
 &=& \left( \omega_q n_1 + \omega_c n_2 + \omega_c n_3 + \omega_q n_4 \right)
      + g \left( \sigma_{1}^{+} a_{2} + \sigma_{1}^{-} a_{2}^{\dagger}
           + a_{3}^{\dagger} \sigma_{4}^{-} + a_{3} \sigma_{4}^{+} \right) \nonumber \\
 &&   - J \left( a_{2}^{\dagger} a_{3} + a_{2} a_{3}^{\dagger} \right) \, ,
\end{eqnarray}
where $\omega_q = 1.0$ ($\omega_c = 1.0$) is the frequency of the qubit
(cavity) and $n_{k}$ are the number operators for the excitations on the corresponding
site $k$. $g = 0.48$ is the coupling constant between cavity and qubit; the
cavities are coupled to each other at a strength $J$ representing the
tunneling of photons. The transition from the ground (excited) state to the
excited (ground) state for the qubits is described with the operator
$\sigma^{+}$ ($\sigma^{-}$). The creation (annihilation) operator for the
photons in the cavity is $a^{\dagger}$ ($a$).
We recall that $n_{1,4} \in \mathbb{R}^{2 \times 2}$ while the
dimension of $n_{2,3}$ varies with the number of photons allowed in the
cavity. We add a spontaneous emission from the excited state to the ground
state for the qubits. The second type of system-environment interaction
is the loss of photons. Both processes assume that the environment is the
vacuum and only absorbs excitations, but cannot provide them to the system.
We can describe this system with the following Lindblad master equation:
\begin{eqnarray}                                                                \label{ed:eq:pjjlindblad}
  \dot{\rho}
 &=& -\mathrm{i} \left[ H_{\mathrm{PJJ}}, \rho \right]
     + \sum_{k \in \{1, 4\}} \gamma_k \sigma_k^{-} \; \rho \; \sigma_k^{+}
     - \frac{\gamma_k}{2} \left \{ \sigma_k^{+} \sigma_k^{-}, \rho \right \}
     + \sum_{k \in \{2, 3\}} \gamma_k a_k \; \rho \; a_k^{\dagger}
     - \frac{\gamma_k}{2} \left \{ a_k^{\dagger} a_k, \rho \right \} \, . \nonumber \\
\end{eqnarray}
Defining the total number of excitations as $N(t) = \sum_{k=1}^{4} n_k$ and
choosing a constant $\gamma = \gamma_k$ for the coupling of the system to
the environment via the Lindblad operator, one can show that the following
relation holds:
\begin{eqnarray}                                                                \label{ed:eq:pjjexp}
  N(t) &=& N(t=0) \exp(-\gamma t) \, .
\end{eqnarray}
We turn to the error analysis. First, we choose to simulate the case $\gamma = 0$
with the open system code, which is, in fact, the evolution according to the
von Neumann equation for density matrices defined in Eq.~\eqref{ed:eq:rhovonneumann}.
This simulation should reproduce the results
for pure state $\ket{\psi}$ and corresponding code. We characterize
the error again in terms of the maximal distance between all single site
density matrices and two site density matrices, the energy, and bond entropy
as defined in Eqs.~\eqref{ed:eq:erhoi} to \eqref{ed:eq:eentr}.
Table~\ref{ed:tab:closedcavities} lists the corresponding errors for our default
settings with a total time $T = 10$ with a time step of $dt = 0.01$. The errors
are for the final measurement at $T = 10$. We start with one photon in the
right cavity and all other sites in the ground state. Therefore, we can cut off
the Hilbert space to a local dimension $d=2$. We remark that the evolution of
the density matrix does not provide the bond entropy and therefore cannot
be compared to the result of the pure state. Furthermore, zero error is to be
interpreted as numerical zero. Table~\ref{ed:tab:opencavities} shows the errors
for the evolution of the open system with $\gamma = 0.05$, where we introduce
also the error in the total number of excitations following
Eq.~\eqref{ed:eq:pjjexp} as
\begin{eqnarray}
  \entim &=&
  \left| N_{\mathrm{EDLib}}(T) - N(t=0) \exp(- \gamma T) \right| \, .
\end{eqnarray}
We recognize the larger error for quantum trajectories in
Table~\ref{ed:tab:opencavities}. This error is due to the finite number
of different trajectories. The standard deviation scales with
$1 / \sqrt{N_{\mathrm{QT}}}$ according to the law of large numbers.
Errors related to the algorithm are at a much lower magnitude for
proper parameters and would only become visible for very large
$N_{\mathrm{QT}}$. We may illustrate this behavior with the following
example: We assume there is one particle and we pick the point in time where
this particle decayed with probability $0.5$ leading to a coin-toss scenario.
The normalized standard deviation for the binomial distribution is then
$(\sqrt{4 N})^{-1}$. For 500 trajectories, we have $0.022$. The standard
deviation in Tab.~\ref{ed:tab:opencavities} is within this order of magnitude.
Errors here are for $T=10$; therefore, the statistics are slightly different.
\begin{table}[t]
  \centering
  \begin{tabular}{@{} cccccc @{}}
    \toprule
    Model              &Error     &Liou (ME)            &Liou (Trotter-2)     &Liou (Krylov-1)      &QT1 (ME)              \\ 
    \cmidrule(r){1-1} \cmidrule(rl){2-2} \cmidrule(rl){3-3} \cmidrule(rl){4-4} \cmidrule(rl){5-5} \cmidrule(l){6-6}
    $H_{\mathrm{PJJ}}$ &$\erhoi$  &$0.0               $ &$1.54\cdot 10^{-10}$ &$2.74\cdot 10^{-13}$ &$0.0               $  \\ 
                       &$\erhoij$ &$0.0               $ &$1.83\cdot 10^{-10}$ &$7.37\cdot 10^{-13}$ &$0.0               $  \\ 
                       &$\eener$  &$6.28\cdot 10^{-14}$ &$6.71\cdot 10^{-07}$ &$5.58\cdot 10^{-13}$ &$5.55\cdot 10^{-15}$  \\ 
                       &$\eentr$  &$-$                  &$-$                  &$-$                  &$1.78\cdot 10^{-15}$  \\ 
    \cmidrule(r){1-1} \cmidrule(rl){2-2} \cmidrule(rl){3-3} \cmidrule(rl){4-4} \cmidrule(rl){5-5} \cmidrule(l){6-6}
    $H_{\mathrm{DW}}$  &$\erhoi$  &$0.0              $  &$5.34\cdot 10^{-10}$ &$8.54\cdot 10^{-14}$ &$9.03\cdot 10^{-14}$  \\ 
                       &$\erhoij$ &$0.0              $  &$7.67\cdot 10^{-10}$ &$8.8\cdot 10^{-14}$  &$2.7\cdot 10^{-13}$   \\ 
                       &$\eener$  &$1.82\cdot 10^{-12}$ &$7.16\cdot 10^{-06}$ &$8.07\cdot 10^{-13}$ &$7.51\cdot 10^{-13}$  \\ 
                       &$\eentr$  &$-$                  &$-$                  &$-$                  &$4.35\cdot 10^{-14}$  \\ 
    \bottomrule
  \end{tabular}
  \caption[Error Analysis for the Density Matrix in the von Neumann Equation.]
  {\emph{Error Analysis for the Density Matrix in the von Neumann
    Equation.} We simulate the closed system with the open system code of the
    library using the von Neumann equation in the Liouville space (Liou). As a
    reference, we use the closed system evolved with the matrix exponential (ME)
    to obtain errors for the maximal distances of one and two site density
    matrices, energy, and bond entropy. Zeros in the errors are to be
    interpreted as numerical zeros. Bond entropies are not calculated for
    density matrices. All of the methods listed demonstrate the correctness of the
    code, where, e.g., the error of the Trotter decomposition originates in the
    Trotter error and is decreased for the fourth order method. The Krylov
    method runs with the first mode, and the quantum trajectories (QT) have
    a single trajectory sufficient for a closed system. Additional data for
    further methods is presented in the Appendix in Table~\ref{ed:tab:closedapp}.
                                                                                \label{ed:tab:closedcavities}}
\end{table}

\begin{table}[t]
  \centering
  \begin{tabular}{@{} cccccc @{}}
    \toprule
    Model              &Error     &Liou (ME)            &Liou (Trotter-2)     &Liou (Krylov-1)      &QT500 (ME)           \\
    \cmidrule(r){1-1} \cmidrule(rl){2-2} \cmidrule(rl){3-3} \cmidrule(rl){4-4} \cmidrule(rl){5-5} \cmidrule(l){6-6}
    $H_{\mathrm{PJJ}}$ &$\entim$  &$8.88\cdot 10^{-16}$ &$3.22\cdot 10^{-07}$ &$3.7\cdot 10^{-13}$  &$1.15\cdot 10^{-02}$ \\
                       &$\erhoi$  & $-$                 &$7.5\cdot 10^{-11}$  &$2.27\cdot 10^{-13}$ &$1.49\cdot 10^{-05}$ \\
                       &$\erhoij$ & $-$                 &$8.13\cdot 10^{-11}$ &$1.25\cdot 10^{-13}$ &$4.48\cdot 10^{-05}$ \\
                       &$\eener$  & $-$                 &$1.32\cdot 10^{-07}$ &$3.71\cdot 10^{-13}$ &$1.15\cdot 10^{-02}$ \\
    \cmidrule(r){1-1} \cmidrule(rl){2-2} \cmidrule(rl){3-3} \cmidrule(rl){4-4} \cmidrule(rl){5-5} \cmidrule(l){6-6}
    $H_{\mathrm{DW}}$  &$\erhoi$  & $-$                 &$1.32\cdot 10^{-11}$ &$0.0            $    &$2.04\cdot 10^{-03}$ \\ 
                       &$\erhoij$ & $-$                 &$2.52\cdot 10^{-11}$ &$0.0            $    &$6.44\cdot 10^{-03}$ \\ 
                       &$\eener$  & $-$                 &$2.53\cdot 10^{-06}$ &$1.16\cdot 10^{-12}$ &$3.75\cdot 10^{-02}$ \\ 
    \bottomrule
  \end{tabular}
  \caption[Error Analysis for the Lindblad Master Equation.]
  {\emph{Error Analysis for the Lindblad Master Equation.} We simulate
    the Lindblad master equation for different time evolution methods and take
    as reference the implementation using the matrix exponential. In addition,
    the simulations with $H_{\mathrm{PJJ}}$ are compared to the exponential
    decay of the number of excitations. Quantum trajectories run with 500
    different trajectories. Zeros in the errors are to be interpreted as
    numerical zeros. Bond entropies are not calculated for density matrices
    and excluded for this reason from the table of errors. Data for additional
    methods can be found in the Appendix in Table~\ref{ed:tab:openapp}.
                                                                                \label{ed:tab:opencavities}}
\end{table}

\subsection{Error Analysis in Case Study 2: Double-Well                        \label{ed:sec:opendoublewell}}

In the second case study, we check the error for simulations with symmetries,
where we choose a Bose-Hubbard model with $\mathcal{U}(1)$ symmetry for the
number conservation. We introduced the double-well problem before in
Eq.~\eqref{ed:eq:HDW} with its Hamiltonian $H_{\mathrm{DW}}$. For the comparison
of the evolution of the von Neumann equation of the density matrix,
see Eq.~\eqref{ed:eq:rhovonneumann}, to the
pure state evolution we choose a system size of $L = 7$ with three bosons over
a time of $T = 20$ and a time step of $dt = 0.01$. The height of the potential
is $V = 1$.
The initial state for the time evolution is the ground state when
applying a potential to the right well of amplitude $1$ in addition to the
potential of the barrier localizing the particles in the left well. Upon
releasing the barrier for the right well at $t = 0$, the particles start
to oscillate between the two wells, i.e., the system is in the Josephson
regime. The errors are shown in Table~\ref{ed:tab:closedcavities}. The open
system is introduced as in \cite{Bonnes2014}
\begin{eqnarray}
  \dot{\rho}
 &=& - \mathrm{i} \left[ H, \rho \right]
     + \sum_{k=1}^{L} \gamma_k n_k \rho \, n_k  
     - \frac{\gamma_k}{2} \left \{ n_k^2, \rho \right \} \, ,
\end{eqnarray}
and we choose a space-independent coupling $\gamma = \gamma_k = 0.05$ in the
following. Table~\ref{ed:tab:opencavities} contains then the open system case
and compares the matrix exponential to the other methods. The setup of this
model is discussed in detail as an example in the Appendix~\ref{ed:app:example}.

\section{Thermalization of the long-range quantum Ising Model with the
  full-spectrum Lindblad Master Equation                                       \label{ed:app:fullspec}}

The derivation of the Lindblad master equation
\cite{BreuerPetruccione,Rivas2012} for a given Hamiltonian
is subtle in the many-body scenario. One possibility, dubbed the full
spectrum Lindblad operators, is the inclusion of the
energy eigenstates of the system \cite{Cuetara2016,Gonzalez2017}. This
approach elegantly handles the secular approximation but is obviously
restricted to many-body systems with approximately six or fewer
qubits, or the size of the Hilbert space equal to six qubits. The
restriction to six qubits stems from the exponential taken in Liouville space
for the Lindblad master equation, defined in Eq.~\eqref{ed:eq:lindbladme}.
Krylov methods or the Trotter approximation are not suitable for this huge
number of non-local Lindblad operators. We use the long-range quantum Ising
model with the transverse field to study the thermalization timescales based
on previous research presenting the complete analysis \cite{JaschkeTQI}, which
uses the full spectrum for the thermalization timescale of the quantum Ising
model with regular nearest-neighbor interactions.

The usage of the full spectrum requires knowledge of the system Hamiltonian
and the interaction Hamiltonian $H_{I}$; the Hamiltonian of the reservoir
$H_{R}$ is assumed to be a set of harmonic oscillators and essential to
derive the actual coupling constants. These parts of the model are defined
in our example in the following as
\begin{eqnarray}
  \HLRQI &=& - \sum_{k=1}^{L-1} \sum_{k'=k+1}^{L}
                 \frac{\cos(\phi)}{(k' - k)^{\alpha}}
                 \sigma_{k}^{z} \sigma_{k'}^{z}
             - \sum_{k=1}^{L} \sin(\phi) \sigma_{k}^{x} \, , \\
  H_{R} &=& \sum_{q} \omega_{q} n_{q} \, , \\
  H_{I} &=& \sum_{k=1}^{L} \sum_{q} \sigma_{k}^{x}
              \left( g_{kq} b_{q} + g_{kq}^{\ast} b_{q}^{\dagger} \right) \, .
\end{eqnarray}
The Hamiltonian of the long-range quantum Ising model $\HLRQI$ depends on
the number of sites $L$, and different long-range interactions are addressed
via a specific choice of $\alpha$. To have similar energy scales across the
complete phase diagram, the coupling of the $zz$-interactions and the external
field in the $x$-direction are defined as a function of $\phi$; the range
$\phi \in [0, \frac{\pi}{2} ]$ describes the phase diagram with the
ferromagnetic limit at $\phi =0$, the critical point in the thermodynamic
limit at $\phi = \frac{\pi}{4}$, and the paramagnetic limit is
$\phi = \frac{\pi}{2}$. The conversion from the standard interaction $J$
and external field coupling $g$ is $\cos(\phi) = J / \sqrt{J^2 + g^2}$
and $\sin(\phi) = g / \sqrt{J^2 + g^2}$; thus, the quantities in $\HLRQI$
are unitless and time and temperature are in units of $1 / \sqrt{J^2 + g^2}$
and $\sqrt{J^2 + g^2}$, respectively. The two limits are the
recovery of the nearest-neighbor quantum Ising model for $\alpha \to \infty$
and the completely interacting model $\alpha = 0$. The reservoir Hamiltonian
$H_{R}$ is a large set of harmonic oscillators indexed by $q$, where the summation
of $q$ is not specified due to the large set of oscillators only required
during the derivation of the final coupling. The interaction $H_{I}$ contains
the creation or annihilation of a mode in the reservoir while flipping
a spin in the system, see $\sigma_{k}^{x}$. The coupling $g_{kq}$ describes
the strength of the interaction based on the site in the spin chain and
the mode of the harmonic oscillator. This choice of $H_{I}$
has interactions of each spin with one reservoir, i.e., the common reservoir
can induce collective long-range interaction in addition to the ones present
in the system's Hamiltonian. Moreover, the choice of the interaction
Hamiltonian conserves the $\mathbb{Z}_2$ symmetry of the quantum Ising model
and we focus on the even symmetry sector in the following.

We skip details of the complete derivation of the structure of the
resulting Lindblad equation; corresponding details can be found in
the original study of the quantum Ising model \cite{JaschkeTQI} or the
references \cite{Cuetara2016,Gonzalez2017}. The Lindblad equation
following the full-spectrum then reads
\begin{eqnarray}                                                                \label{ed:eq:fullspec}
  \dot{\rho} = - \frac{\mathrm{i}}{\hbar} \left[ H_{S}', \rho \right]
  + \sum_{abcd} \mathcal{C}_{abcd} \left( L_{ab} \rho L_{cd}^{\dagger}
  - \frac{1}{2} \left \{ L_{cd}^{\dagger} L_{ab} , \rho \right \} \right)
  \, .
\end{eqnarray}
The Lindblad operators $L_{ab} = \ket{a} \!\bra{b}$ are transitions between
eigenstates $\ket{a}$ and $\ket{b}$ of the system's Hamiltonian, i.e.,
in our example $\HLRQI$. The coupling $\mathcal{C}_{abcd}$
carries the dependence on the type of reservoir, the interaction strength
between system and reservoir, the operator $\sigma_{k}^{x}$ acting in the
interaction Hamiltonian $H_{I}$ on the system in terms of the overlaps
$\bra{a} \sigma_{k}^{x} \ket{b}$, and the temperature of the reservoir.
The Hamiltonian $H_{S}'$ can have corrections with regards
to the system Hamiltonian $H_{S}$. We choose the statistics of the reservoir
following a 3D electromagnetic field \cite{JaschkeTQI}; we truncate the
corrections in $H_{S}'$ which are related to the imaginary part of the
principal value integral while resolving the integration of the
electromagnetic field.

\begin{figure}[t]
  \begin{center}
    \vspace{0.8cm}\begin{minipage}{0.47\linewidth}
      \begin{overpic}[width=1.0 \columnwidth,unit=1mm]{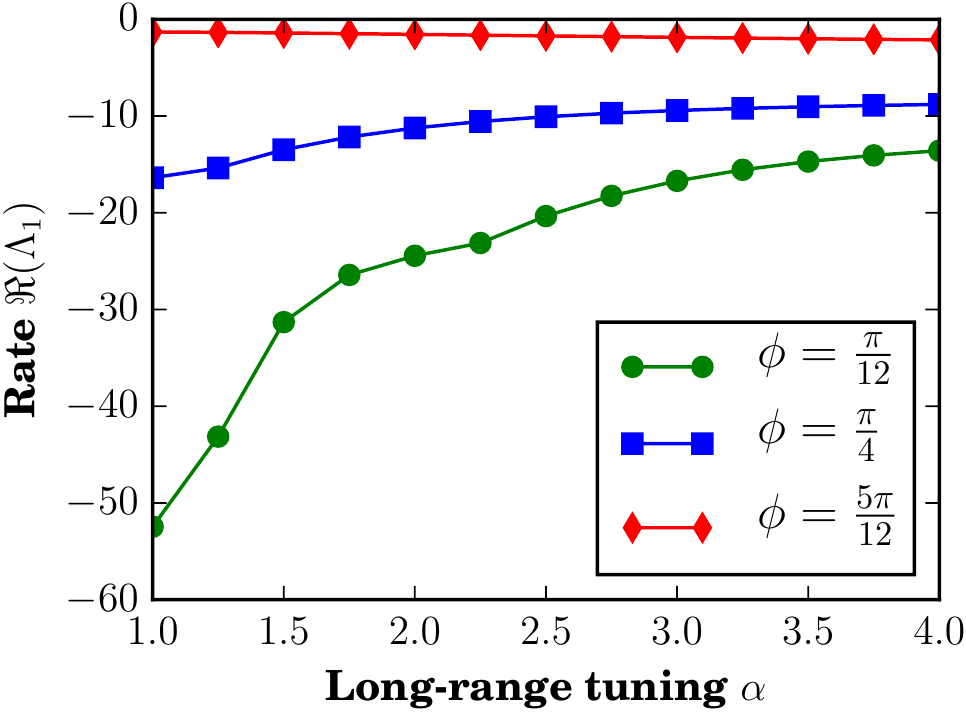}
        \put( 0,78){(a)}
      \end{overpic}
    \end{minipage}\hfill
    \begin{minipage}{0.47\linewidth}
      \begin{overpic}[width=1.0 \columnwidth,unit=1mm]{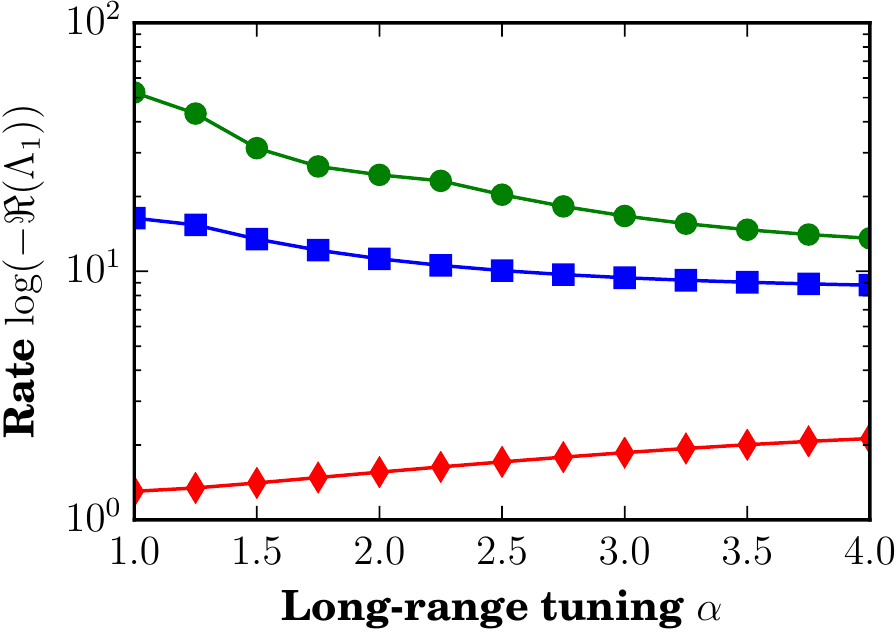}
        \put( 0,78){(b)}
      \end{overpic}
    \end{minipage}\vspace{0.4cm}
    \begin{minipage}{0.47\linewidth}
      \begin{overpic}[width=1.0 \columnwidth,unit=1mm]{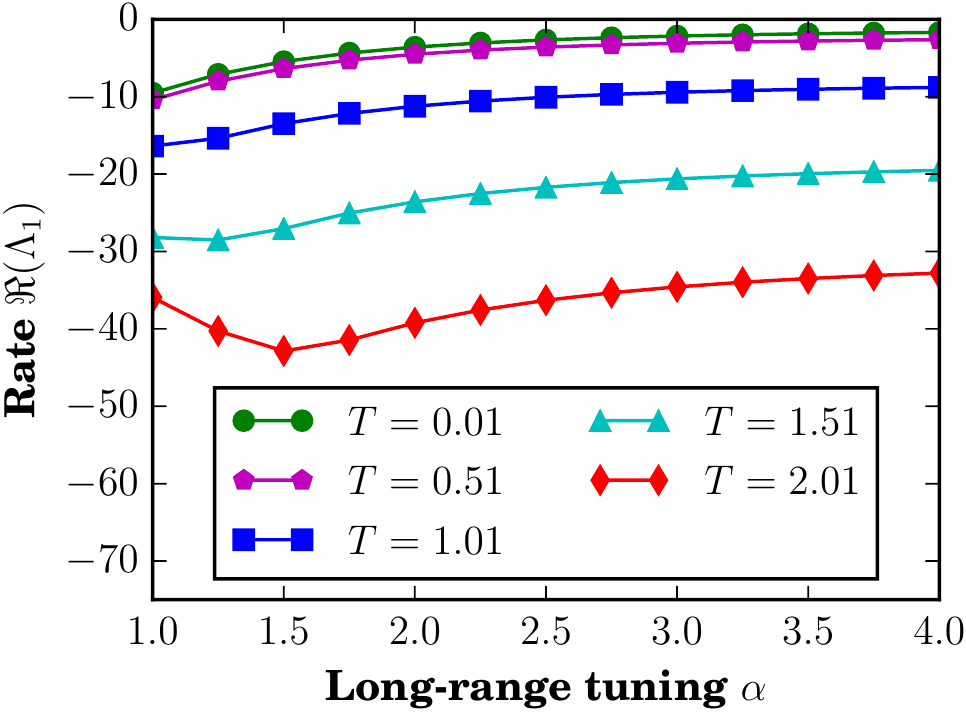}
        \put( 0,78){(c)}
      \end{overpic}
    \end{minipage}\hfill
    \begin{minipage}{0.47\linewidth}
      \begin{overpic}[width=1.0 \columnwidth,unit=1mm]{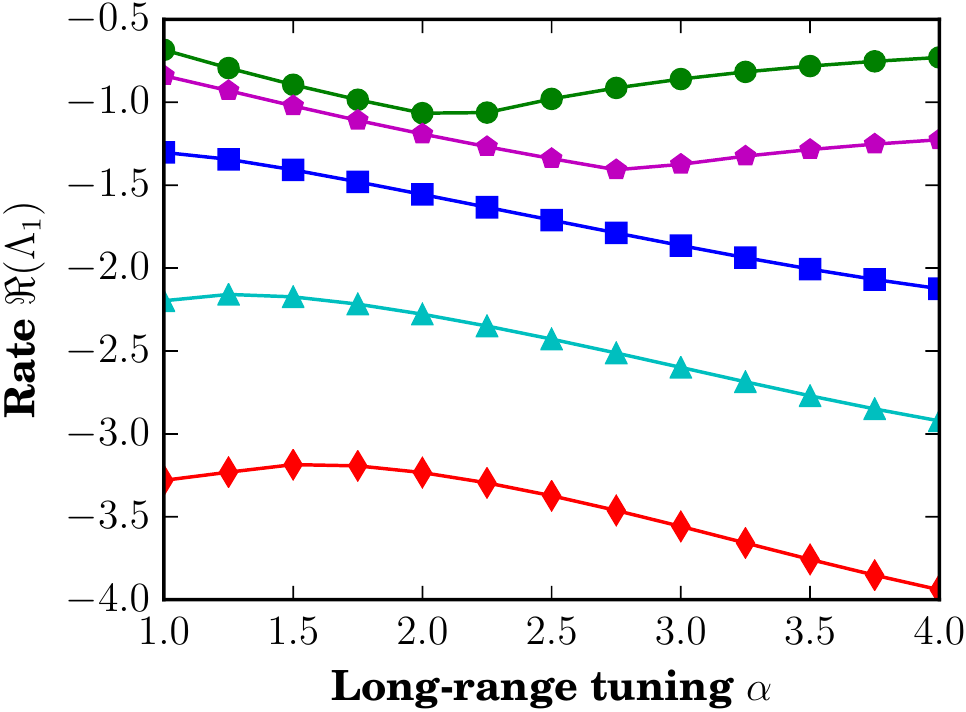}
        \put( 0,78){(d)}
      \end{overpic}
    \end{minipage}
    \caption[Thermalization of the long-range quantum Ising model.]
    {\emph{Thermalization of the long-range quantum Ising model.}
      The thermalization timescale depends on the coefficient of the power-law
      decay of the long-range interactions and temperature, where a change in
      $\alpha$ can both increase and decrease decoherence depending on the
      value $\phi$ setting the position in the phase diagram.
      (a,b)~The slowest thermalization timescale obtained from the eigenvalues
      $\Lambda$ of the Liouville operator $\mathcal{L}$ for a temperature
      $T = 1.01$. The same legend applies to (a) and (b).
      (c)~The timescale for $\phi = \frac{\pi}{4}$ for different temperatures $T$.
      (d)~The timescale for $\phi = \frac{5 \pi}{12}$ for different temperatures
      $T$. The same legend as in (c) applies.
                                                                                \label{ed:fig:full}}
  \end{center}
\end{figure}

We are interested in how the thermalization timescale changes with the
tuning of the long-range interactions $\alpha$; the unique steady
state of the even symmetry sector corresponds to the thermal state. One
can extract the slowest possible timescale from the eigenvalues $\Lambda_i$ of the
Liouville operator $\mathcal{L}$, defined in Eq.~\eqref{ed:eq:liousuperket},
constructed based on
Eq.~\eqref{ed:eq:fullspec}. The real part identifies the steady states
of the open system whenever $\Re(\Lambda_i)$ equals zero. The first
non-zero, and negative, $\Re(\Lambda_i)$ defines the slowest
possible timescale in the system; this timescale is labeled with
subscript $1$ in the following. Depending on the initial state,
this timescale may not be present. We observe for a temperature
$T = 1.01$ in Fig.~\ref{ed:fig:full}(a) and (b) that for small
$\phi$ there is a stronger decoherence while decreasing $\alpha$,
i.e., $|\Re(\Lambda_i)|$ is increasing when moving away from the
limit of nearest-neighbor interactions. In contrast, the effects
of decoherence are smaller for long-range interactions close to the
paramagnetic limit, which is better visible in the log-representation
of $\Re(\Lambda_1)$. The system's thermalization timescale depends
on $\phi$ in the nearest-neighbor model, with no thermalization in
the paramagnetic limit due to the choice of $\sigma_{k}^{x}$ as
the operator acting on the system in the interaction Hamiltonian $H_{I}$.
Figure~\ref{ed:fig:full}(c) and (d) shows that this behavior just
described for a specific temperature $T$ depends as well on the
temperature of the reservoir. Thus, experimental systems such as
trapped ions with tunable $\alpha$ \cite{Kim2009,Blatt2012,Wilson2014}
can exhibit a variety of timescales across the $(\alpha, T)$ parameters
space.

These results illustrate why the full spectrum is worth being included
in a library as it allows for the study of thermalization via a reservoir.
This type of simulation is one key example where ED is preferable over tensor
network methods. The method requires a knowledge of all eigenstates. The
structure of the Lindblad operators $L_{ab}$ combining two eigenstates $\ket{a}$
and $\ket{b}$ via an outer product to an operator leads to a bond dimension
without truncation of $2^{L}$ in the case of qubits. Disregarding the huge
amount of Lindblad operators of this kind for a possible system and the
problems arising from it, the compression without truncation can only
begin after contracting a single Lindblad operator to the state with an
intermediate bond dimension of $2^{\frac{3L}{2}}$; the advantages of
handling such a problem with vectors given by the ED methods and adding
Lindblad operators directly to the Liouville operator are apparent.

\section{Comparison to QuTip Package for Benchmarking                          \label{ed:sec:benchmarking}}

We compare single features of the ED package of OSMPS with comparable
features of the QuTip library \cite{Johansson2013,QuTiP} under version
$4.0.2$. The QuTip library provides a far more general framework to simulate
closed and open quantum systems. We focus on the benchmark of the following tasks:
\begin{itemize}
\item{Task (1): Partial trace over a pure state $\ket{\psi}$ retaining
  the reduced density matrix for the first site $\rho_{1}$. We point out that
  this case does not need a permutation in \OSMPS{}. With regards to the
  implementation in QuTip, the many-body case does not seem to be the focus,
  and the density matrix of the complete pure state is built first and then
  traced out over $\rho$, which leads to some overhead.}
\item{Task (2): Partial trace over a density matrix $\rho$ retaining
  the reduced density matrix for the first site $\rho_{1}$. Since both
  libraries start from the density matrix, this comparison is more valuable
  than tracing over pure states in (1).}
\item{Task (3): Time evolution of the quantum Ising model with a
  time-independent Hamiltonian. The QuTip library uses an ODE solver to evolve
  a given initial state in time, which we compare to the Krylov and fourth-order
  Trotter method from \OSMPS{}. For QuTip we set up a single measurement
  of the identity matrix. \OSMPS{} takes the default measurements including
  the measurement of the energy (Hamiltonian) and the overlap to the initial
  state, where the first one contains the major part of the measurement time
  in \OSMPS{}. For \OSMPS{} we consider the two approximate methods with the
  fourth order Trotter (3a) and the Krylov (3b). We do not use the
  $\mathbb{Z}_{2}$ symmetry of the system here.
}
\end{itemize}

We run the benchmark simulations on an \hppaviliondv{} and present the CPU times
measured with \py{}'s \texttt{time.clock} package in Table.~\ref{ed:tab:QuTip}. The
run times of the trace over a density matrix are comparable and show that our
implementation scales as well as QuTip. The time evolutions are more difficult
to compare with regards to the order of the method etc., but the time of the
Trotter decomposition for large systems shows that we can offer a valid
alternative. The Krylov evolution, as expected from the previous scaling
analysis, e.g., in Fig.~\ref{ed:fig:ising_scal_timestep}, takes longer.

\begin{table}[t]
  \centering
  \begin{tabular}{@{} cccccccc @{}}
    \toprule
    Task                              & $\Tcpu(4)$         & $\Tcpu(6)$         & $\Tcpu(8)$         & $\Tcpu(10)$        & $\Tcpu(12)$        & $\Tcpu(14)$        & $\Tcpu(26)$    \\
    \cmidrule(r){1-1}                 \cmidrule(rl){2-2} \cmidrule(rl){3-3} \cmidrule(rl){4-4} \cmidrule(rl){5-5} \cmidrule(rl){6-6} \cmidrule(rl){7-7} \cmidrule(l){8-8}
    (1), QuTip                        & $0.15$           & $0.16$           & $0.20$           & $0.42$           & $2.76$           & $-$              & $-$          \\
    (1), \OSMPS{}                     & $0.01$           & $0.01$           & $0.01$           & $0.01$           & $0.01$           & $0.01$           & $22.00$      \\
    \cmidrule(r){1-1}                 \cmidrule(rl){2-2} \cmidrule(rl){3-3} \cmidrule(rl){4-4} \cmidrule(rl){5-5} \cmidrule(rl){6-6} \cmidrule(rl){7-7} \cmidrule(l){8-8}
    (2), QuTip                        & $0.12$           & $0.14$           & $0.18$           & $0.39$           & $2.71$           & $-$              & $-$          \\
    (2), \OSMPS{}                     & $0.01$           & $0.01$           & $0.02$           & $0.16$           & $2.61$           & $-$              & $-$          \\
    \cmidrule(r){1-1}                 \cmidrule(rl){2-2} \cmidrule(rl){3-3} \cmidrule(rl){4-4} \cmidrule(rl){5-5} \cmidrule(rl){6-6} \cmidrule(rl){7-7} \cmidrule(l){8-8}
    (3), QuTip                        & $3.19$           & $4.50$           & $7.16$           & $17.00$          & $57.05$          & $248.60$         & $-$          \\
    (3a), \OSMPS{}                    & $9.72$           & $14.71$          & $20.86$          & $33.20$          & $72.54$          & $232.75$         & $-$          \\
    (3b), \OSMPS{}                    & $16.84$          & $20.20$          & $25.60$          & $132.69$         & $314.91$         & $973.38$         & $-$          \\
    \bottomrule
  \end{tabular}
  \caption[Benchmarking QuTip and \OSMPS{}.]
  {\emph{Benchmarking QuTip and \OSMPS{}.} The comparison of run times
    includes (1) the partial trace of a pure state and (2) a density matrix,
    both tracing over the sites $2$ to $L$ where $L$ is the system size.
    Furthermore, (3) contains the time evolution for the quantum Ising
    model for a time-independent Hamiltonian. For \OSMPS{} we distinguish
    between the 4$^{\mathrm{th}}$ Trotter decomposition, i.e., (3a), and the Krylov
    method labeled as (3b). QuTip uses the default ODE solver. All CPU times $\Tcpu(L)$ are
    given in seconds as a function of the system size $L$. The times are
    calculated cumulatively for 100 iterations of the corresponding task.
                                                                                \label{ed:tab:QuTip}}
\end{table}

\section{Conclusions                                                           \label{ed:sec:conclusion}}

In this paper, we have shown the approaches for exact diagonalization methods including the
use of symmetries and other evolution techniques without truncating
entanglement, including the Trotter decomposition and the Krylov subspace method. The
methods can be extended from closed to open systems.

We have described efficient methods for time evolutions and
measurements without symmetries when the Hilbert space is built from a tensor
product of local Hilbert spaces. Therefore, the state vector representing the wave
function can be rewritten as a rank-$L$ tensor. The methods include
especially approaches for the Trotter time evolution, matrix-vector
multiplication necessary for the Krylov time evolution, and measurements via
reduced density matrices. For symmetry-adapted methods, the key is the mapping
into the basis. However, methods for Trotter decomposition or
matrix-vector multiplication cannot avoid looping over entries and cannot
use faster matrix algebra. Reduced density matrices, in particular, suffer from
this limitation and we have presented a feasible approach here, scaling with
$\mathcal{O}(D_S L) + \mathcal{O}(d^{2l} D_S)$.

Open system methods according to the Lindblad master equation follow two
known approaches via the quantum trajectories
and the Liouville space mapping the density matrix to a vector. The evolution
of the vector representing the density matrix follows a Schr\"odinger-like
equation. While quantum trajectories can use system sizes and methods equivalent
to pure states sampling over many realizations, the methods for Liouville space
have to be adapted slightly for the use with density matrices. This adaption refers to
the matrix exponential or its approximation. While the Hamiltonian was hermitian, the Liouville
operator is not necessarily hermitian. For any propagation of the whole density
matrix, we can use the rule of thumb that about half the number of sites of
the pure system can be simulated. The disadvantage of quantum trajectories
is the error, which was of the order $10^{-2}$ for the total number of
excitations in the photon Josephson junctions for $500$ trajectories. We
point out that the number of trajectories needed might differ for various
Hamiltonians.
The application of the full-spectrum Lindblad operators to the long-range
quantum Ising model shows that the thermalization timescale exhibits a rich
landscape as a function of the strength of the long-range interaction, the
temperature of the reservoir, and the position in the phase diagram.

The exact diagonalization methods presented help to develop new methods and
can serve as a convergence test and benchmark. One application is the
exploration of highly entangled states, which cannot be simulated with tensor
network methods with the examples specified in the following. The full
spectrum Lindblad equation presented in Sec.~\ref{ed:app:fullspec} is too inefficient to be used
in tensor network methods due to the vast amount of Lindblad operators to
be stored while ED methods can add them directly to the Liouville operator
stored as a matrix. Methods such as the Krylov time evolution work for both,
but tensor networks need to variational fit states in the case of long-range
interaction or add MPSs and compress them; in contrast, ED can sum the
vectors directly without overhead. Although tensor networks can tackle in
theory equal system sizes in comparison to ED when not truncating entanglement,
ED can be a more convenient approach and with less parameters to tune.
Other Hamiltonians exploring only a small part of the Hilbert space might be
scalable in size to more than the $25$ qubits of the Ising model, where we
show the scaling of Trotter decomposition up to 27 qubits. Models
like the Bose-Hubbard model in $k$-space have long-range interactions and
a large bond dimension in the Matrix Product Operators of the Hamiltonian
when using tensor network methods such as MPS, and thus
find in exact diagonalization methods an attractive alternative.
The rule sets allow for the flexible construction of the Hamiltonians without the
restriction of the MPO representation with tensor network methods, e.g.,
any infinite function does not have to be fitted to exponential rules. This argument
might apply as well to systems with randomized interactions such as spin glasses
\cite{Sherrington1975,Wu1991}.
Finally, open quantum systems are another attractive field for exact
diagonalization libraries despite the limitations to fewer sites
due to the Liouville space. Any formulation of a master equation for $\rho$
could be explored with the easy to program tools within our exact
diagonalization codes, e.g., if Lindblad operators are non-local and/or
long-range including a coupling through the environment. Especially if the
full spectrum of the Hamiltonian is necessary for the Lindblad master equation,
exact diagonalization is an ideal test case for any approximation
truncating the number of states in the spectrum.

Therefore, these methods are not only useful as problems set in classes, but
are worth being improved and optimized for exploring new frontiers in research.

\section*{Acknowledgments}

We thank I.~de Vega for many discussions and insights on the full spectrum
approach to the Lindblad master equation.
We gratefully appreciate contributions from and discussions with L.~Hillberry,
S.~Montangero, H.~North, G.~Shchedrin, and M.~L.~Wall. The calculations were
carried out using the high performance computing resources provided by the
Golden Energy Computing Organization at the Colorado School of Mines. This
work has been supported by the NSF under the grants PHY-1520915, and
OAC-1740130, and the AFOSR under grant FA9550-14-1-0287. This work was
performed in part at the Aspen Center for Physics, which is supported by
the US National Science Foundation grant PHY-1607611. We acknowledge
support of the U.K. Engineering and Physical Sciences Research Council
(EPSRC) through the ``Quantum Science with Ultracold Molecules'' Programme
(Grant No. EP/P01058X/1).


%

\appendix

\section{Rule sets for exact diagonalization                                   \label{ed:app:rules}}

As \OSMPS{} is a
many-body library, Hamiltonians are not built by the user, but through
\emph{rule sets}. The following rule sets are currently enabled in our
open source diagonalization libraries:
%
\begin{eqnarray}
  \textrm{Site rule}&:& \sum_{k=1}^{L} w \cdot c_k \cdot O_k                    \label{ed:eq:appsiters} \\
  \textrm{Bond rule}&:&
  \sum_{k=1}^{L-1} w \cdot c_k \cdot O_{k} \otimes O_{k+1}'                     \label{ed:eq:appbondrs} \\
  \textrm{Exp rule}&:&
  \sum_{k < j}^{L} \frac{w \cdot c_k}{d_p^{(j - k - 1)}} O_{k} \otimes \left(
              \bigotimes_{q=k+1}^{j-1} Q_q \right) \otimes O_{j}'               \label{ed:eq:appexprs} \\
  \textrm{InfiniteFunc}&:&
  \sum_{k < j}^{L} w \cdot c_k \cdot f(|j - k|) O_{k} \otimes \left(
              \bigotimes_{q=k+1}^{j-1} Q_q \right) \otimes O_{j}'               \label{ed:eq:appinfrs} \\
  \textrm{MBString}&:&
  \sum_{k=1}^{L-j} w \cdot c_k \bigotimes_{j=1}^{W_{\mathrm{MB}}} O_{j}^{[k+j-1]}
  \, ,                                                                          \label{ed:eq:appmbsrs}
\end{eqnarray}
%
where $L$ is the number of sites in the one-dimensional chain, $w$ is a
general coupling, and $c_k$ is a possibly space-dependent coupling. $O_{k}$,
$O_{k}'$, and $Q_{k}$ are operators acting on the local Hilbert space.
$Q_{k}$ targets fermionic systems, which have a phase operator due to the
Jordan-Wigner transformation \cite{Jordan1928,SachdevQPT} mapping
fermions onto the lattice. The Jordan-Wigner transformation
is a mapping between spins and fermions where the transformation ensures that
the corresponding commutation relations hold. The exponential rule includes
a decay parameter $d_p \le 1$. The function $f(\cdot)$ describes the coupling
of the two operators at an arbitrary distance for the InfiniteFunc rule.
The Many-Body String (MBString)
term has a width of $W_{\mathrm{MB}}$ sites and uses the tensor product $\otimes$ to indicate
the construction on the composite Hilbert space. Synonyms for the tensor
product are "outer product" or "Kronecker product".

\section{Scaling of memory resources                                           \label{ed:app:ram}}

\begin{figure}[t]
  \begin{center}
    \vspace{0.8cm}
    \begin{minipage}{0.47\linewidth}
      \begin{overpic}[width=1.0 \columnwidth,unit=1mm]{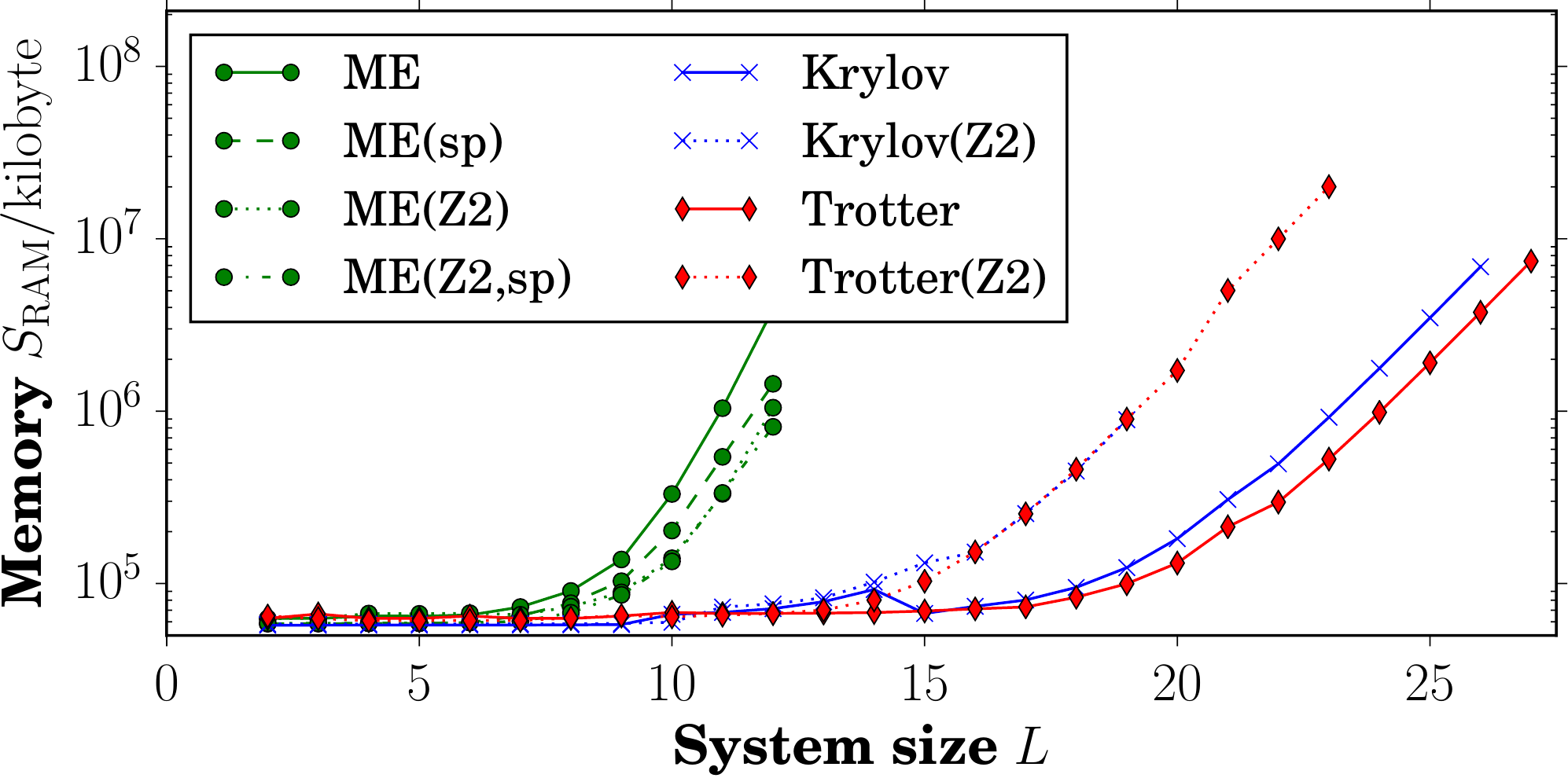}
        \put( 0,53){(a)}
      \end{overpic}
    \end{minipage}\hfill
    \begin{minipage}{0.47\linewidth}
      \begin{overpic}[width=1.0 \columnwidth,unit=1mm]{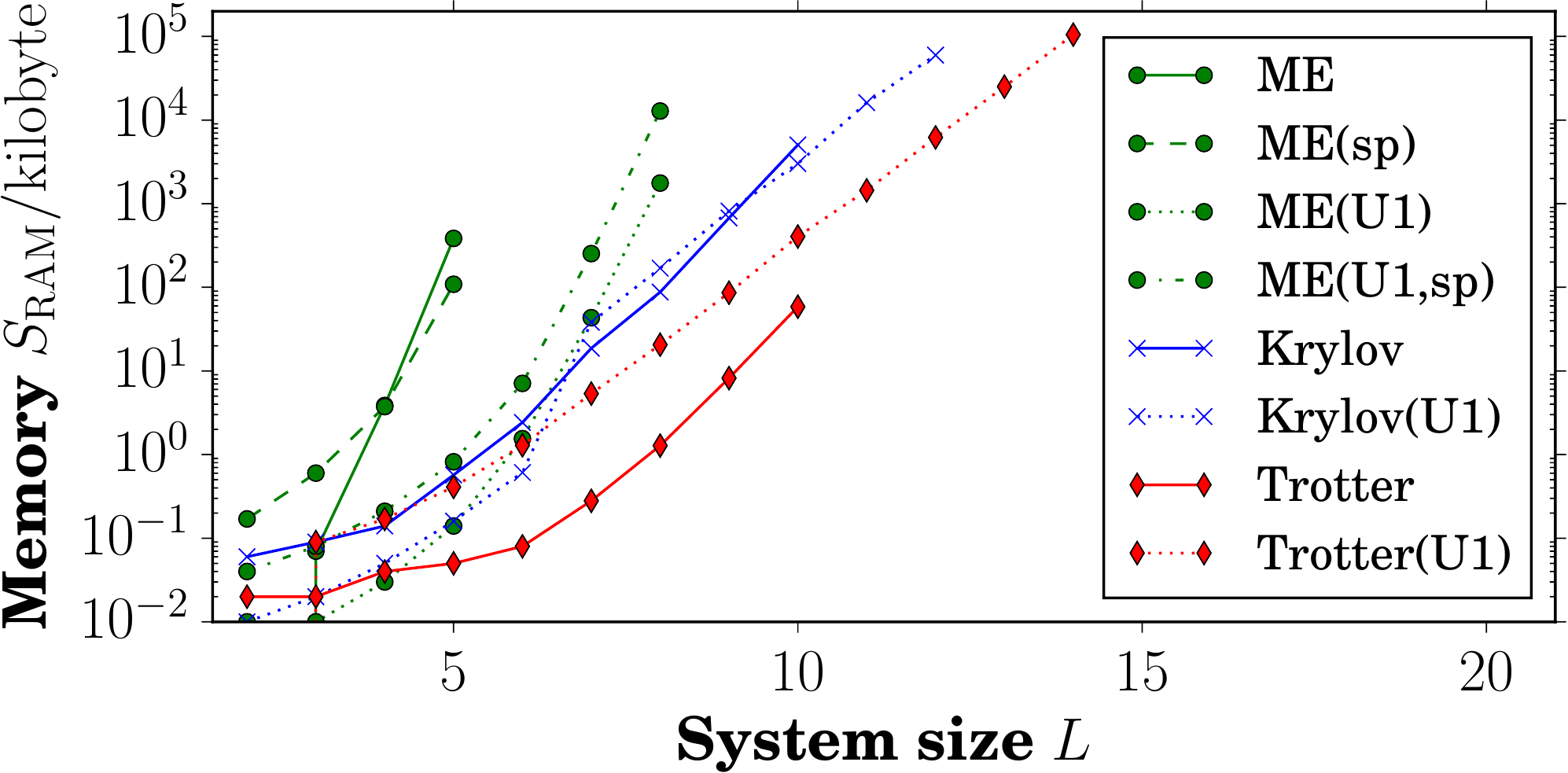}
        \put( 0,53){(b)}
      \end{overpic}
    \end{minipage}\vspace{0.5cm}
    \begin{minipage}{0.47\linewidth}
      \begin{overpic}[width=1.0 \columnwidth,unit=1mm]{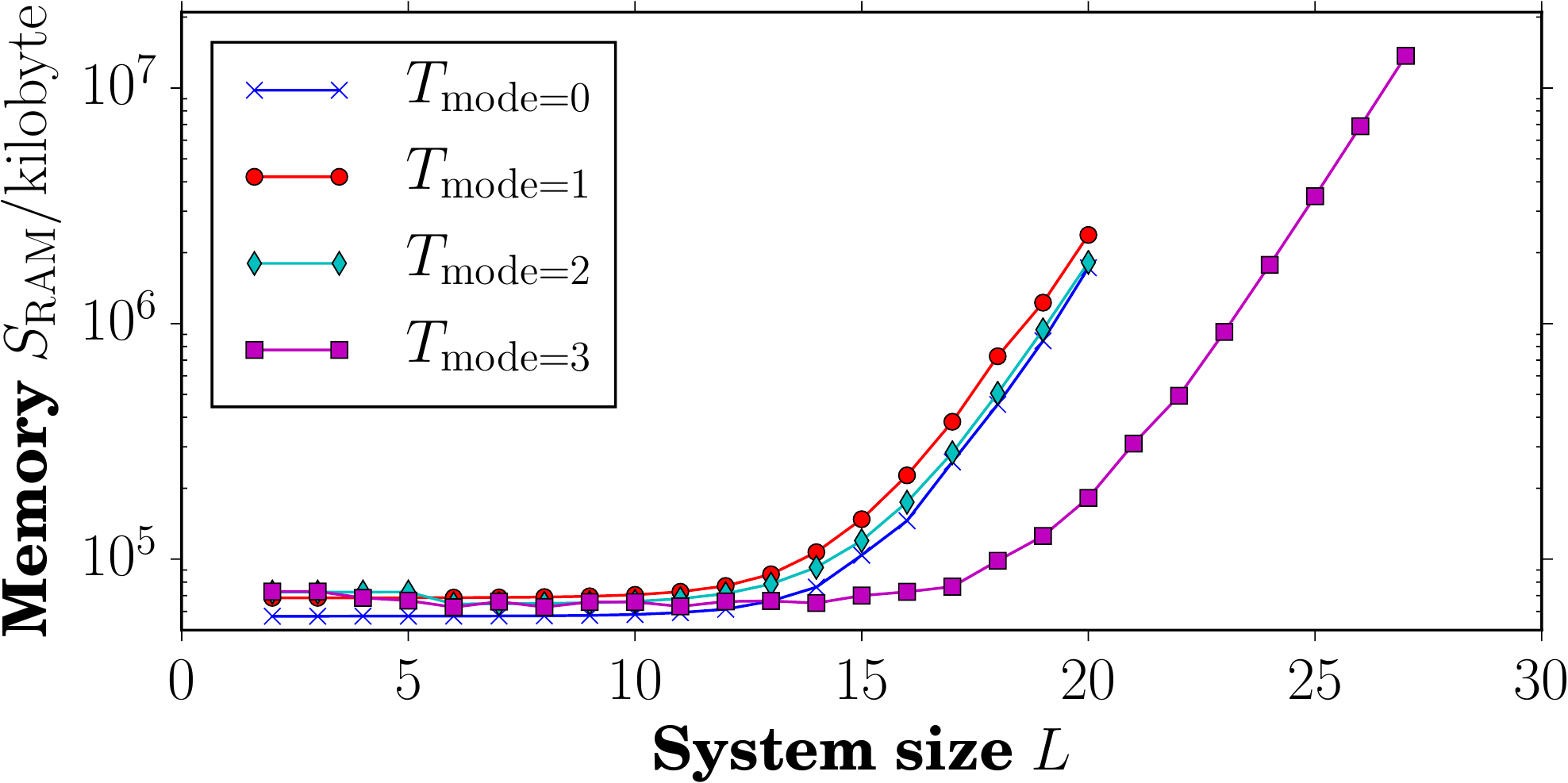}
        \put( 0,53){(c)}
      \end{overpic}
    \end{minipage}\hfill
    \begin{minipage}{0.47\linewidth}
      \begin{overpic}[width=1.0 \columnwidth,unit=1mm]{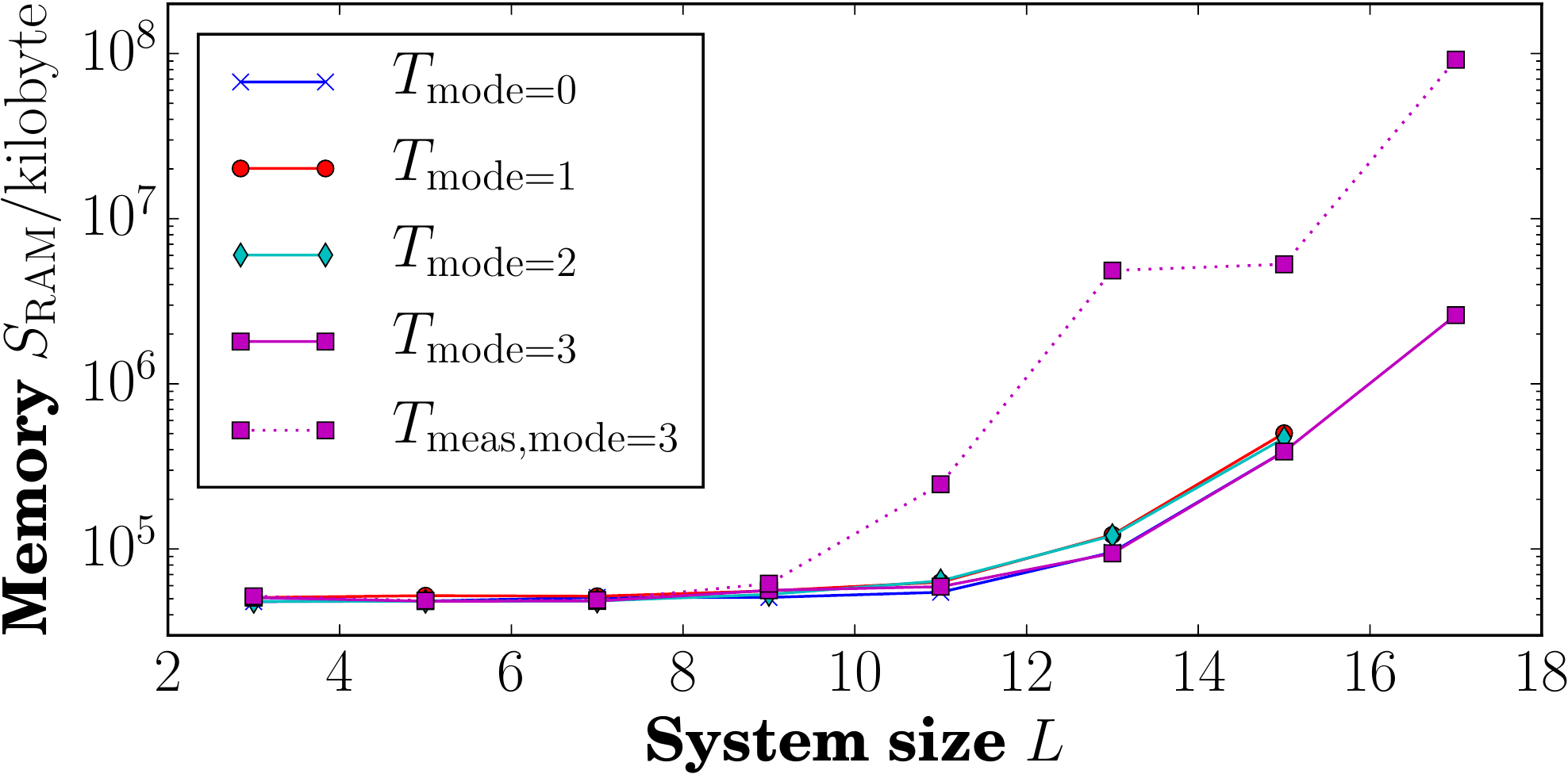}
        \put( 0,53){(d)}
      \end{overpic}
    \end{minipage}
    \caption[Scaling of the Memory for Time Evolution Methods.]
    {\emph{Scaling of the Memory for Time Evolution Methods.} We
      profile five time steps for different system sizes $L$ comparing the
      matrix exponential (ME) with sparse matrices (sp) and with dense
      matrices to the Trotter decomposition and to the default Krylov
      method. The maximal memory of the process in kilobytes is plotted on a
      lin-log-scale.
      (a)~The nearest neighbor quantum Ising model has a $\mathbb{Z}_{2}$
      symmetry, which we use in comparison to the methods without symmetry.
      (b)~The Bose-Hubbard model is considered with a local dimension of
      $d = 5$, and the $\mathcal{U}(1)$ symmetry is used with unit filling
      in the simulations marked with $U1$.
      (c)~The scaling of the RAM for the different Krylov modes in the
      quantum Ising model without using its $\mathbb{Z}_{2}$ symmetry.
      The memory-optimized mode 3 clearly saves resources for large
      systems.
      (d)~The memory requirements for the Bose-Hubbard model in a
      double-well at filling $N = (L - 1) / 2$ with $N$ odd are
      dominated by the measurement of the site and bond entropy. Without
      measurements, the Krylov modes show different RAM requirements.
                                                                                \label{ed:fig:ising_scal_ramstep}}
  \end{center}
\end{figure}

We have discussed in Fig.~\ref{ed:fig:ising_scal_timestep} the scaling of the
computation time for the different time evolution methods. In this appendix,
we discuss the corresponding scaling of the memory resources, shown in
Fig.~\ref{ed:fig:ising_scal_ramstep}. We use the python package \texttt{resource}
to track the maximal memory usage over the complete calculation.

The matrix exponential has the highest demands on memory as the complete
Hamiltonian has to be stored, see Fig.~\ref{ed:fig:ising_scal_ramstep}(a). The
two different set of curves distinguish the matrix exponential with and without
$\mathbb{Z}_{2}$ symmetry. The symmetry has a smaller Hilbert space by a
factor of $2$ and thus lower memory requirements. The difference between
dense and sparse methods is minimal within the quantum Ising model.

In contrast, both the Krylov and the Trotter method show higher memory
requirements for the symmetry-conserving. We explain this trend with the
fact that the mapping for the symmetry must be stored, overcoming the trend
that the state vector itself is smaller for systems with $\mathbb{Z}_{2}$
symmetry. The lower memory for the Trotter without symmetry is due to the
Krylov methods storing a set of the Krylov vectors.
Figure~\ref{ed:fig:ising_scal_ramstep}(b) shows the same trend for the
Bose-Hubbard model, where we recall that the $\mathcal{U}(1)$ symmetry
takes a much smaller fraction of the Hilbert space in comparison to the
quantum Ising model with and without symmetry.

We have the memory-optimized algorithms for the Krylov algorithm in our library
replacing the scipy implementation. Therefore, we take a look at the memory
consumption of the different Krylov modes; Fig.~\ref{ed:fig:ising_scal_ramstep}(c)
considers the quantum Ising model without using the $\mathbb{Z}_{2}$ symmetry.
For example, the memory requirements for $24$ qubits with mode 3 are below the
memory used for $20$ qubits in the other modes. This increased number of qubits
is a clear improvement.
The Bose-Hubbard double-well problem, see Fig.~\ref{ed:fig:ising_scal_ramstep}(d),
is the only simulation pushing the limits past laptop and desktop computers.
Peaking at around 91GB, such simulations are limited to HPC environments. This high memory
demand is related to the measurements. Therefore, we concentrate on simulations
without the measurement of the site and bond entropy to distinguish the different
Krylov modes. We conclude that the third mode a priori saves memory for systems
without symmetry.

\section{Quantum Cellular Automata and Quantum Gates                           \label{ed:app:gates}}

The algorithms described so far evolve the quantum state or density
matrix according to the differential equation presented in
Eqs.~\eqref{ed:eq:schroedinger}, \eqref{ed:eq:lindbladme}, or \eqref{ed:eq:rhovonneumann}.
The methods using the full matrix exponential and the Trotter decomposition
calculate the propagator, which we apply to the quantum state. But some
problems in quantum physics are directly formulated in terms of propagators,
i.e., quantum cellular automata~\cite{Watrous1995,Hillberry2016} and quantum
gates. Quantum gates are especially common in
quantum information theory: most researchers know the quantum Fourier
transformation defined by their gates~\cite{NielsenChuang}, but not
the corresponding Hamiltonian.

\begin{figure}[t]
  \begin{center}
      \begin{overpic}[width=0.5 \columnwidth,unit=1mm]{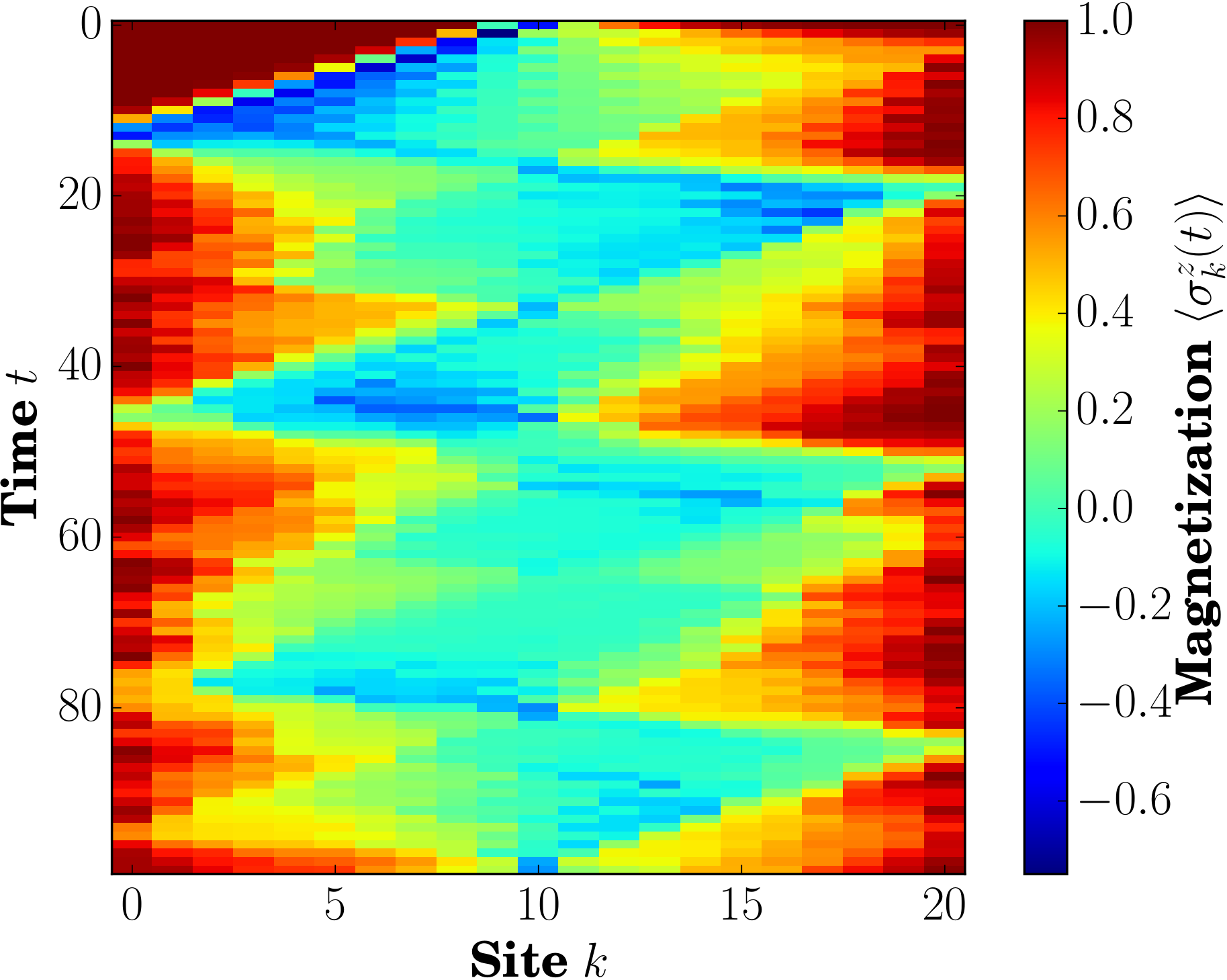}
      \end{overpic}
    \caption[Quantum Cellular Automata example for Quantum Circuits.]
    {\emph{Quantum Cellular Automata example for Quantum Circuits.}
      Asymmetric SWP Quantum Cellular Automata rule for $L = 21$ sites is one
      example for the application of quantum circuits.
                                                                                \label{ed:fig:qeca_swp}}
  \end{center}
\end{figure}

As the major steps are already implemented with the application of the
propagator to a state, we also provide the tools to formulate dynamics
solely in terms of quantum gates defined on quasi-local gates. The time
evolution is then formulated as one quantum circuit consisting of multiple
sub-circuits. We measure after each sub-circuit, which can be
different. Each sub-circuit is a series of gates $G_1$ to $G_{K}$:
\begin{eqnarray}
  \ket{\psi'} = G_{K} G_{K-1} \cdots G_{2} G_{1} \ket{\psi} \, ,
\end{eqnarray}
where $\ket{\psi'}$ is the wave-function after the sub-circuit. We set up
a QECA rule, i.e., the SWP-rule 6 for $L = 21$ sites \cite{Hillberry2016}.
The initial state is a single spin down in the middle of the system.
Figure~\ref{ed:fig:qeca_swp} describes the evolution according to the rule.
We waive a detailed study of the scaling for the quantum circuits or gates.
On the one hand, such an evolution without symmetries can be easily estimated
by the contraction of the gates to the state. On the other hand, even gates
obeying a symmetry and acting on two sites have an upper bound of the Trotter
time evolution up to some factor for the number of applications.

In conclusion, this evolution tool comes at almost no additional cost for the
implementation. Although we have at present only gates in the OSMPS library,
it can be generalized to Kraus operators \cite{NielsenChuang} as the next step,
and QECA evolutions with Kraus operators have already been suggested in
\cite{Brennen2003}.

\section{Best-Practice Example                                                 \label{ed:app:example}}

\begin{figure}[t]
  \begin{center}
    \vspace{0.5cm}
    \begin{overpic}[width=0.75 \columnwidth,unit=1mm]{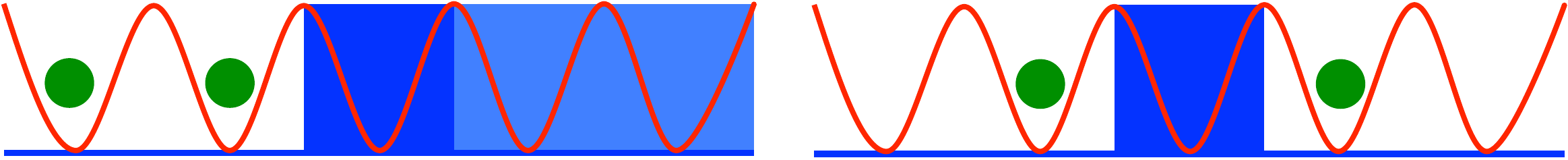}
      \put(-1, 12){(a) $\quad t = 0$}
      \put(50, 12){(b) $\quad t > 0$}
    \end{overpic}
    \caption[Bose-Hubbard Double-Well Problem.]
    {\emph{Bose-Hubbard Double-Well Problem.} Two bosons
      occupy five sites. The first two sites represent the left well, the
      center site represents the barrier, and the two last sites are the
      right well.
      (a)~The initial state at time $t = 0$ is the ground state of the system
      with a potential on the middle site (3rd site, dark blue) and the right
      well (4th and 5th site, light blue).
      (b)~For times $t > 0$, we switch of the potential on the right well
      and the bosons oscillate between the left and right well.
                                                                                \label{ed:fig:DoubleWell}}
  \end{center}
\end{figure}

In this appendix, we concentrate on the setup of an actual simulation for the
exact diagonalization library within \OSMPS{}. We choose the double-well
problem using the Bose-Hubbard model with the Hamiltonian introduced in
Eq.~\eqref{ed:eq:HDW} and compare the behavior of the closed system versus the
open system. Figure~\ref{ed:fig:DoubleWell} represents a sketch of the problem.
The first step of any \py{} program is importing the
necessary packages. We need the \py{} packages for MPS, the EDLib,
numpy, a function for copying classes, the interface to the system parameters,
and the plotting library. We show the corresponding code snippet always in the
following Listing. Listing~\ref{py:import} shows the importing of the packages.

\pyfrag{./code/DoubleWell.py}
       {Import the packages.}
       {py:import}{1}{6}

The next parts of the code are in the main function. First, we generate the
operators for the Bose-Hubbard model. In addition to the set of bosonic
operators, \OSMPS{} can provide fermionic operators and spin operators. We
choose a maximum of four bosons per site leading to a local dimension of $d = 5$.
Furthermore, we have to construct the interaction operator $n (n - \1)$.
Matrices within \texttt{Operators} are accessed via string keys following
the dictionary structure of \py{}. Similar new operators are defined as a
new entry in the dictionary.

\pyfrag{./code/DoubleWell.py}
       {Generate the operators for the Bose-Hubbard model.}
       {py:operators}{18}{22}

In the next step, we build the Hamiltonian consisting of the tunneling term
$b_{k} b_{k+1}^{\dagger} + h.c.$, the interaction $n (n - \1)$, and a site-dependent
potential to model the double-well and its action through the number
operator $n$. The argument \texttt{hparam} is the coupling for each term
in the Hamiltonian defined as a string. The actual value is specified later on
for each simulation allowing for flexibility, which is then shown in
Listing~\ref{py:cparams}. We define all values, which are constant for
all simulations directly as variables. We make a copy of the MPO for the
open system and add the Lindblad term.

\pyfrag{./code/DoubleWell.py}
       {Create the Hamiltonian via rule sets.}
       {py:mporuleset}{24}{37}

As a next step, the measurements are defined. We restrict ourselves to the local
number operator, which allows us to show the oscillations between the two
wells. The arguments in \texttt{AddObservable} specify the type of
measurement, if applicable, the operators used, and a string identifier to
access the results in a dictionary later on.

\pyfrag{./code/DoubleWell.py}
       {Define the observables.}
       {py:observables}{39}{41}

We define the dynamics as objects of the \texttt{QuenchList} class, one
for the closed and one for the open system. We start with the potential
with the barrier in the middle of the system and define a function, which
returns the time-independent potential. For the time evolution, we use
the Krylov method. The time-dependency \texttt{timedep} must be set by hand
in this case.

\pyfrag{./code/DoubleWell.py}
       {Define the dynamics via the QuenchList class.}
       {py:quenches}{43}{60}

The simulations to be carried out are specified in a list of dictionaries.
We start with the dictionary for the closed system. The necessary keys can
be found in the documentation, but we point out that we now specify the
\texttt{hparam} used in the definition of the rule sets for the Hamiltonian.

\pyfrag{./code/DoubleWell.py}
       {Add the closed system to the list of simulations.}
       {py:cparams}{62}{79}

Since we do not iterate over different system sizes, we can construct the
basis states for the $\mathcal{U}(1)$ symmetry once prior to running the
simulations. Although EDLib would build the symmetry sector for each
simulation automatically, it would use a general method, which is not
necessary in this case. The class containing the basis is constructed via
the following Listing~\ref{py:symmsec}.

\pyfrag{./code/DoubleWell.py}
       {Construct the basis states for a single $\mathcal{U}(1)$ symmetry.}
       {py:symmsec}{81}{83}

In the next step, we iterate two different coupling for the Lindblad
operators. We can append the additional dictionaries to the list.

\pyfrag{./code/DoubleWell.py}
       {Add open system simulations with two different couplings to the list
       of simulations.}
       {py:oparams}{85}{105}

Since we use different rule sets for the MPO in the simulations, we need a
list of MPOs with the same length as the simulation list. Afterwards, we can
execute the simulations.

\pyfrag{./code/DoubleWell.py}
       {Build a list of Hamiltonians and run simulations.}
       {py:mpolistexecute}{107}{113}

The post-processing imports the results as dictionaries. For the statics, the
list contains the dictionaries directly, for dynamics the first list iterates
over the different simulations and the inner list iterates over the
dictionaries for each time step.

\pyfrag{./code/DoubleWell.py}
       {Reading the results and plotting them.}
       {py:postprocess}{115}{156}

\begin{figure}[t]
  \begin{center}
    \vspace{0.5cm}
    \begin{overpic}[width=0.8 \columnwidth]{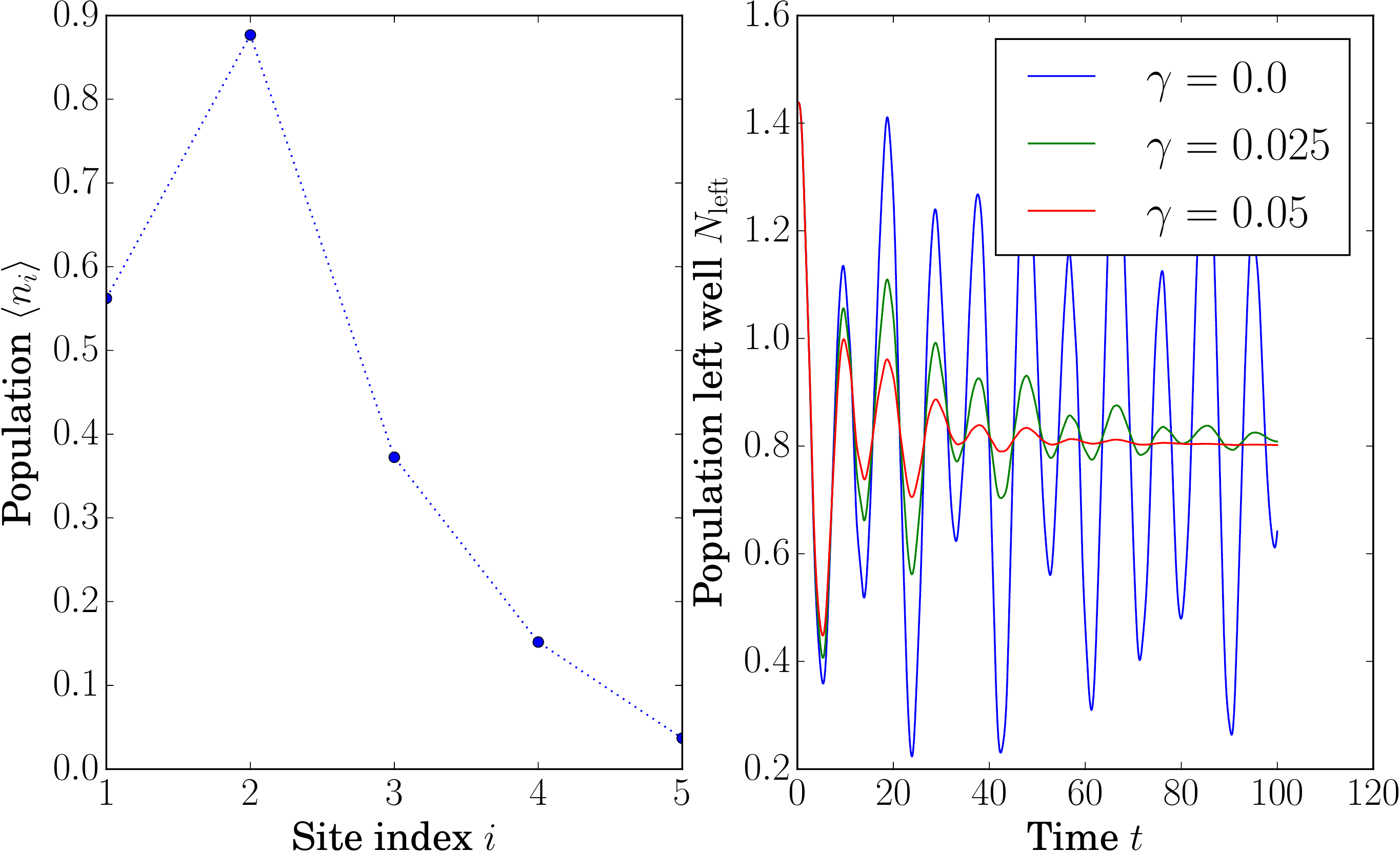}
      \put(0,  62){(a)}
      \put(50, 62){(b)}
    \end{overpic}
    \caption[Closed and Open System Double-Well Problem.]
    {\emph{Closed and Open System Double-Well Problem.}
      (a)~The initial state is characterized  by the population per site
      on the left. We obtain it via an eigendecomposition of the
      Hamiltonian, where only the left well has a zero chemical potential.
      (b)~The oscillations between the two wells damp faster
      for larger coupling $\gamma$ to the reservoir of the Lindblad master
      equation. The limit $\gamma = 0$ is the closed system. The time $t$ is
      in units of the tunneling $J$.
                                                                                \label{ed:fig:A1_Example}}
  \end{center}
\end{figure}
       
The plot obtained from this simulation is presented in Fig.~\ref{ed:fig:A1_Example}
and shows that the coupling considered in the Lindblad master equation leads
to a damping of the oscillations between the left and right well. The last
lines of the code call the main function in case the script is called, which
allows us in addition to import the function from another module.

\pyfrag{./code/DoubleWell.py}
       {Calling the main function.}
       {py:callmain}{159}{166}

\section{Details of Convergence Studies                                        \label{ed:app:addons}}

\begin{figure}[t]
  \begin{center}
    \vspace{0.8cm}\begin{minipage}{0.47\linewidth}
      \begin{overpic}[width=1.0 \columnwidth,unit=1mm]{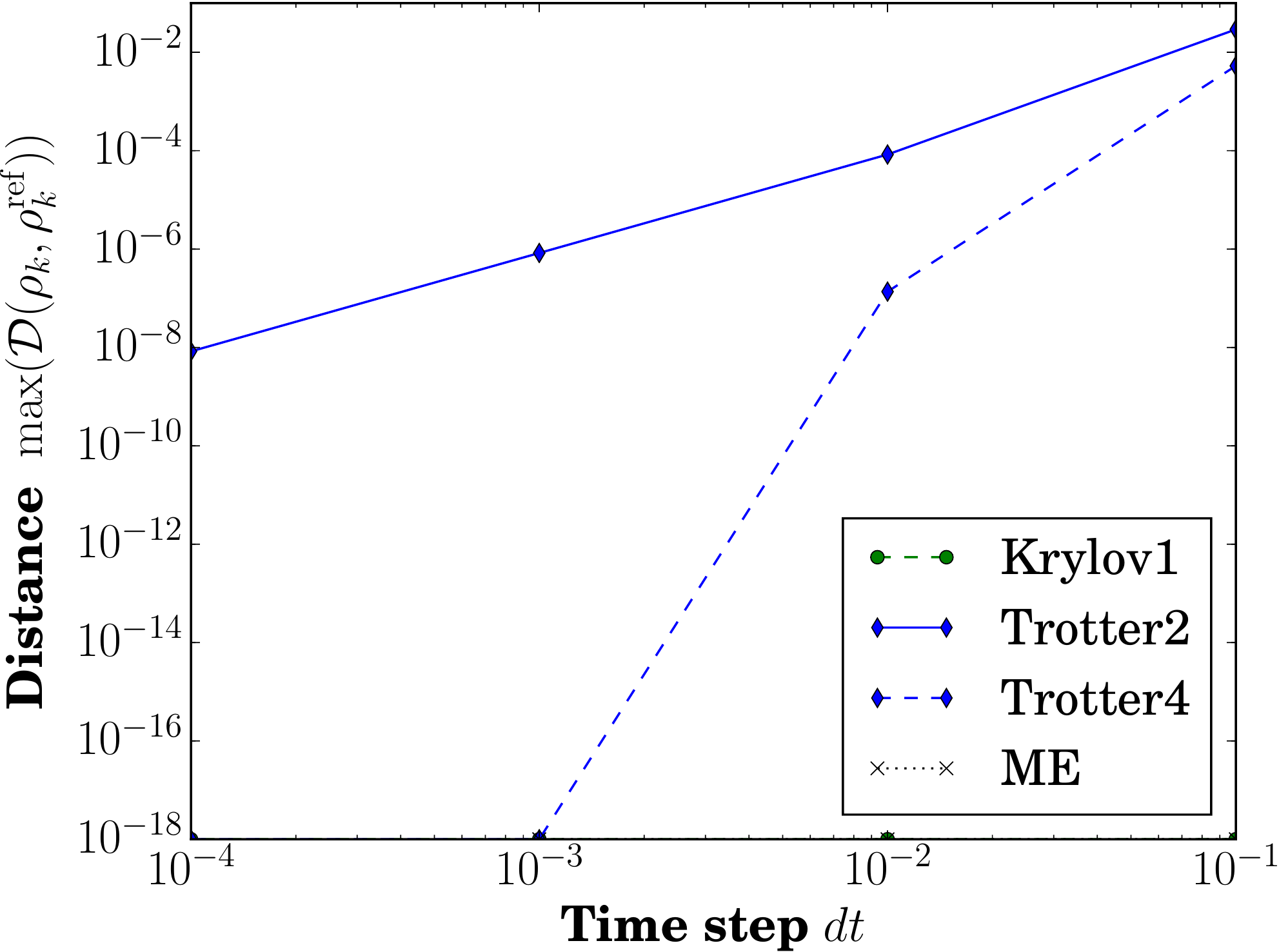}
        \put( 0,78){(a)}
      \end{overpic}
    \end{minipage}\hfill
    \begin{minipage}{0.47\linewidth}
      \begin{overpic}[width=1.0 \columnwidth,unit=1mm]{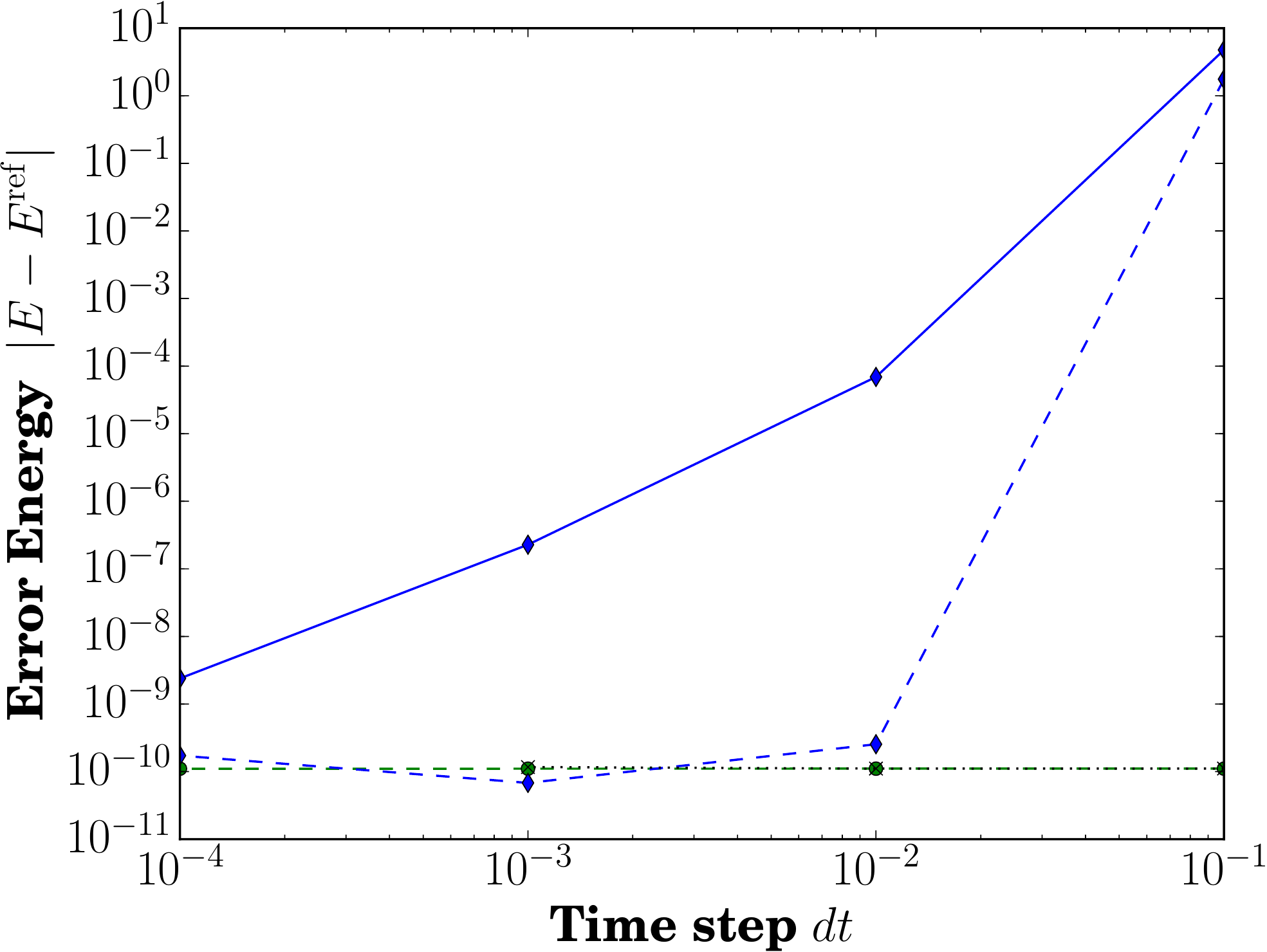}
        \put( 0,78){(c)}
      \end{overpic}
    \end{minipage}\vspace{0.4cm}
    
    \begin{minipage}{0.47\linewidth}
      \begin{overpic}[width=1.0 \columnwidth,unit=1mm]{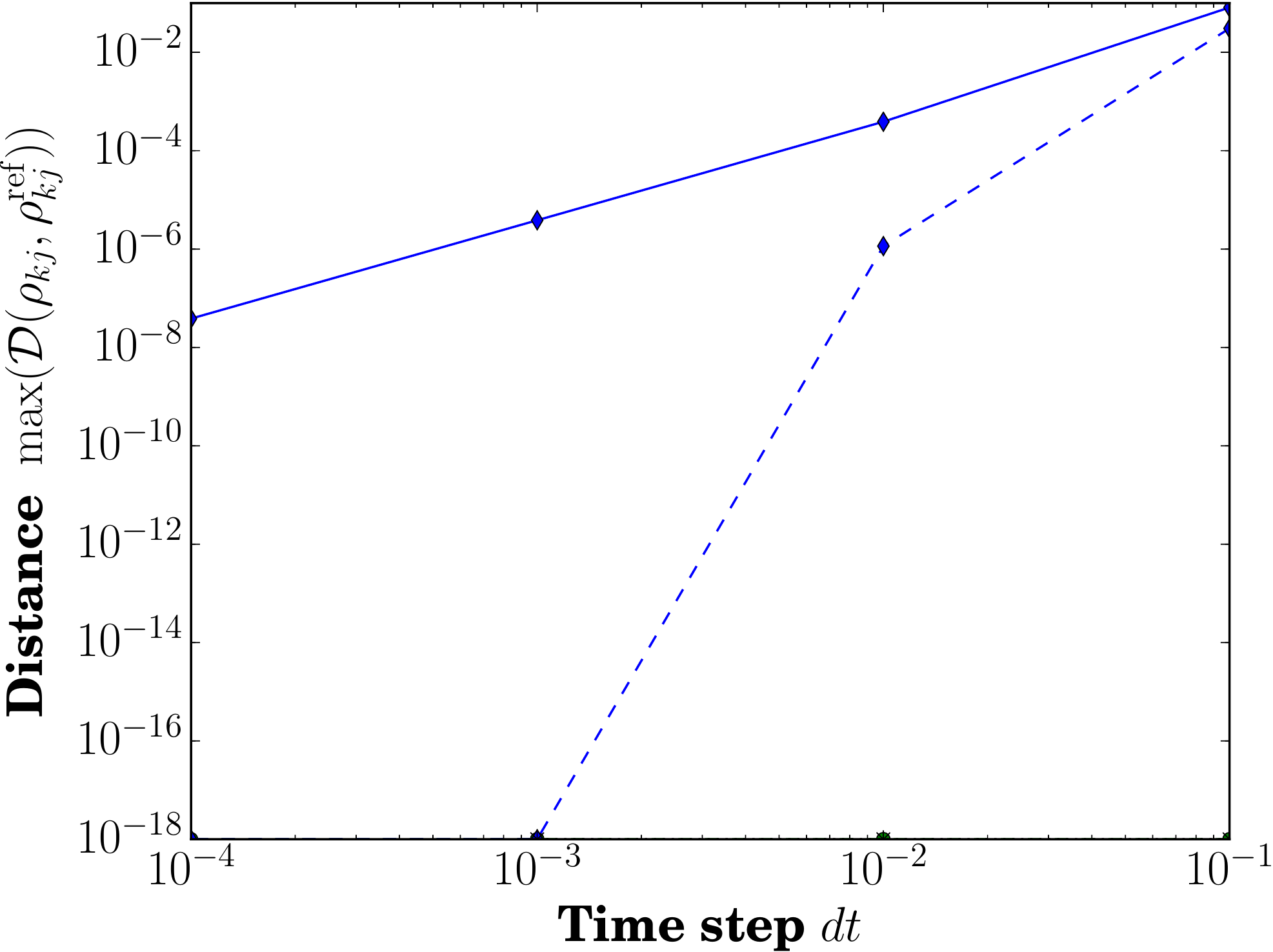}
        \put( 0,78){(b)}
      \end{overpic}
    \end{minipage}\hfill
    \begin{minipage}{0.47\linewidth}
      \begin{overpic}[width=1.0 \columnwidth,unit=1mm]{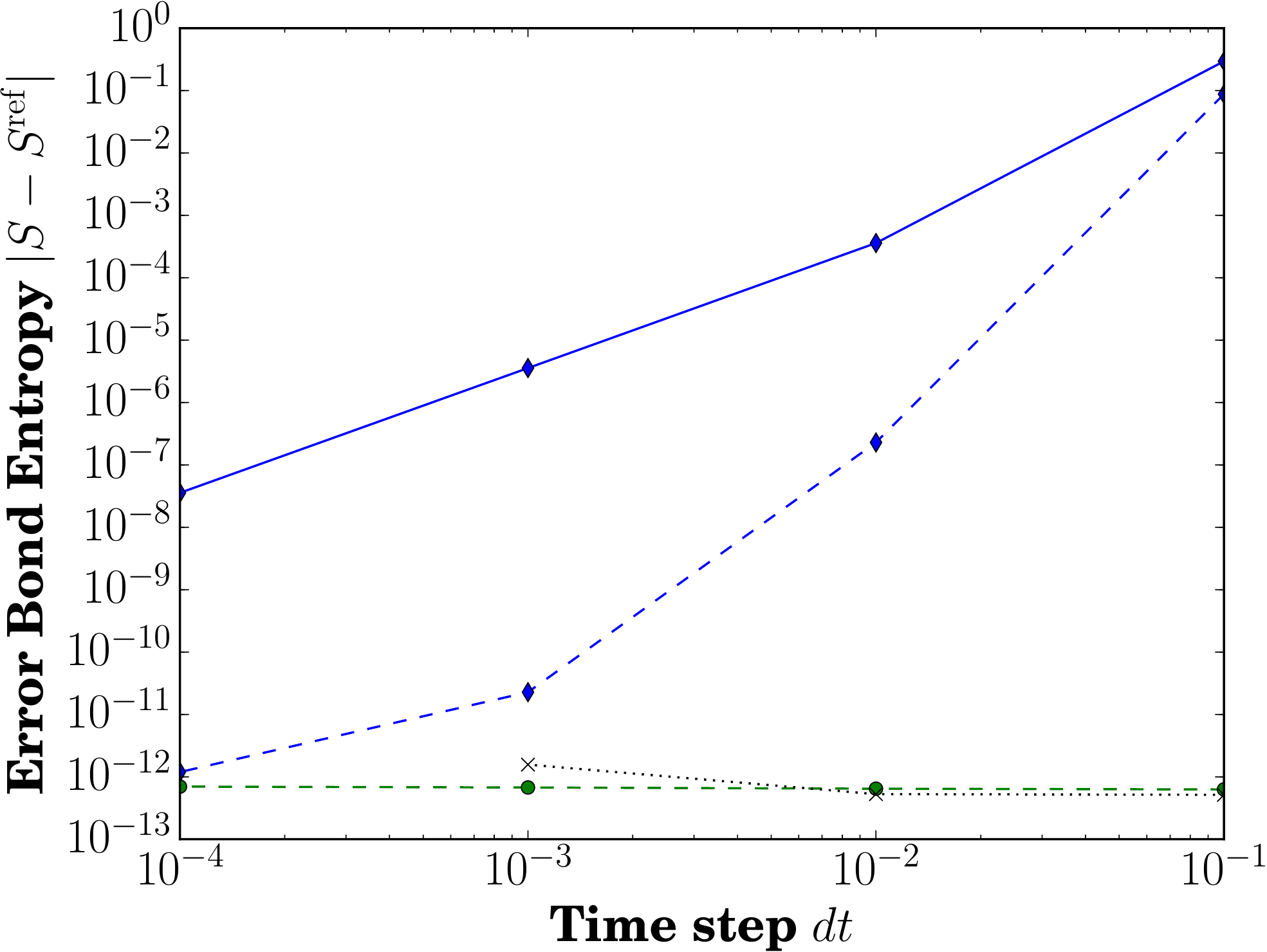}
        \put( 0,78){(d)}
      \end{overpic}
    \end{minipage}
    \caption[Convergence of Methods for the time-independent Bose-Hubbard Model.]
    {\emph{Convergence of Methods for the time-independent
      Bose-Hubbard Model.} We compare the 2$^{\mathrm{nd}}$
      and 4$^{\mathrm{th}}$ order Trotter time evolution and the Krylov method
      with mode $1$ against the reference (ref) taken as the matrix exponential (ME) with
      $dt = 0.0001$ for various measures. We consider the Bose-Hubbard model
      in a linear quench from on-site interaction strength $U(t=0) = 10.0$ to
      $U(t=0.5) = 8.0$ for a system size of $L = 6$ and unit filling. The
      reference value is the matrix exponential for the smallest time step
      $dt = 0.0001$. The minimal error is set to $10^{-18}$ indicating machine
      precision. The measures are (a) the minimal distance over all
      single-site reduced density matrices, (b) the minimal distance over all
      two-site density matrices, (c) the error in energy, and (d) the error
      in the bond entropy for the splitting in the middle of the system.
                                                                                \label{ed:fig:bose_ssu}}
  \end{center}
\end{figure}

\begin{figure}[t]
  \begin{center}
    \vspace{0.8cm}\begin{minipage}{0.47\linewidth}
      \begin{overpic}[width=1.0 \columnwidth,unit=1mm]{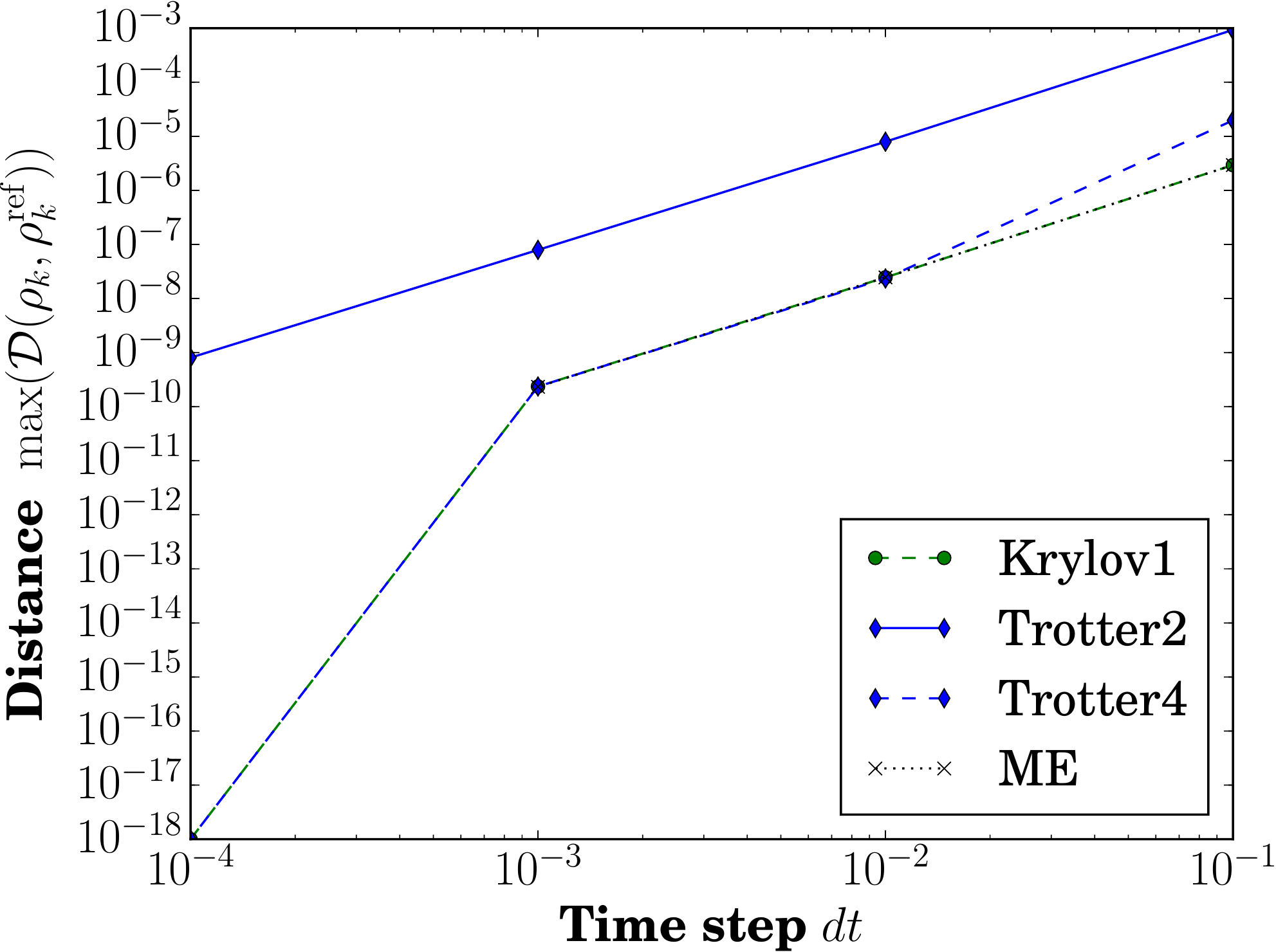}
        \put( 0,78){(a)}
      \end{overpic}
    \end{minipage}\hfill
    \begin{minipage}{0.47\linewidth}
      \begin{overpic}[width=1.0 \columnwidth,unit=1mm]{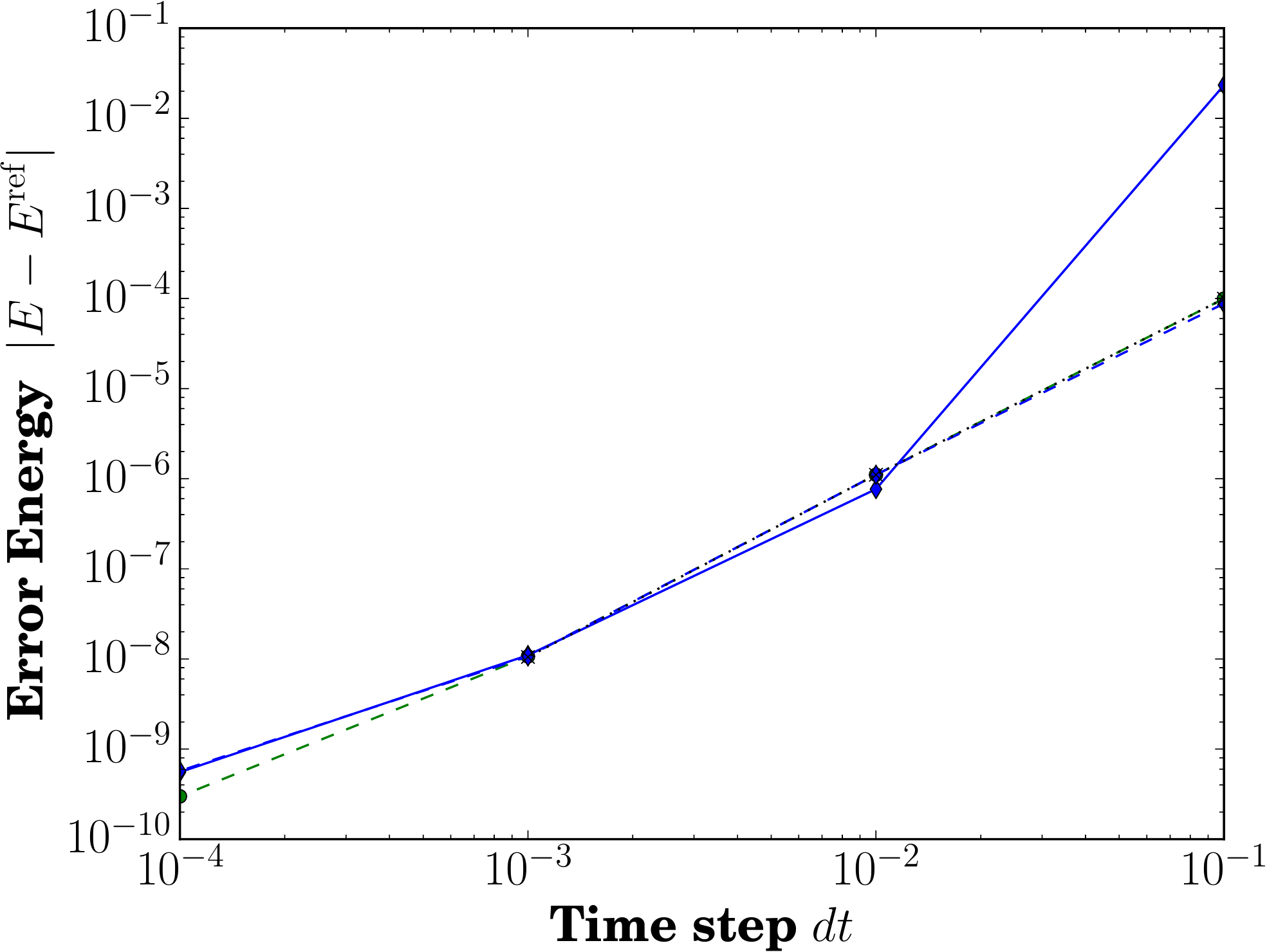}
        \put( 0,78){(c)}
      \end{overpic}
    \end{minipage}\vspace{0.4cm}
    
    \begin{minipage}{0.47\linewidth}
      \begin{overpic}[width=1.0 \columnwidth,unit=1mm]{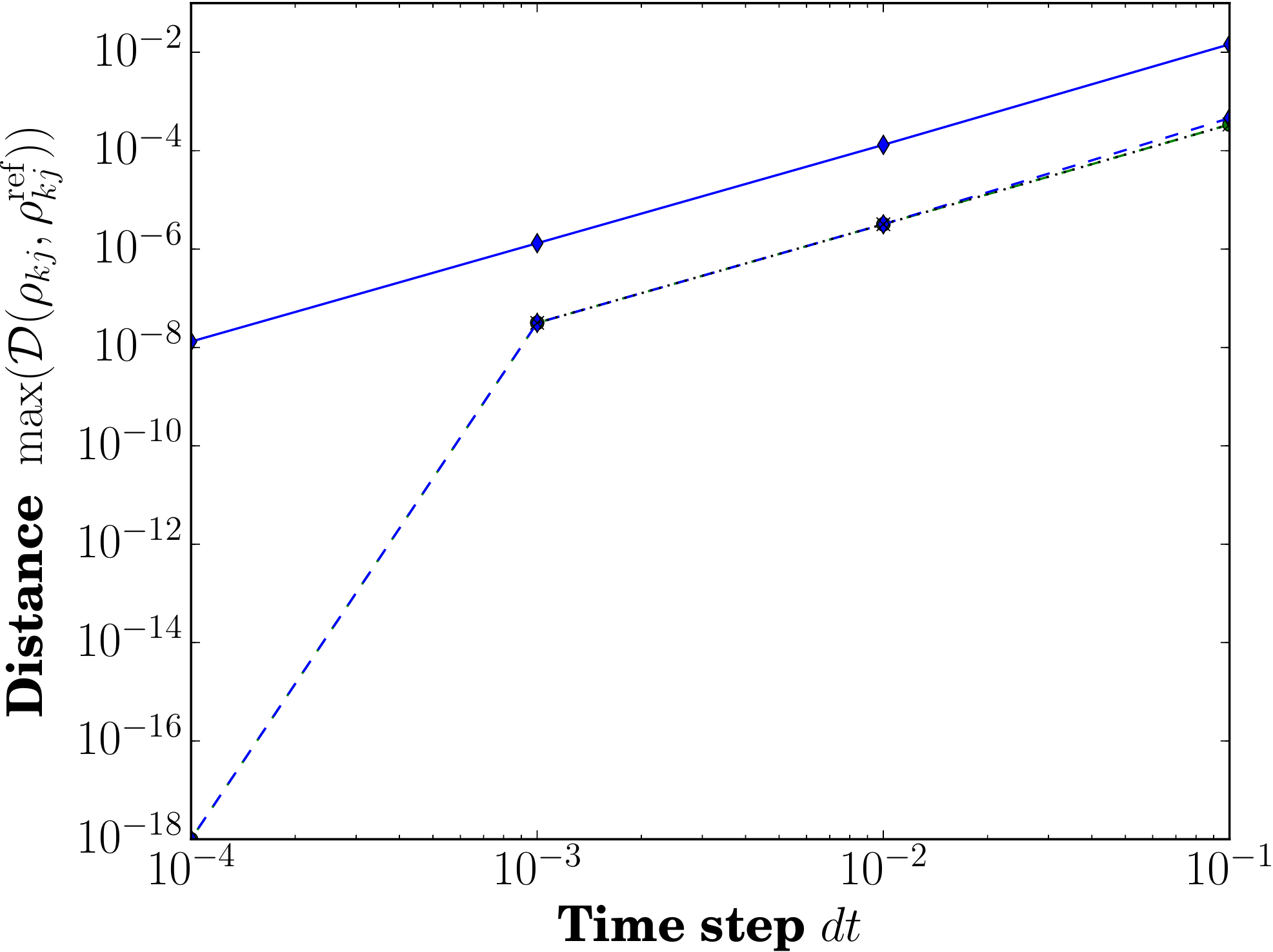}
        \put( 0,78){(b)}
      \end{overpic}
    \end{minipage}\hfill
    \begin{minipage}{0.47\linewidth}
      \begin{overpic}[width=1.0 \columnwidth,unit=1mm]{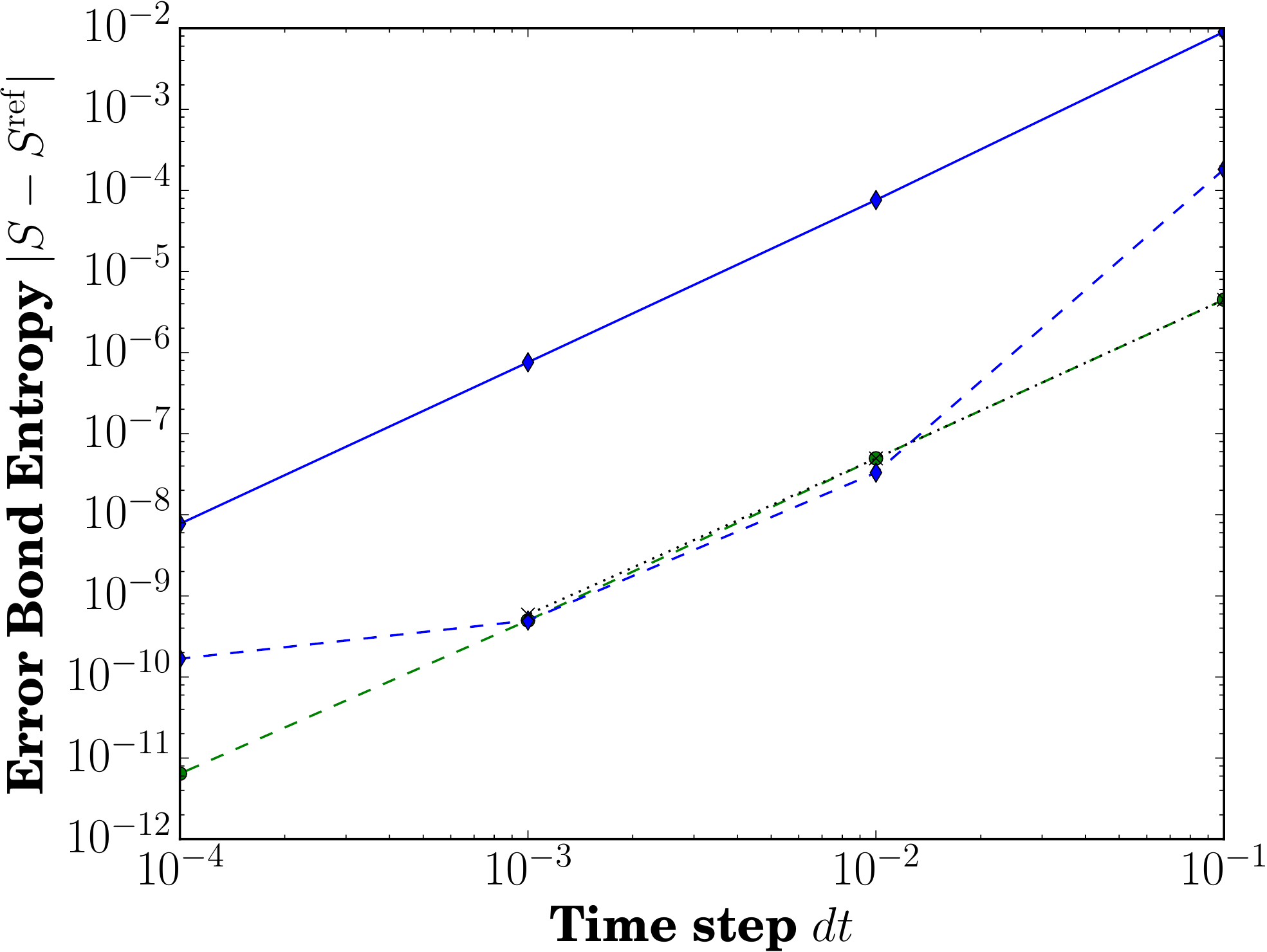}
        \put( 0,78){(d)}
      \end{overpic}
    \end{minipage}
    \caption[Convergence of Methods for the Quantum Ising Model.]
    {\emph{Convergence of Methods for the Quantum Ising Model.} We
      compare the 2$^{\mathrm{nd}}$ and 4$^{\mathrm{th}}$ order Trotter time
      evolution and the Krylov method in mode 1 and the matrix exponential
      (ME). The reference (ref) is the matrix exponential with $dt = 0.001$ for all
      four measures. We consider the quantum Ising model using its
      $\mathbb{Z}_{2}$ symmetry in a linear quench from external field
      $h(t=0) = 5.0$ to $h(t=0.5) = 4.5$ for a system size of $L = 10$.
      The measures are (a) the minimal distance over all
      single-site reduced density matrices, (b) the minimal distance over all
      two-site density matrices, (c) the error in energy, and (d) the error
      in the bond entropy for the splitting in the middle of the system.
                                                                                \label{ed:fig:ising_qpz}}
  \end{center}
\end{figure}

In this appendix, we extend the convergence studies of the main body of this
article, i.e., Sec.~\ref{ed:sec:conv}. We showed the dependence of the error on
the method and the time step
in Fig.~\ref{ed:fig:bose_qsu} for a time-dependent Hamiltonian in the
Bose-Hubbard model. Figure~\ref{ed:fig:bose_ssu} shows the same kind of study for
the time-independent Hamiltonian corresponding to a sudden quench within the
Mott insulator as described in the second scenario in Sec.~\ref{ed:sec:conv}. The
error from the second order time ordering vanishes and only the Trotter
decompositions remain with an error depending on the time step $dt$. In
addition, we present the first scenario with a time-dependent Ising
Hamiltonian in Fig.~\ref{ed:fig:ising_qpz}. We calculate again the rate of
convergence according to Eq.~\eqref{ed:eq:convrate} and the maximal distance
between the single site density matrices. Those rates are $2.07$, $2.0$
for the second order Trotter decomposition, $2.93$, $1.97$ for the fourth
order Trotter decomposition, $2.08$, $2.0$ for the Krylov in mode 1, and
the matrix exponential has rates of convergence of $2.08$, $2.02$. The first
value corresponds to the pair $(dt = 0.01, dt' = 0.1)$ and the second value
$(dt = 0.001, dt' = 0.01)$. We confirm that the convergence rate of $dt^2$
is reproduced in the Ising model due to the error in time-ordering.

For the open system analysis, we studied the error of the evolution of the
von Neumann equation for the density matrix in a closed system and the
Lindblad master equation for selected time evolution schemes in
Tables~\ref{ed:tab:closedcavities} and \ref{ed:tab:opencavities}.
Tables~\ref{ed:tab:closedapp} and \ref{ed:tab:openapp} show the errors of
different settings. We conclude that the 4$^{\mathrm{th}}$ order Trotter
evolution reduces the error to almost machine precision. The built-in
scipy Krylov method delivers a better slightly better result than the
Krylov methods from the EDLib of \OSMPS{}, e.g., the error in energy
is about two orders (one order) of magnitude smaller for the photon
Josephson junctions in the von Neumann (Lindblad master) equation,
defined in Eq.~\eqref{ed:eq:pjjlindblad}. In
Table~\ref{ed:tab:openapp} we point out that the errors for the quantum
trajectories do not depend on the method. We explain this by the fact
that the seed of the pseudo-random number generator is fed with the
ID of the trajectory and therefore the trajectories for each method are
equal. Also, the error from the quantum trajectory approach is bigger than
the error from the time evolution method and, therefore, the latter error is
not visible in the table.

\begin{table}[t]
  \centering
  \begin{tabular}{@{} cccccccc @{}}
    \toprule
    Model              &Error     &Liou (Trotter-4)     &Liou (Krylov-0)      &Liou (Krylov-2)      &Liou (Krylov-3)      &QT1 (Trotter-4)      &QT1 (Krylov-0)       \\ 
    \cmidrule(r){1-1} \cmidrule(rl){2-2} \cmidrule(rl){3-3} \cmidrule(rl){4-4} \cmidrule(rl){5-5} \cmidrule(rl){6-6} \cmidrule(rl){7-7} \cmidrule(l){8-8}
    $H_{\mathrm{PJJ}}$ &$\erhoi$  &$0.0              $  &$0.0              $  &$2.71\cdot 10^{-13}$ &$2.71\cdot 10^{-13}$ &$0.0               $ &$0.0              $  \\ 
                       &$\erhoij$ &$0.0              $  &$0.0              $  &$7.34\cdot 10^{-13}$ &$7.34\cdot 10^{-13}$ &$0.0               $ &$0.0              $  \\ 
                       &$\eener$  &$2.71\cdot 10^{-13}$ &$7.33\cdot 10^{-15}$ &$5.52\cdot 10^{-13}$ &$5.52\cdot 10^{-13}$ &$1.16\cdot 10^{-13}$ &$4.0\cdot 10^{-15}$  \\ 
                       &$\eentr$  &$-$                  &$-$                  &$-$                  &$-$                  &$2.45\cdot 10^{-12}$ &$8.88\cdot 10^{-16}$ \\ 
    \cmidrule(r){1-1} \cmidrule(rl){2-2} \cmidrule(rl){3-3} \cmidrule(rl){4-4} \cmidrule(rl){5-5} \cmidrule(rl){6-6} \cmidrule(rl){7-7} \cmidrule(l){8-8}
    $H_{\mathrm{DW}}$  &$\erhoi$  &$6.34\cdot 10^{-13}$ &$8.87\cdot 10^{-14}$ &$8.56\cdot 10^{-14}$ &$8.54\cdot 10^{-14}$ &$9.0\cdot 10^{-14}$  &$8.99\cdot 10^{-14}$ \\ 
                       &$\erhoij$ &$0.0              $  &$2.21\cdot 10^{-13}$ &$7.95\cdot 10^{-14}$ &$1.11\cdot 10^{-13}$ &$9.46\cdot 10^{-14}$ &$3.91\cdot 10^{-13}$ \\ 
                       &$\eener$  &$3.53\cdot 10^{-11}$ &$7.81\cdot 10^{-13}$ &$8.04\cdot 10^{-13}$ &$8.13\cdot 10^{-13}$ &$3.48\cdot 10^{-11}$ &$7.74\cdot 10^{-13}$ \\ 
                       &$\eentr$  &$-$                  &$-$                  &$-$                  &$-$                  &$2.12\cdot 10^{-10}$ &$2.38\cdot 10^{-14}$ \\ 
    \bottomrule
  \end{tabular}
  \caption[Additional Error Analysis for the Density Matrix in the
    von Neumann Equation.]
  {\emph{Additional Error Analysis for the Density Matrix in the
    von Neumann Equation.} We show additional methods extending
    Table~\ref{ed:tab:closedcavities} for simulations of the closed system
    with the open system code of the library and the reference result
    is the closed system solving the Schr\"odinger equation.
                                                                                \label{ed:tab:closedapp}}
\end{table}

\begin{table}[t]
  \centering
  \begin{tabular}{@{} cccccccc @{}}
    \toprule
    Model              &Error     &Liou (Trotter-4)     &Liou (Krylov-0)      &Liou (Krylov-2)      &Liou (Krylov-3)      &QT500 (Trotter-4)    &QT500 (Krylov-0)     \\
    \cmidrule(r){1-1} \cmidrule(rl){2-2} \cmidrule(rl){3-3} \cmidrule(rl){4-4} \cmidrule(rl){5-5} \cmidrule(rl){6-6} \cmidrule(rl){7-7} \cmidrule(l){8-8}
    $H_{\mathrm{PJJ}}$ &$\entim$  &$1.32\cdot 10^{-13}$ &$1.78\cdot 10^{-14}$ &$3.76\cdot 10^{-13}$ &$3.76\cdot 10^{-13}$ &$1.15\cdot 10^{-02}$ &$1.15\cdot 10^{-02}$ \\
                       &$\erhoi$  &$4.88\cdot 10^{-15}$ &$0.0\cdot 10^{+00}$  &$2.31\cdot 10^{-13}$ &$2.31\cdot 10^{-13}$ &$1.49\cdot 10^{-05}$ &$1.49\cdot 10^{-05}$ \\
                       &$\erhoij$ &$8.99\cdot 10^{-15}$ &$0.0\cdot 10^{+00}$  &$5.21\cdot 10^{-13}$ &$5.21\cdot 10^{-13}$ &$4.48\cdot 10^{-05}$ &$4.48\cdot 10^{-05}$ \\
                       &$\eener$  &$2.09\cdot 10^{-14}$ &$1.71\cdot 10^{-14}$ &$3.77\cdot 10^{-13}$ &$3.77\cdot 10^{-13}$ &$1.15\cdot 10^{-02}$ &$1.15\cdot 10^{-02}$ \\
    \cmidrule(r){1-1} \cmidrule(rl){2-2} \cmidrule(rl){3-3} \cmidrule(rl){4-4} \cmidrule(rl){5-5} \cmidrule(rl){6-6} \cmidrule(rl){7-7} \cmidrule(l){8-8}
    $H_{\mathrm{BH}}$  &$\erhoi$  &$4.85\cdot 10^{-14}$ &$0.0            $    &$0.0            $    &$0.0            $    &$2.04\cdot 10^{-03}$ &$2.04\cdot 10^{-03}$ \\ 
                       &$\erhoij$ &$4.87\cdot 10^{-14}$ &$0.0            $    &$0.0            $    &$0.0            $    &$6.44\cdot 10^{-03}$ &$6.44\cdot 10^{-03}$ \\ 
                       &$\eener$  &$1.15\cdot 10^{-10}$ &$4.09\cdot 10^{-13}$ &$1.17\cdot 10^{-12}$ &$1.16\cdot 10^{-12}$ &$3.75\cdot 10^{-02}$ &$3.75\cdot 10^{-02}$ \\ 
    \bottomrule
  \end{tabular}
  \caption[Additional Error Analysis for the Density Matrix in the
    Lindblad Master Equation.]
  {\emph{Additional Error Analysis for the Density Matrix in the
    Lindblad Master Equation.} We show additional methods extending
    Table~\ref{ed:tab:opencavities} for simulations of the open quantum system
    and the reference result is the open system using the matrix exponential.
                                                                                \label{ed:tab:openapp}}
\end{table}

\section{Add-ons for Open Quantum Systems                                      \label{ed:app:openaddons}}

In the following, we derive the transformation into the Liouville space
given via the equation and already introduced in Eq.~\eqref{ed:eq:lioutrafo}
\begin{eqnarray}
  O' \rho O \rightarrow O' \otimes O^T \Lket{\rho} \, ,
\end{eqnarray}
which we used to build the Liouville operator $\mathcal{L}$ for the time
evolution of an open system. First, we rewrite the equation in terms of
indices\footnote{The indices are not related to the same symbols introduced
previously.} where the elements can be permuted:
\begin{eqnarray}
  O' \rho O
  &=& O_{ij}' \rho_{jk} O_{kl}
   =  O_{ij}' O_{kl} \rho_{jk}
   =  O_{ij}' O_{lk}^{T} \rho_{jk} \, .
\end{eqnarray}
$\rho$ can be transformed into a vector with a combined index $(jk)$. From the
two matrices the combination of indices can be achieved using the tensor
product:
\begin{eqnarray}
  O' \rho O
 &=& O_{ij}' O_{lk}^{T} \Lket{\rho}_{(jk)}
  =  (O' \otimes O^{T})_{(il), (jk)} \Lket{\rho}_{(jk)} \, . 
\end{eqnarray}
To have the index $k$ of $O$ in the columns of the result of the
tensor product, we had transposed $O$ before. We point out that this
definition works independently of row- or column-major order of the
memory.

\end{document}